\documentclass[aps,prd,reprint,superscriptaddress,floatfix,altaffilletter,nofootinbib]{revtex4-2} 
\usepackage{color}
\usepackage{graphicx}
\usepackage{amsmath}
\usepackage{amssymb}
\usepackage{upgreek}  
\usepackage{url}
\usepackage{here}

\begin{document}

\title{Solar neutrino measurements using the full data period \\ of Super-Kamiokande-IV}

\newcommand{\AFFicrr}{\affiliation{Kamioka Observatory, Institute for Cosmic Ray Research, University of Tokyo, Kamioka, Gifu 506-1205, Japan}}
\newcommand{\AFFkashiwa}{\affiliation{Research Center for Cosmic Neutrinos, Institute for Cosmic Ray Research, University of Tokyo, Kashiwa, Chiba 277-8582, Japan}}
\newcommand{\AFFicrronly}{\affiliation{Institute for Cosmic Ray Research, University of Tokyo, Kashiwa, Chiba 277-8582, Japan}}
\newcommand{\AFFipmu}{\affiliation{Kavli Institute for the Physics and
Mathematics of the Universe (WPI), The University of Tokyo Institutes for Advanced Study,
University of Tokyo, Kashiwa, Chiba 277-8583, Japan }}
\newcommand{\AFFmad}{\affiliation{Department of Theoretical Physics, University Autonoma Madrid, 28049 Madrid, Spain}}
\newcommand{\AFFubc}{\affiliation{Department of Physics and Astronomy, University of British Columbia, Vancouver, BC, V6T1Z4, Canada}}
\newcommand{\AFFbu}{\affiliation{Department of Physics, Boston University, Boston, MA 02215, USA}}
\newcommand{\AFFuci}{\affiliation{Department of Physics and Astronomy, University of California, Irvine, Irvine, CA 92697-4575, USA }}
\newcommand{\AFFcsu}{\affiliation{Department of Physics, California State University, Dominguez Hills, Carson, CA 90747, USA}}
\newcommand{\AFFcnm}{\affiliation{Institute for Universe and Elementary Particles, Chonnam National University, Gwangju 61186, Korea}}
\newcommand{\AFFduke}{\affiliation{Department of Physics, Duke University, Durham NC 27708, USA}}
\newcommand{\AFFgifu}{\affiliation{Department of Physics, Gifu University, Gifu, Gifu 501-1193, Japan}}
\newcommand{\AFFgist}{\affiliation{GIST College, Gwangju Institute of Science and Technology, Gwangju 500-712, Korea}}
\newcommand{\AFFuh}{\affiliation{Department of Physics and Astronomy, University of Hawaii, Honolulu, HI 96822, USA}}
\newcommand{\AFFicl}{\affiliation{Department of Physics, Imperial College London , London, SW7 2AZ, United Kingdom }}
\newcommand{\AFFkek}{\affiliation{High Energy Accelerator Research Organization (KEK), Tsukuba, Ibaraki 305-0801, Japan }}
\newcommand{\AFFkobe}{\affiliation{Department of Physics, Kobe University, Kobe, Hyogo 657-8501, Japan}}
\newcommand{\AFFkyoto}{\affiliation{Department of Physics, Kyoto University, Kyoto, Kyoto 606-8502, Japan}}
\newcommand{\AFFliv}{\affiliation{Department of Physics, University of Liverpool, Liverpool, L69 7ZE, United Kingdom}}
\newcommand{\AFFminn}{\affiliation{School of Physics and Astronomy, University of Minnesota, Minneapolis, MN  55455, USA}}
\newcommand{\AFFmiyagi}{\affiliation{Department of Physics, Miyagi University of Education, Sendai, Miyagi 980-0845, Japan}}
\newcommand{\AFFnagoya}{\affiliation{Institute for Space-Earth Environmental Research, Nagoya University, Nagoya, Aichi 464-8602, Japan}}
\newcommand{\AFFkmi}{\affiliation{Kobayashi-Maskawa Institute for the Origin of Particles and the Universe, Nagoya University, Nagoya, Aichi 464-8602, Japan}}
\newcommand{\AFFpol}{\affiliation{National Centre For Nuclear Research, 02-093 Warsaw, Poland}}
\newcommand{\AFFsuny}{\affiliation{Department of Physics and Astronomy, State University of New York at Stony Brook, NY 11794-3800, USA}}
\newcommand{\AFFokayama}{\affiliation{Department of Physics, Okayama University, Okayama, Okayama 700-8530, Japan }}
\newcommand{\AFFosaka}{\affiliation{Department of Physics, Osaka University, Toyonaka, Osaka 560-0043, Japan}}
\newcommand{\AFFox}{\affiliation{Department of Physics, Oxford University, Oxford, OX1 3PU, United Kingdom}}
\newcommand{\AFFqmul}{\affiliation{School of Physics and Astronomy, Queen Mary University of London, London, E1 4NS, United Kingdom}}
\newcommand{\AFFregina}{\affiliation{Department of Physics, University of Regina, 3737 Wascana Parkway, Regina, SK, S4SOA2, Canada}}
\newcommand{\AFFseoul}{\affiliation{Department of Physics, Seoul National University, Seoul 151-742, Korea}}
\newcommand{\AFFsheff}{\affiliation{Department of Physics and Astronomy, University of Sheffield, S3 7RH, Sheffield, United Kingdom}}
\newcommand{\AFFshizuokasc}{\affiliation{Department of Informatics in
Social Welfare, Shizuoka University of Welfare, Yaizu, Shizuoka, 425-8611, Japan}}
\newcommand{\AFFstfc}{\affiliation{STFC, Rutherford Appleton Laboratory, Harwell Oxford, and Daresbury Laboratory, Warrington, OX11 0QX, United Kingdom}}
\newcommand{\AFFskk}{\affiliation{Department of Physics, Sungkyunkwan University, Suwon 440-746, Korea}}
\newcommand{\AFFtodai}{\affiliation{Department of Physics, University of Tokyo, Bunkyo, Tokyo 113-0033, Japan }}
\newcommand{\AFFtit}{\affiliation{Department of Physics,Tokyo Institute of Technology, Meguro, Tokyo 152-8551, Japan }}
\newcommand{\AFFtus}{\affiliation{Department of Physics, Faculty of Science and Technology, Tokyo University of Science, Noda, Chiba 278-8510, Japan }}
\newcommand{\AFFtoronto}{\affiliation{Department of Physics, University of Toronto, ON, M5S 1A7, Canada }}
\newcommand{\AFFtriumf}{\affiliation{TRIUMF, 4004 Wesbrook Mall, Vancouver, BC, V6T2A3, Canada }}
\newcommand{\AFFtokai}{\affiliation{Department of Physics, Tokai University, Hiratsuka, Kanagawa 259-1292, Japan}}
\newcommand{\AFFtsinghua}{\affiliation{Department of Engineering Physics, Tsinghua University, Beijing, 100084, China}}
\newcommand{\AFFynu}{\affiliation{Department of Physics, Yokohama National University, Yokohama, Kanagawa, 240-8501, Japan}}
\newcommand{\AFFllr}{\affiliation{Ecole Polytechnique, IN2P3-CNRS, Laboratoire Leprince-Ringuet, F-91120 Palaiseau, France }}
\newcommand{\AFFbari}{\affiliation{ Dipartimento Interuniversitario di Fisica, INFN Sezione di Bari and Universit\`a e Politecnico di Bari, I-70125, Bari, Italy}}
\newcommand{\AFFnapoli}{\affiliation{Dipartimento di Fisica, INFN Sezione di Napoli and Universit\`a di Napoli, I-80126, Napoli, Italy}}
\newcommand{\AFFroma}{\affiliation{INFN Sezione di Roma and Universit\`a di Roma ``La Sapienza'', I-00185, Roma, Italy}}
\newcommand{\AFFpadova}{\affiliation{Dipartimento di Fisica, INFN Sezione di Padova and Universit\`a di Padova, I-35131, Padova, Italy}}
\newcommand{\AFFkeio}{\affiliation{Department of Physics, Keio University, Yokohama, Kanagawa, 223-8522, Japan}}
\newcommand{\AFFwinnipeg}{\affiliation{Department of Physics, University of Winnipeg, MB R3J 3L8, Canada }}
\newcommand{\AFFkcl}{\affiliation{Department of Physics, King's College London, London, WC2R 2LS, UK }}
\newcommand{\AFFwarwick}{\affiliation{Department of Physics, University of Warwick, Coventry, CV4 7AL, UK }}
\newcommand{\AFFral}{\affiliation{Rutherford Appleton Laboratory, Harwell, Oxford, OX11 0QX, UK }}
\newcommand{\AFFwu}{\affiliation{Faculty of Physics, University of Warsaw, Warsaw, 02-093, Poland }}
\newcommand{\AFFbcit}{\affiliation{Department of Physics, British Columbia Institute of Technology, Burnaby, BC, V5G 3H2, Canada }}
\newcommand{\AFFtohoku}{\affiliation{Department of Physics, Faculty of Science, Tohoku University, Sendai, Miyagi, 980-8578, Japan }}
\newcommand{\AFFicise}{\affiliation{Institute For Interdisciplinary Research in Science and Education, ICISE, Quy Nhon, 55121, Vietnam }}
\newcommand{\AFFilance}{\affiliation{ILANCE, CNRS - University of Tokyo International Research Laboratory, Kashiwa, Chiba 277-8582, Japan}}
\newcommand{\AFFibs}{\affiliation{Center for Underground Physics, Institute for Basic Science (IBS), Daejeon, 34126, Korea}}
\newcommand{\AFFglasgow}{\affiliation{School of Physics and Astronomy, University of Glasgow, Glasgow, Scotland, G12 8QQ, United Kingdom}}
\newcommand{\AFFoecu}{\affiliation{Media Communication Center, Osaka Electro-Communication University, Neyagawa, Osaka, 572-8530, Japan}}
\newcommand{\AFFsilesia}{\affiliation{August Che\l{}kowski Institute of Physics, University of Silesia in Katowice, 75 Pu\l{}ku Piechoty 1, 41-500 Chor\
z\'{o}w, Poland}}

\AFFicrr
\AFFkashiwa
\AFFicrronly
\AFFmad
\AFFbcit
\AFFbu
\AFFuci
\AFFcsu
\AFFcnm
\AFFduke
\AFFllr
\AFFgifu
\AFFgist
\AFFglasgow
\AFFuh
\AFFibs
\AFFicise
\AFFicl
\AFFbari
\AFFnapoli
\AFFpadova
\AFFroma
\AFFilance
\AFFkeio
\AFFkek
\AFFkcl
\AFFkobe
\AFFkyoto
\AFFliv
\AFFminn
\AFFmiyagi
\AFFnagoya
\AFFkmi
\AFFpol
\AFFsuny
\AFFokayama
\AFFoecu
\AFFosaka
\AFFox
\AFFral
\AFFseoul
\AFFsheff
\AFFshizuokasc
\AFFsilesia
\AFFstfc
\AFFskk
\AFFtohoku
\AFFtokai
\AFFtodai
\AFFipmu
\AFFtit
\AFFtus
\AFFtoronto
\AFFtriumf
\AFFtsinghua
\AFFubc
\AFFwu
\AFFwarwick
\AFFwinnipeg
\AFFynu

\author{K.~Abe}
\AFFicrr
\AFFipmu
\author{C.~Bronner}
\AFFicrr
\author{Y.~Hayato}
\author{K.~Hiraide}
\AFFicrr
\AFFipmu
\author{K.~Hosokawa}
\AFFicrr
\author{K.~Ieki}
\author{M.~Ikeda}
\AFFicrr
\AFFipmu
\author{S.~Imaizumi}
\author{K.~Iyogi}
\AFFicrr
\author{J.~Kameda}
\AFFicrr
\AFFipmu
% ...

\collaboration{The Super-Kamiokande Collaboration}
\noaffiliation

\date{December 20, 2023}

\begin{abstract}

An analysis of solar neutrino data from the fourth phase of Super-Kamiokande~(SK-IV) from October 2008 to May 2018 is performed and the results are presented. The observation time of the data set of SK-IV corresponds to $2970$~days and the total live time for all four phases is $5805$~days. For more precise solar neutrino measurements, several improvements are applied in this analysis:
lowering the data acquisition threshold in May 2015,
further reduction of the spallation background using neutron clustering events,
precise energy reconstruction considering the time variation of the PMT gain.
The observed number of solar neutrino events in $3.49$--$19.49$~MeV electron kinetic energy region 
during SK-IV is 
$65,443^{+390}_{-388}\,(\mathrm{stat.})\pm 925\,(\mathrm{syst.})$ 
events.
Corresponding $\mathrm{^{8}B}$ solar neutrino flux is
$(2.314 \pm 0.014\, \rm{(stat.)} \pm 0.040 \, \rm{(syst.)}) \times 10^{6}~\mathrm{cm^{-2}\,s^{-1}}$,
assuming a pure electron-neutrino flavor component without neutrino oscillations. 
The flux combined with all SK phases up to SK-IV is
$(2.336 \pm 0.011\, \rm{(stat.)} \pm 0.043 \, \rm{(syst.)}) \times 10^{6}~\mathrm{cm^{-2}\,s^{-1}}$.
Based on the neutrino oscillation analysis from all solar experiments, including the SK $5805$~days data set, the best-fit neutrino oscillation parameters are
$\rm{sin^{2} \theta_{12,\,solar}} = 0.306 \pm 0.013 $ and 
$\Delta m^{2}_{21,\,\mathrm{solar}} = (6.10^{+ 0.95}_{-0.81})  \times 10^{-5}~\rm{eV}^{2}$,
with a deviation of about 
1.5$\sigma$
from the $\Delta m^{2}_{21}$ parameter obtained by KamLAND.
The best-fit neutrino oscillation parameters obtained from all solar experiments and KamLAND are
$\sin^{2} \theta_{12,\,\mathrm{global}} = 0.307 \pm 0.012 $ and 
$\Delta m^{2}_{21,\,\mathrm{global}} = (7.50^{+ 0.19}_{-0.18})  \times 10^{-5}~\rm{eV}^{2}$.

\end{abstract}

\maketitle

\section{Introduction}

The existence of neutrino oscillations~\cite{Maki:1962mu,Pontecorvo:1967fh}, which is a consequence of neutrino masses and mixing, is experimental evidence of elementary particle physics `beyond the standard model'. Observations of solar neutrinos were first made by the Homestake experiment~\cite{Davis:1968cp} using a radiochemical method, and then followed by real-time measurement with KAMIOKANDE-II~\cite{Hirata:1989zj} and other radiochemical experiments using gallium by SAGE and GALLEX/GNO~\cite{Abazov:1991rx, Anselmann:1992um, Altmann:2000ft}.
An initial indication of solar neutrino oscillations was obtained from the difference between the $\mathrm{^{8}B}$ solar neutrino fluxes as measured in the elastic-scattering channel at Super-Kamiokande~(SK) and the charged-current channel at the Sudbury Neutrino Observatory~(SNO) in 2001~\cite{Fukuda:2001nj, Ahmad:2001an}. Solar neutrino oscillation was subsequently established by including neutral-current measurements from SNO~\cite{Ahmad:2002jz}. 
Solar neutrino oscillations were confirmed using reactor 
anti-neutrinos by KamLAND~\cite{eguchi2003first}. Since these discoveries, Borexino and KamLAND experiments have measured the neutrino fluxes from different solar nuclear fusion processes, such as $pp$, $pep$, $\mathrm{^{7}Be}$, and carbon–nitrogen–oxygen~(CNO) cycle~\cite{Agostini:2017ixy,Agostini:2018uly, Gando:2014wjd, Agostini:2020mfq}. 

All measurements to date are naturally explained by neutrino flavor change due to neutrino oscillations with matter effects predicted by Mikheyev, Smirnov, and Wolfenstein~\cite{Mikheev:1986gs, Wolfenstein:1977ue}, termed the MSW effect: higher energy neutrinos undergo adiabatic conversion from the electron flavor state to the second mass eigenstate. While neutrino oscillations and MSW effect is consistent with all current solar neutrino measurements, two distinctive predictions are yet to be observed: the characteristic energy dependence of the solar neutrino electron-flavor survival probability $P_{ee}(E_\nu)$ distortion due to the MSW effect in the Sun and the day/night flux asymmetry induced by the matter effect in the Earth~\cite{Baltz:1986hn, Bouchez:1986kb, Carlson:1986ui, Cribier:1986ak, Petcov:1998su, Bakhti:2020tcj}.
One of the interests of solar neutrino experiments is to determine the neutrino oscillation parameters of $\varDelta m^{2}_{21}$ and $\sin^{2}\theta_{12}$. 
Independent of solar neutrino measurements, the KamLAND experiment used reactor anti-neutrinos to measure the same oscillation parameters, assuming CPT symmetry holds~\cite{Gando:2013nba}. 
Our previous paper on solar neutrino measurements~\cite{Abe:2016nxk} reported consistency in $\sin^{2}\theta_{12}$ while observing about $2\sigma$  tension in $\varDelta m^{2}_{21}$ between the solar global fit and the KamLAND result. 
Further precise measurements of neutrino oscillation parameters are required to test the framework of three-neutrino oscillation as well as the conservation of CPT in the neutrino sector~\cite{Bahcall:2002ia}.

In this article, the results of solar neutrino analysis from the full observation period of the fourth phase of SK~(SK-IV) are described. Moreover, the combined results together with the earlier phases of SK~(SK-I, SK-II, and SK-III) are also presented. 
This article is organized as follows: Sec.~\ref{sec_det} provides an overview of the SK detector and summarises the simulations. In Sec.~\ref{sec_recon}, improvements to the event reconstruction methods are explained. In Sec.~\ref{sec_calib}, the calibration methods and the detector performance are presented. In Sec.~\ref{sec_data_ana}, the data set of SK-IV, data reduction, and its systematic uncertainties are described. In Sec.~\ref{sec_result}, 
observed solar neutrino results from SK are presented. 
In Sec.~\ref{sec_osc}, \ref{sec:osc_dn}, and \ref{sec:osc_spec}, the oscillation analysis from SK and other experiments are discussed. In the final section, we conclude this study and give future prospects.

\section{Super-Kamiokande-IV detector} \label{sec_det}

The Super-Kamiokande detector consists of about $50,000$~tonnes 
of purified water in a stainless steel cylindrical water tank with
photomultiplier tubes~\cite{Fukuda:2002uc}. The inner detector~(ID) holds
$32,000$~tonnes of water as physics target and its standard fiducial volume
for solar neutrino analysis is $22,500$~tonnes.
The outer detector~(OD) is optically separated
from the ID and used to veto cosmic-ray muons penetrating the
mountain above~\cite{Abe:2013gga}.
The ID is viewed by 20-inch photomultiplier tubes~(PMTs) and
the OD is viewed by 8-inch PMTs.
SK was originally started in April 1996, and observation in phase IV was
finished in May 2018 to refurbish the detector in preparation for dissolution of gadolinium into the water~\cite{Super-Kamiokande:2021cuo}.
Table~\ref{tb:phase} shows a summary of the experimental phases of SK.
\begin{table}[htp]
    \begin{center}
    \caption{Experimental phases of SK. The live times are the total duration of the good observation periods for solar neutrino analysis. The energy threshold is based on the electron kinetic energy.} 
    \label{tb:phase}
        \begin{tabular}{lrrrr}
        \hline
        \hline
        Phase & SK-I & SK-II & SK-III & SK-IV \\ \hline 
        Period (Start) & Apr. `96 & Oct. `02 & Jul. `06 & Sep. `08\\
        Period (End) & Jul. `01 & Oct. `05 & Aug. `08 & May `18 \\
        Livetime~[days] & $1,496$ & $791$ & $548$ & $2,970$ \\      
        ID PMTs & $11,146$ & $5,182$ & $11,129$ & $11,129$ \\  
        OD PMTs & $1,885$ & $1,885$ & $1,885$ & $1,885$ \\
        PMT coverage~[$\%$] & $40$ & $19$ & $40$ & $40$ \\
        Energy thr.~[MeV] & $4.49$ & $6.49$ & $3.99$ & $3.49$ \\
        \hline
        \hline
        \end{tabular}
    \end{center}
\end{table}
Starting with SK-II, an acrylic cover and fiber-reinforced plastic~(FRP) case were installed around the ID PMTs to avoid a chain reaction of implosions~\cite{Cravens:2008aa}. The start of SK-IV corresponds to the installation of new `QBEE' front-end electronics~\cite{Nishino:2009zu,Yamada:2010zzc}.  The new electronics is capable of high-speed signal processing and records every hit of all the PMTs including hits from PMTs' dark current. 

The coordinate of the SK detector is defined the same way as in the previous publication~\cite{Abe:2016nxk}. The origin is at the center of the detectors, with $X$- and $Y$- axes lying in the horizontal plane and the $Z$-axis vertically upwards. We categorize the cylindrical ID surfaces into ``top", ``barrel", and ``bottom" regions. 

\subsection{Water system and water quality} \label{sec:water_sys}

Water purification is extremely important for the experiment, not only to improve the attenuation length of Cherenkov photons but also to reduce the radioactive backgrounds due to dissolved isotopes. For these purposes, the SK-IV water is continuously recirculated through a water purification system and returned to the detector~\cite{Fukuda:2002uc, Nakano:2019bnr}.  The water quality, which affects the absorption and scattering of photons during their propagation, is monitored by using a laser system~\cite{Abe:2013gga}, and the water transparency~(water attenuation length) is measured using cosmic-ray muon data.

\begin{figure}[htp]
    \begin{center}
        \includegraphics[width=\linewidth]{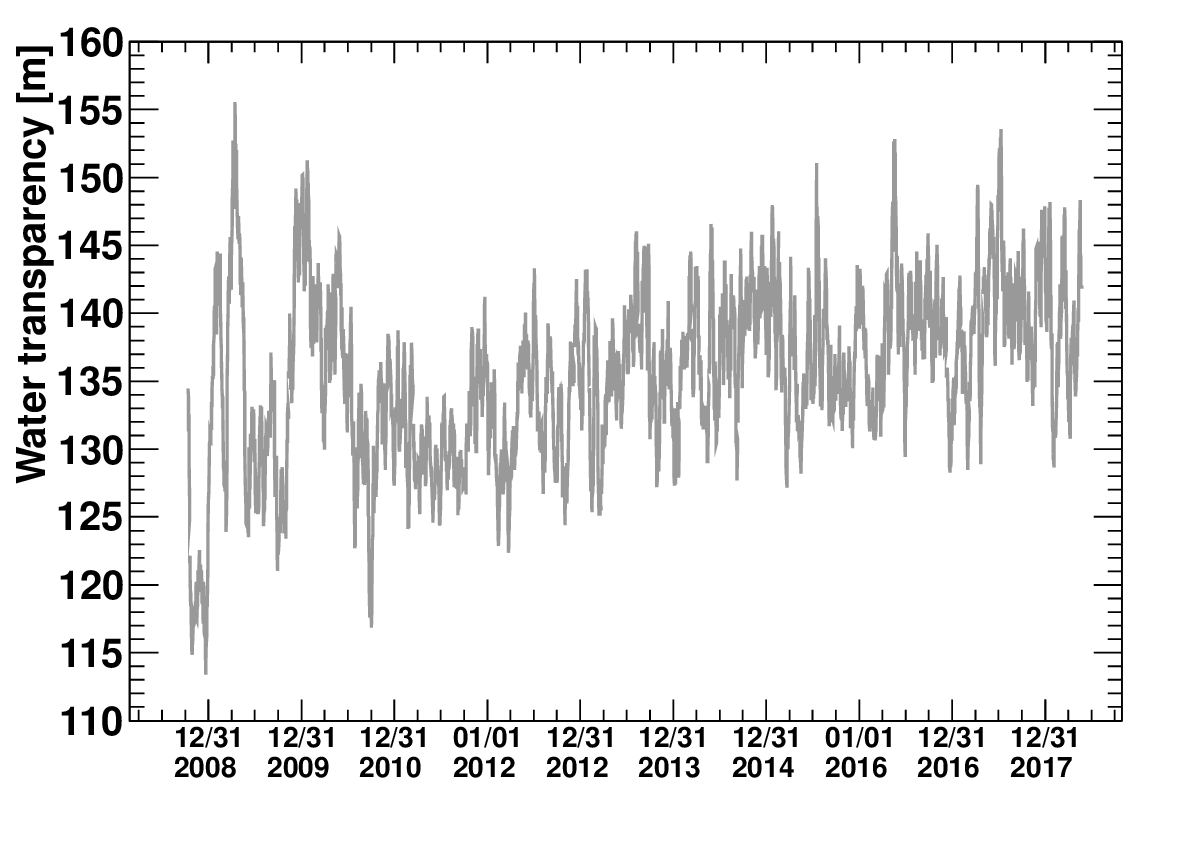}
    \end{center}
    \caption{The time variation of water transparency measured by decay electrons. 
    This parameter is described as $L$ in Appendix~\ref{sec:app_erec}.
    \label{fig:water_t}
    }
\end{figure}

The electrons~(or positrons) from decays of stopped muons are used to monitor the variation over time of the water attenuation length. 
The light intensity observed by PMTs depends on the distance from the position of the decay electrons; light reaching more distant PMTs will be more attenuated. 
Using many events, the correlation between light observed and distance-to-PMT is used to measure the water transparency in the water tank~\cite{Hosaka:2005um}. 
Figure~\ref{fig:water_t} shows the time variation of water attenuation length over the data sample described in this article. 

\begin{figure}[htp]
    \begin{center}
        \includegraphics[width=\linewidth]{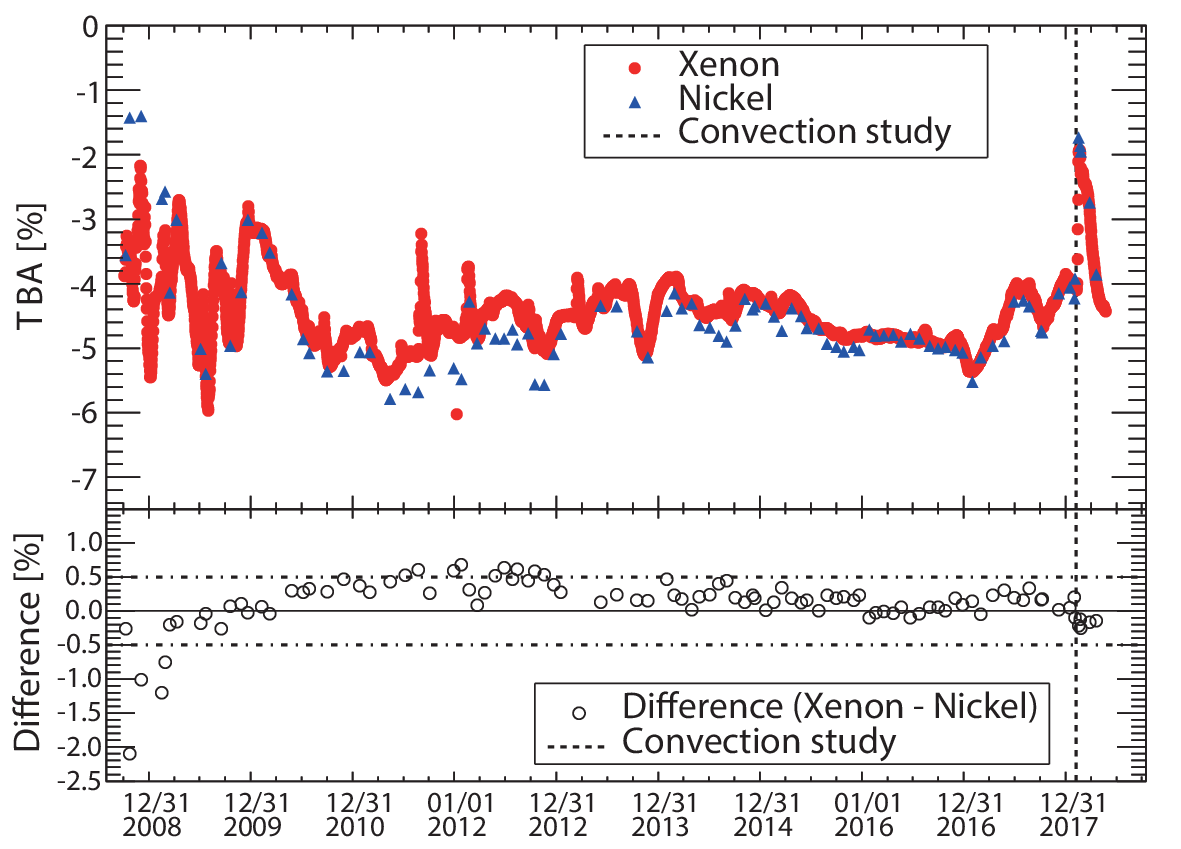}
    \end{center}
    \caption{(Color online) Top: The time variation of the top-bottom asymmetry~(TBA) throughout SK-IV full period. The red dots~(blue upward-pointing triangle) show the TBA value measured by the auto-Xenon calibration system~(Ni-Cf calibration source)~\cite{Abe:2013gga}. The black vertical line shows when the temperature of the supply water was changed, as described in the main text. Bottom: The difference between two TBA measurements, which is normally within $\pm0.5\%$ during the SK-IV data set.}
 \label{fig:tba}
\end{figure}

Purified water is supplied from the bottom of the detector, and drained from the top of the detector. The temperature of the supply water is carefully controlled as this affects the amount of convection in the detector. Normally, convection inside the detector occurs below $Z=-11$~m~\cite{Abe:2013gga}.  This water flow results in a small asymmetry in the water transparency as a function of vertical position $Z$, and is responsible for a top-bottom asymmetry~(TBA).
In order to model the $Z$-dependence of photon absorption, we introduce the parameter $\alpha_{\mathrm{TBA}}$, which is defined as
\begin{equation}
    \alpha_{\mathrm{TBA}} = \frac{\langle N_{\mathrm{top}} \rangle - \langle N_{\mathrm{bottom}} \rangle}{\langle N_{\mathrm{barrel}} \rangle},
\end{equation}
where $\langle N_{\mathrm{top}} \rangle$,  $\langle N_{\mathrm{bottom}} \rangle$, and $\langle N_{\mathrm{barrel}} \rangle$ are the averages of the hit probabilities of top, bottom, and barrel of the ID. 
The TBA parameter, $\alpha_{\mathrm{TBA}}$, is measured by two calibration devices. One is the auto-Xenon lamp, which continuously injects light into the detector every $1$~s. The other is a Ni-Cf calibration source, which is inserted into the detector to take calibration data about once per month, as shown in Fig.~\ref{fig:tba}.
Due to the water temperature distribution in the detector, a TBA of about $-5\%$ has been observed by the auto-Xenon lamp, indicating a higher hit rate in the bottom region.
The TBA measured by the Ni-Cf calibration source has a similar value to that measured by the auto-Xenon lamp. 
The difference over the whole data set is around $\pm0.5\%$ level.

In February 2018, we changed the temperature of the supply water injected at the bottom of the ID region, from $+13.06\mathrm{^{\circ}C}$ to $+13.52\mathrm{^{\circ}C}$.
This lowered the density of the water in the bottom region  and evoked large scale convection. 
The water temperature measured at various places in the detector became consistent after about two weeks of re-circulation, 
this indicated that water is fully mixed at that time and hence water quality should be uniform across the entire detector volume.
After achieving this uniform water condition, the TBA value became significantly smaller, but a residual $1$--$2\%$ asymmetry remained. 
This remaining asymmetry is interpreted as the asymmetry of photon detection efficiency of the PMTs, and is corrected for in the energy reconstruction as described in Sec.~\ref{sec:energy}.
Another convection test was done in May 2018. 
A consequence of these 'water convection tests' was increased Rn contamination (see Sec.~\ref{sec_runselection}).

\subsection{Radioactive contamination in the water} \label{sec:radon}
A large part of the low-energy intrinsic background of the SK-IV detector comes from radon~(hereafter Rn) remaining in the purified water. 
To prevent Rn gas from the mine entering into the water tank, the detector is tightly sealed and a low-Rn buffer gas is continuously supplied to the air layer between the top of the tank and the water surface. 
To keep the detector clean, the radon concentrations in the buffer gas, as well as the experimental site of the SK detector, are monitored by several kinds of Rn detectors~\cite{Nakano:2017rsy, Pronost:2018ghn}. 
The measured Rn concentrations in the buffer gas supply is $0.08\pm0.07~\mathrm{mBq/m^{3}}$ while that in the air layer is $28.8\pm1.7~\mathrm{mBq/m^{3}}$. 
This indicates Rn sources exist inside the detector.

In addition to the buffer gas monitoring system, another Rn measurement system was developed in 2013~\cite{Nakano:2019bnr} which evaluates the Rn concentration in purified water in the detector directly.
Since then, water has been sampled from the various positions in the detector and its Rn concentrations have been measured continuously. 
The Rn concentrations in the center and bottom regions of the SK-IV detector in 2015 are $<0.23~\mathrm{mBq/m^{3}}$ and $2.63\pm0.22~\mathrm{mBq/m^{3}}$, respectively. 
Details of the Rn study will be given in a future publication~\cite{nakano2023}.

\subsection{Time variation of PMT gain and dark rate}
\label{sec:gain}
\begin{figure}[htp]
    \begin{center}
        \includegraphics[width=\linewidth]{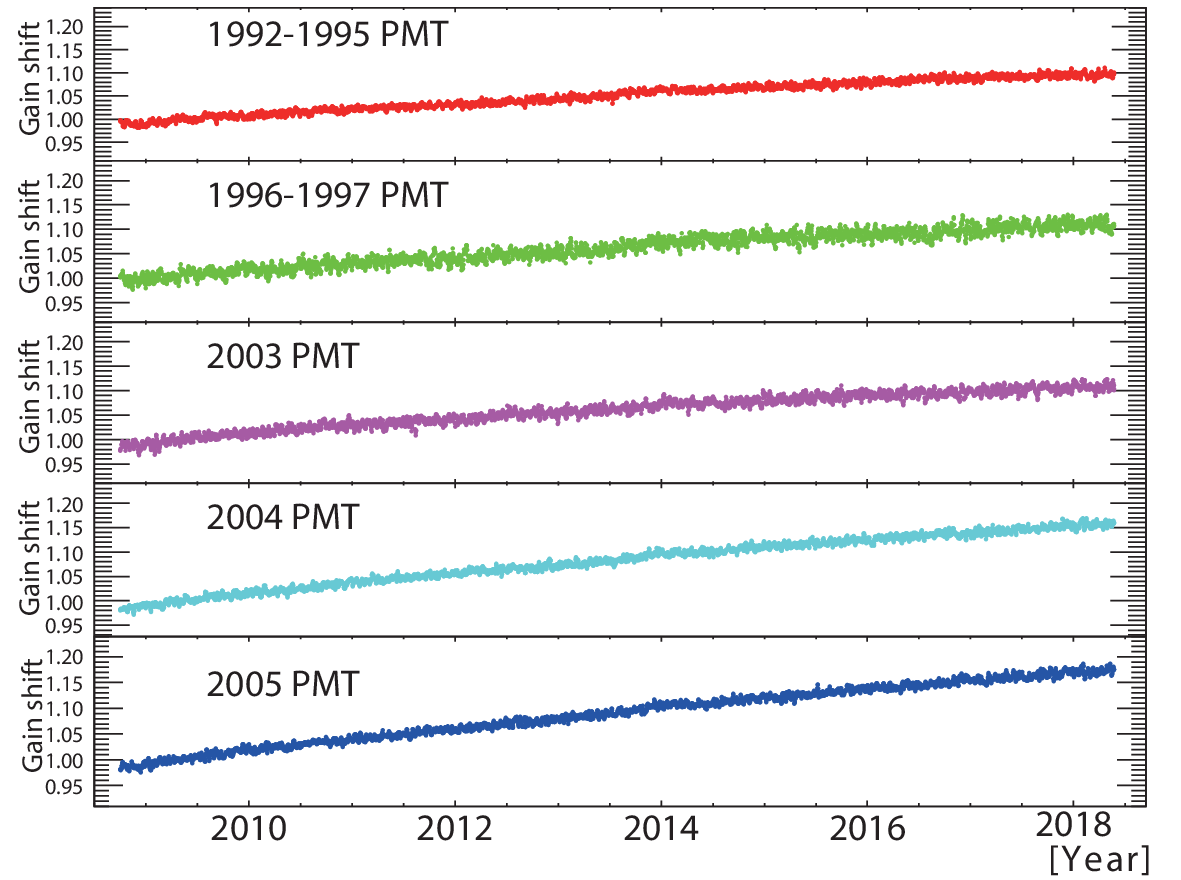}
    \end{center}
    \caption{(Color online) The time variation of the relative PMT gain. 
    Each panel shows PMTs from different production years.}
    \label{fig:gain}
\end{figure}

\begin{figure}[htp]
    \begin{center}
        \includegraphics[width=\linewidth]{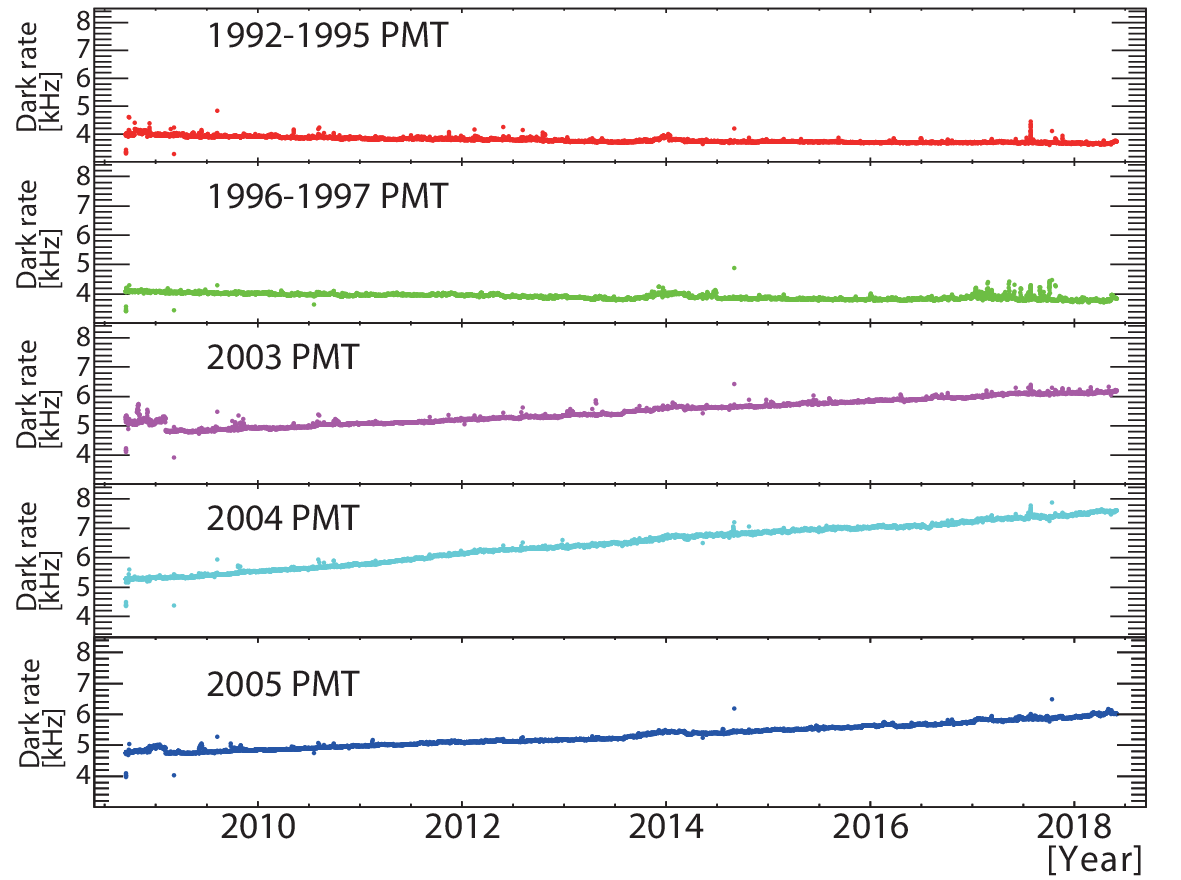}
    \end{center}
    \caption{(Color online) The time variation of the average dark rate of PMTs in each run. 
    Each panel shows PMTs from different production years.} 
    \label{fig:darkrate}
\end{figure}

\begin{figure}[htp]
    \begin{center}
        \includegraphics[width=\linewidth]{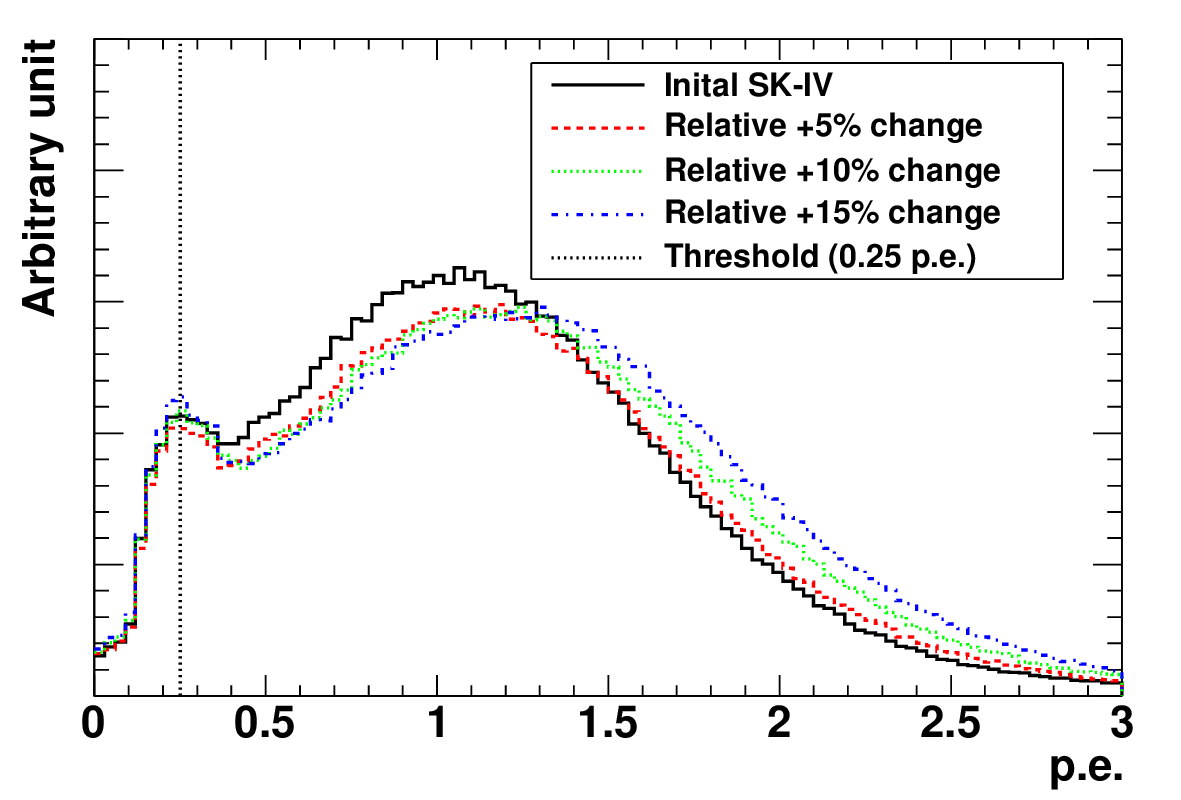}
    \end{center}
    \caption{(Color online) A comparison of the shape of the $1$~photo-electron~(p.e.). the distribution used in the detector simulation. 
    The original PMT gain~(black-solid) is shifted by $+5\%$~(in red-dashed), $+10\%$~(in green-dotted), $+15\%$~(in blue-dotted-dashed). 
    The vertical line shows the threshold set at $0.25$~p.e.}
    \label{fig:1pe}
\end{figure}

During SK-IV, we observed a time variation of the gain and dark rate of PMTs, as shown in Fig.~\ref{fig:gain}~and Fig.~\ref{fig:darkrate}.
The average PMT gain increased by around $+10\%{\sim}+15\%$ during SK-IV,
depending on the production year of PMTs.
Since the hardware threshold at which a PMT's signal is recorded was fixed during the SK-IV
period, this gain shift resulted in a gradual increase in the detection efficiency of a photon hit.
Figure~\ref{fig:1pe} shows a comparison of the $1$~p.e. charge distributions
used in the detector simulation, which illustrates this effect.
The dark rate of the PMTs also changed over time depending on the PMT production year, as shown in Fig.~\ref{fig:darkrate}. 
In our previous report, based on the first 1664 days of SK-IV~\cite{Abe:2016nxk}, the effect from these variations was small and within the energy scale systematic uncertainty.
However, since the full SK-IV is substantially longer than earlier SK phases, the effect cannot be ignored in the full SK-IV period. 
A correction factor is therefore included in the SK-IV solar neutrino energy reconstruction to account for this time variation. 
Some improvement in gain over time is often seen in large PMTs. We are not sure why our PMTs continue to improve after such a long period of operation, but we believe that our empirical correction factor will account for it within our stated systematic uncertainties.
This is described in more detail in Sec.~\ref{sec:energy} and Appendix~\ref{sec:app_erec}.

\subsection{SK-IV detector simulation} \label{sec:simulation}
A Monte Carlo~(MC) simulation that reproduces physics events is used to understand the SK-IV detector performance.
This simulation is based on \textsc{Geant3} package~\cite{Brun:1994aa}, and customized for the SK-IV detector.

In order to reproduce the data more accurately, the following parameters are evaluated, typically day-by-day,
then used in the solar neutrino event simulations in SK-IV:  
water transparency,
TBA of the water transparency,
PMT dark rate, and PMT gain.
For the $^{8}$B solar neutrino event generation, the Winter
spectrum~\cite{Winter:2004kf} is used for the initial $^{8}$B solar neutrino energy.
For the $hep$ solar neutrinos, the Bahcall spectrum~\cite{hep} is used.
We consider the actual detector live time when simulating the direction of solar neutrinos. 
Ejection of a recoil electron by neutrino-electron elastic
scattering is simulated using the cross section including radiative correction~\cite{Bahcall:1995mm}.
The expected $\mathrm{{^8}B}$ and $hep$ solar neutrino event rates in the whole SK ID volume ($32.5$~kt) without neutrino oscillation
are $294.7$~events/day and $0.6375$~events/day, respectively.
Flux values used in this calculation are based on the neutral current results from SNO~\cite{Aharmim:2011vm} for $\mathrm{{^8}B}$ solar neutrino (($5.25 \pm 0.20) \times 10^{6}~\mathrm{cm^{-2}\,s^{-1}}$), and BP2004~\cite{Bahcall:2004fg} standard solar model~(SSM) for $hep$ solar neutrino~($7.88 \times 10^{3}~\mathrm{cm^{-2}\,s^{-1}}$).
We generate about $10^5$~simulation events/day each for  $\mathrm{{^8}B}$ and $hep$ solar neutrino fluxes.

\section{Event reconstruction} \label{sec_recon}
\subsection{Vertex and Direction} \label{sec_verdir}
\begin{figure}[htp]
    \begin{center}
\includegraphics[width=\linewidth]{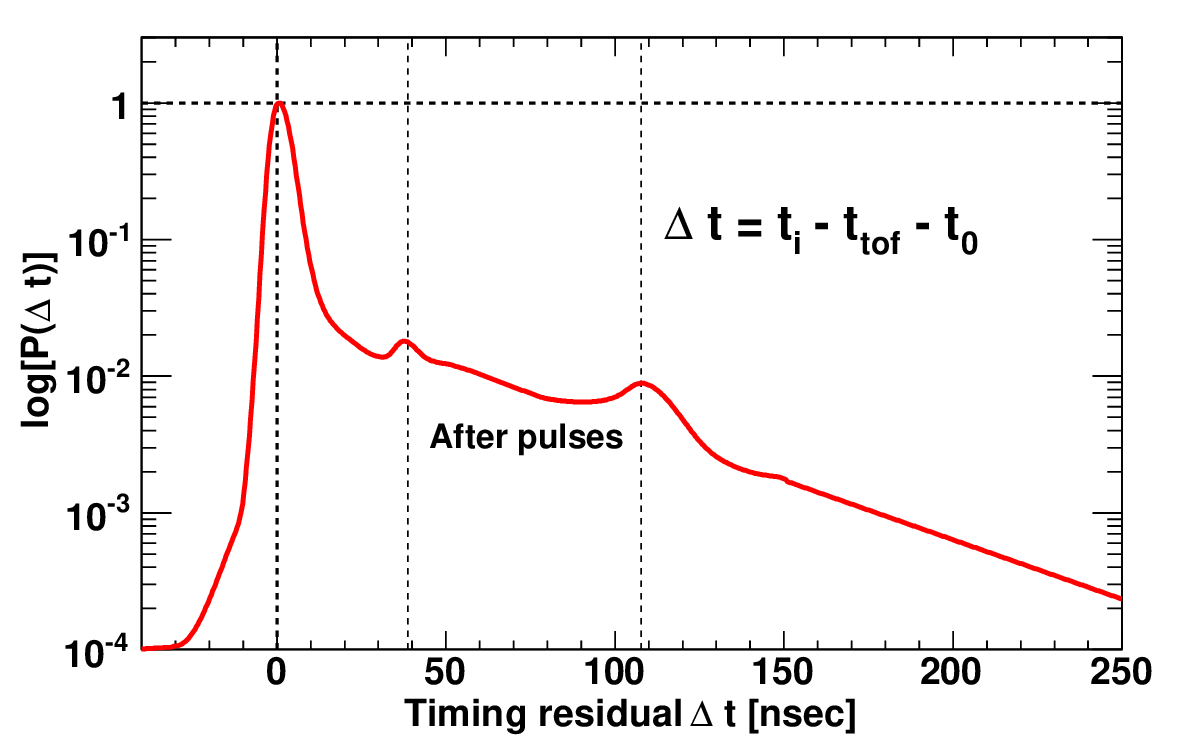}
    \end{center}
    \caption{(Color online) The probability density function of the timing residual~($t_{i}-t_{\mathrm{tof}}-t_{0} = \Delta t$) for the single photo-electron signal. 
    The second and the third peaks around $40$~ns and $110$~ns are caused by the PMT's after pulses.} 
    \label{fig:time_residual}
\end{figure}

The event vertex and direction are reconstructed using the same method as the previous analysis~\cite{Abe:2016nxk}. 
The electrons scattered by the solar neutrinos can travel only a few cm in water and thus the location of the Cherenkov emission is recognized as a point. 
Under this assumption, the vertex position is reconstructed with a maximum likelihood fit of the photon arrival time on the PMTs in the solar neutrino analysis~\cite{Smy:2007maa}. 
The log-likelihood function is defined as,
\begin{equation}
    L(\vec{x},t_{0}) = \sum_{i=1}^{N_{\mathrm{hit}}}\log \left[ P(t_{i}-t_{\mathrm{tof}}-t_{0}) \right],
\end{equation}
where $\vec{x}$ is the vertex position, being tested,
$N_{\mathrm{hit}}$ is the number of PMTs that have received light (hereafter hit-PMT),
$t_{i}$ is the hit time of $i$-th hit-PMT,
$P(\Delta t)$ is the probability density function of the timing residual~($t_{i}-t_{\mathrm{tof}}-t_{0} = \Delta t$) for a single photo-electron signal, 
$t_{\mathrm{tof}}$ is the timing after subtracting the time-of-flight~(hereafter TOF), 
and $t_{0}$ is the initial time of the event. 
The probability density function for the timing residual is extracted from LINAC calibration data~\cite{Nakahata:1998pz},
and is shown in Fig.~\ref{fig:time_residual}. 
Although small secondary peaks corresponding to the PMTs' after pulses can be seen in the residual density function, the likelihood function is strongly peaked at $\Delta t = 0$. 

\begin{figure}[htp]
    \begin{center}
        \includegraphics[width=\linewidth]{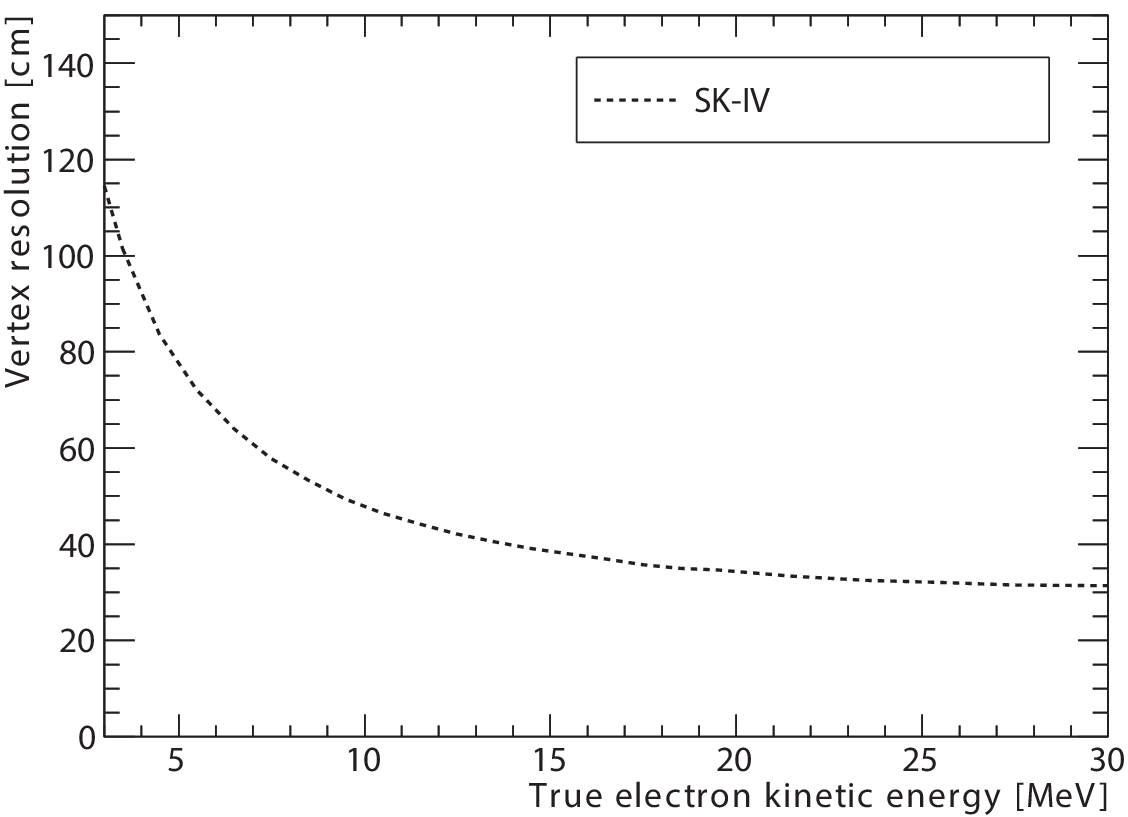}
    \end{center}
    \caption{The vertex resolution as a function of the true electron kinetic energy. 
    The events whose reconstructed vertex is in the standard fiducial volume (more than $200$~cm from the ID wall) are sampled.
    } 
    \label{fig:ver_reso}
\end{figure}

Figure~\ref{fig:ver_reso} shows the vertex resolution of an electron event in the standard fiducial volume. 
Here, the vertex resolution is defined as the root mean square of the residual difference between the generated position and the reconstructed vertex position. 
To evaluate this, electrons are generated uniformly across the whole ID volume with a random direction.
Then events reconstructed within the standard fiducial volume are selected and the difference between the vertex positions is calculated. Typical vertex resolutions are $101.5$~cm, $64.0$~cm, $49.3$~cm, and $39.1$~cm at $3.49$~MeV, $6.49$~MeV, $9.49$~MeV, and $14.49$~MeV of electron kinetic energy, respectively.

\begin{figure}[htp]
    \begin{center}
        \includegraphics[width=\linewidth]{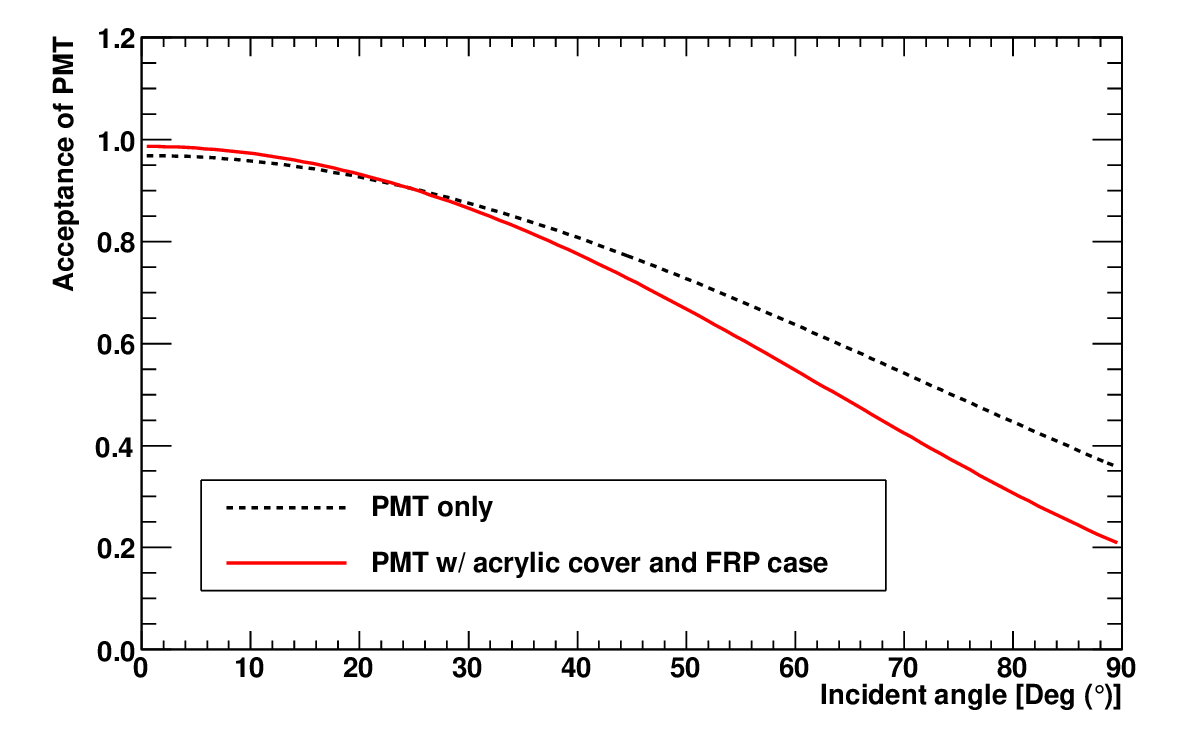}
    \end{center}
    \caption{(Color online) The PMT acceptance as a function of the incident angle of the photon. 
    The dashed-black line shows the acceptance in the case of the PMT itself while the solid-red line shows that of the PMTs equipped with the acrylic cover and FRP case used after SK-II.} 
    \label{fig:pmt_acc}
\end{figure}

\begin{figure}[htp]
    \begin{center}
        \includegraphics[width=\linewidth]{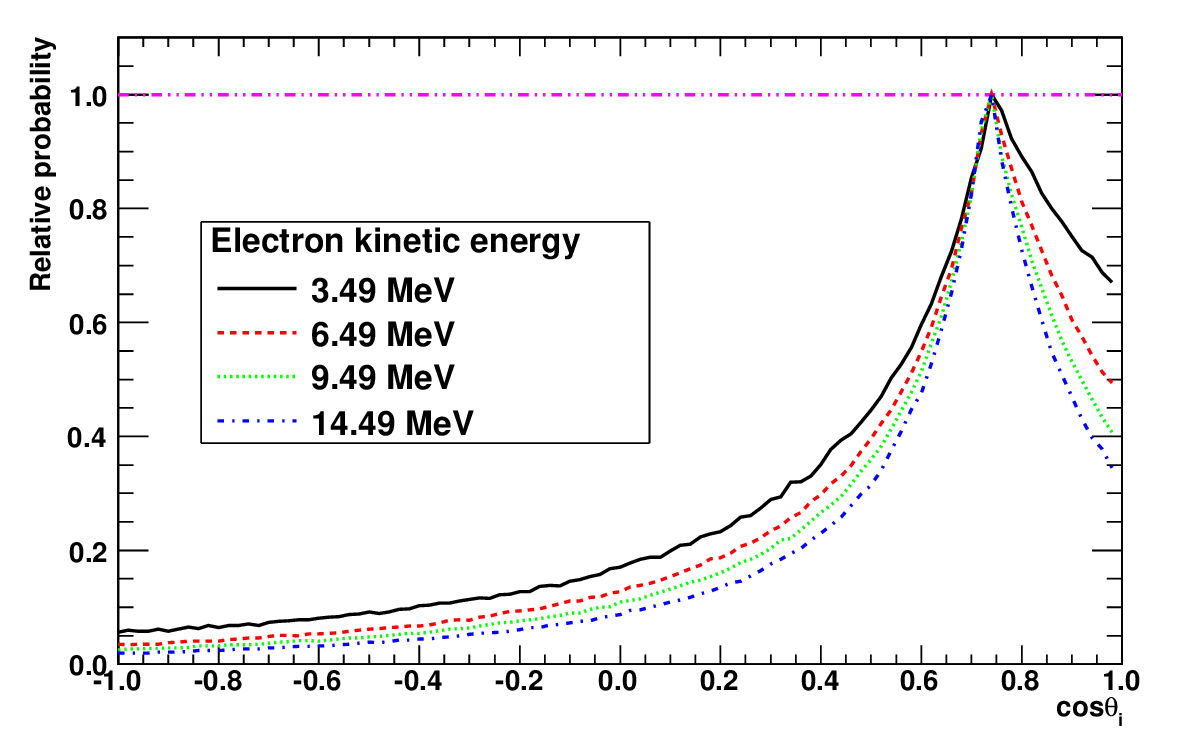}
    \end{center}
    \caption{(Color online) The distribution of the opening angle between the direction of the generated electron and the vector from the reconstructed vertex to each hit-PMT's position. 
    The black solid, red dashed, green dotted, and blue dash-dotted lines show the distribution of $3.49$~MeV, $6.49$~MeV, $9.49$~MeV, and $14.49$~MeV in electron kinetic energy, respectively.} 
    \label{fig:f_cos_dist}
\end{figure}

The direction of the event is reconstructed with a maximum likelihood method to find a Cherenkov ring that best matches the position of the hit PMTs.
The log-likelihood function is defined as
\begin{equation}
    L(\vec{d}) = \sum^{N_{30}}_{i} \log \left[f(\cos\theta_{\mathrm{dir},i}, E) \right] \times \frac{\cos\theta_{i}}{a(\theta_{i})},
\end{equation}
where $N_{30}$ is the number of hit-PMTs with residual time~($t_{i}-t_{\mathrm{tof}}$) within $30$~ns;
$\theta_{\mathrm{dir},i}$ is the angle between the electron direction and the vector from the reconstructed vertex to the position of $i$-th hit-PMT;
$E$ is the reconstructed energy of the electron and
$f(\cos\theta_{\mathrm{dir},i},E)$ is the probability of simulated electron events with $\theta_{\mathrm{dir},i}$ and $E$.
The logarithm is weighted by an acceptance term based on $\theta_{i}$, which is the angle between the vector from the reconstructed vertex to the $i$-th hit-PMT position and the direction that the $i$-th hit-PMT is facing.
The parameter $a(\theta_{i})$ is the typical acceptance of a PMT, which is shown in Fig.~\ref{fig:pmt_acc}.
The use of the acrylic cover and FRP cases since SK-II has little effect on the acceptance of photons at small angles to the tube facing, but the decrease of acceptance at large angles is noticeable, 
so this effect is corrected in this solar analysis. 
Figure~\ref{fig:f_cos_dist} shows typical distributions of the function of $f(\cos\theta_{\mathrm{dir},i},E)$.
Because of the multiple scattering of electrons in water, the distribution peaks at $42^{\circ}$, but is quite broad, with the width depending on the electron energy.

\begin{figure}[htp]
    \begin{center}
        \includegraphics[width=\linewidth]{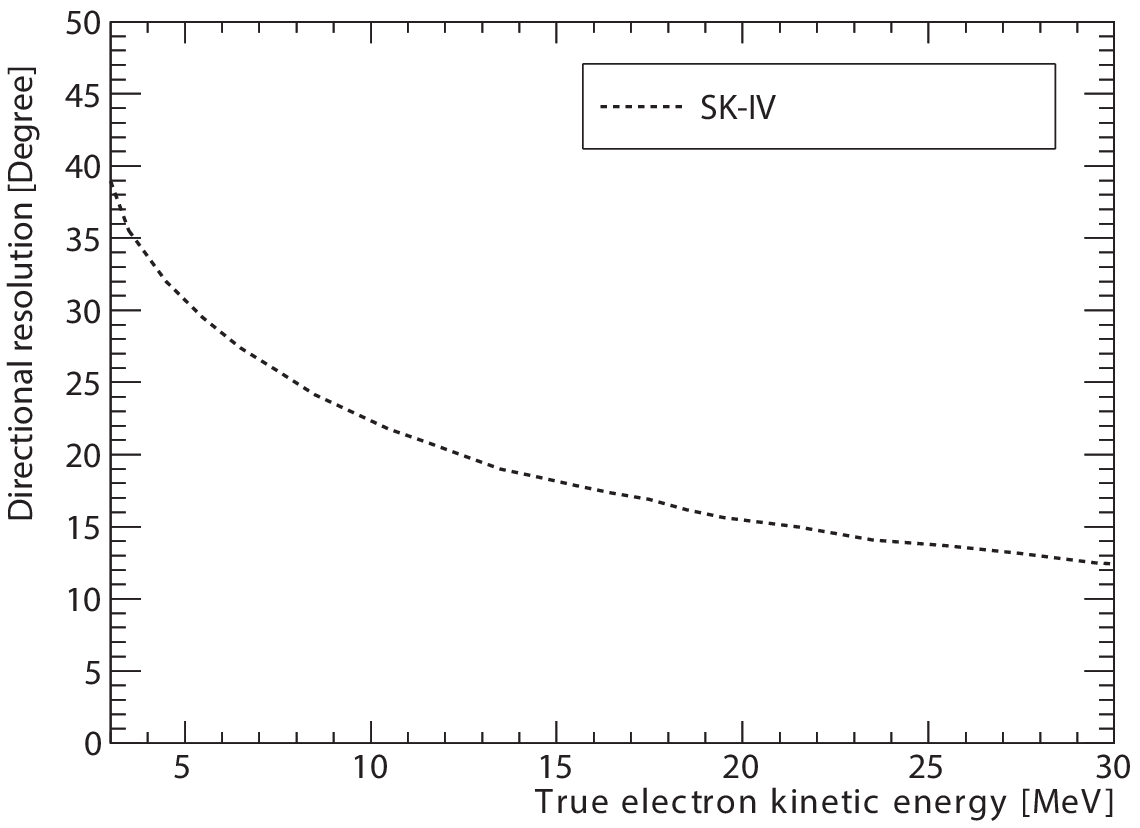}
    \end{center}
    \caption{The average directional resolution as a function of the true electron kinetic energy, for standard fiducial events. The typical angular resolutions are $35.5^{\circ}$, $27.4^{\circ}$, $23.0^{\circ}$, and $18.5^{\circ}$ at $3.49$~MeV, $6.49$~MeV, $9.49$~MeV, and $14.49$~MeV of electron kinetic energy, respectively.} 
    \label{fig:dir_reso}
\end{figure}

The $1\sigma$ directional resolution is defined as the angle that includes $68.3\%$ of events in the distribution of the angular difference between their reconstructed direction and their true direction.
It is estimated from simulated electron events that are generated uniformly in the whole ID volume with random direction, and reconstructed in the standard fiducial volume.    
Figure~\ref{fig:dir_reso} shows the energy dependence of the directional resolution. 
Details of the vertex and direction reconstruction methods and their performances can be found elsewhere~\cite{Abe:2010hy,Abe:2016nxk}.

\subsection{Energy} \label{sec:energy}
Electron energy is reconstructed based on the number of hit PMTs within
$50$~ns of the expected photon arrival time from the reconstructed 
vertex. Then, an effective number of hits, $N_{\mathrm{eff}}$, is calculated by
applying a correction for the relative difference of PMT performance,
contributions of dark rate and late arrival hits, the effective surface area of PMTs, 
and light attenuation in the water.
Finally, $N_{\mathrm{eff}}$ is converted to a recoil
electron energy using a function derived from simulations of mono-energetic electrons in the standard fiducial volume in the SK-IV detector.

\begin{figure}[htp]
    \begin{center}
        \includegraphics[width=\linewidth]{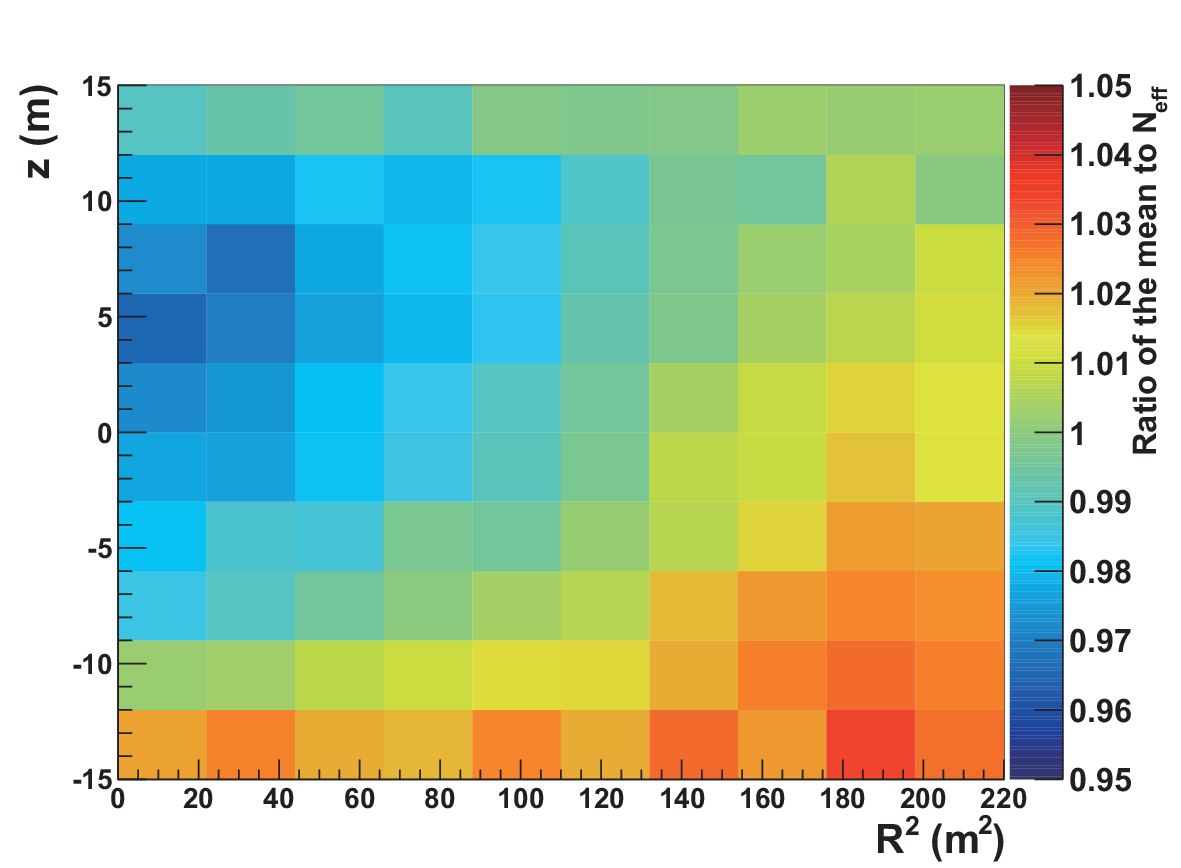}
        \includegraphics[width=\linewidth]{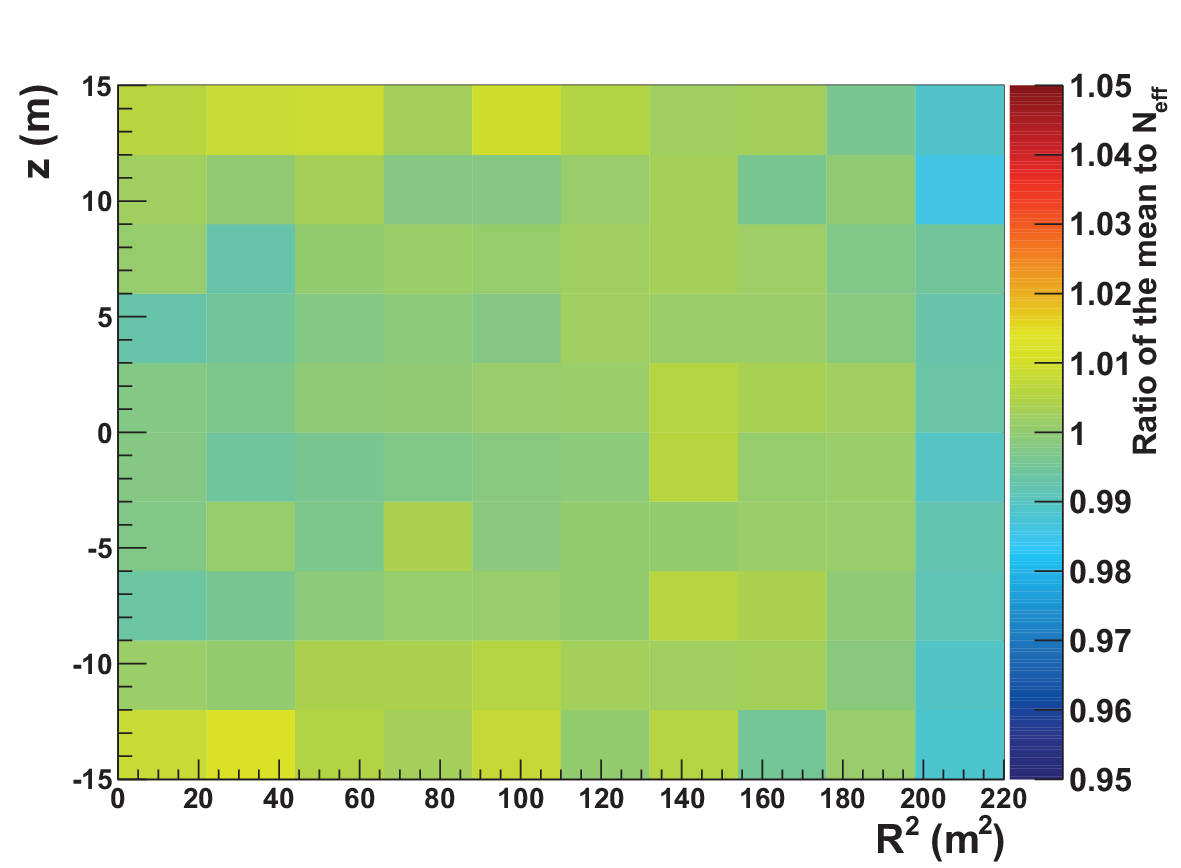}
    \end{center}
    \caption{(Color online) Relative size of reconstructed $N_{\mathrm{eff}}$ for simulated electrons with $10$~MeV/$c$ momentum generated uniformly in the entire ID volume. 
    The top~(bottom) panel shows results with the analysis method used for the previous~(this) analysis. 
    The vertical and horizontal axis represents reconstructed $Z$ and $R^2 = X^2 + Y^2$ positions, respectively.}
    \label{fig:neff_pos_dep}
\end{figure}

The method of calculating $N_{\mathrm{eff}}$  was improved since the previous analysis~\cite{Abe:2016nxk} and is expressed as
\begin{multline}
  \label{eq:neff}
  N_{\mathrm{eff}} = \sum_i^{N_{50}} 
  \left[ 
    \vphantom{\frac{(}{)}}  (X_{i} + \varepsilon_{\mathrm{tail}} -\varepsilon_{\mathrm{dark}}^{i})\,C_{i}
    \right. \\
    \left.
    \times \frac{N_{\mathrm{all}}}{N_{\mathrm{alive}}}
    \times \frac{S(0,0)}{S(\theta_{i},\phi_{i})}  \times \frac{1}{P_i} \times \frac{1}{QE_{i}}
  \right],
\end{multline}
where $N_{50}$ is the number of hit PMTs included in the $50$~ns window, $i$ is an index running over these hit PMTs and $N_{\mathrm{all}}~(N_{\mathrm{alive}})$ is the number of total~(live) ID PMTs at the time the event was recorded.
Descriptions of the other variables in the above $N_{\mathrm{eff}}$ definition are following.
\begin{description}

\item[$X_i$]
Hit occupancy correction calculated for the $i$-th hit-PMT and its eight surrounding PMTs. 
Assuming the Poisson distribution, the correction is calculated as
\begin{equation}
    X_{i} = 
    \begin{cases}
\log \left(\frac{1}{1-x_{i}} \right)/{x_{i}} & (x_{i} < 1) \\ 
        3.0 & (x_{i} = 1) 
    \end{cases},
\end{equation}
where $x_{i}$ is the fraction of hit-PMTs in the $3 \times 3$ PMT block surrounding the $i$-th hit-PMT. 
The details are explained in Appendix B in Ref.~\cite{Koshio:1998}.

\item[$\varepsilon_{\mathrm{tail}}$] 
Correction for late hits due to reflected and scattered light. 
It is calculated as
\begin{equation}
    \varepsilon_{\mathrm{tail}} = \frac{N_{100} - N_{50} - \left(N_{\mathrm{alive}} \times R_{\mathrm{dark}} \times t_{50}\right)}{N_{50}},
\end{equation}
where $N_{100}$ is the number of hit-PMTs in a $100$~ns window, $R_\mathrm{dark}$ is the average dark rate overall live PMTs, and $t_{50}$ is the interval of $50$~ns. 
Since the latter $50$~ns is dominated by scattered light and background hits, it is not appropriate to apply the corrections in Eq.~(\ref{eq:neff}). 
So, we only correct statistically for the extra light outside the first $50$~ns.

\item[$\varepsilon^{i}_{\mathrm{dark}}$] 
Correction for dark hit contribution. This parameter is calculated using the measured dark rates for the corresponding PMT as,
\begin{equation}
  \varepsilon^{i}_{\mathrm{dark}} = N_{\mathrm{alive}} \times R_{\mathrm{dark}} \times t_{50}
  \times \frac{r^{i}_{\mathrm{dark}}}{\sum_{j}^{N_{50}}r^{j}_{\mathrm{dark}}},
\end{equation}
where $r^{i}_{\mathrm{dark}}$ is measured dark rate for $i$-th hit-PMTs, and $t_{50}$ is the interval of $50$~ns.

\item[$C_i$]
Correction for PMT gain increase. This correction for the observed data was empirically determined and expressed as 
\begin{equation}
C_i = \frac{1}{1 + 0.226 \times G_{i}},
\end{equation}
where $G_{i}$ is relative gain of $i$-th hit-PMT with respect to the gain in October 2008. 
Those relative gains are evaluated for five PMT groups separated by their production time, as shown in Fig.~\ref{fig:gain}. 
A similar correction is also applied for the simulated data, since the relative gain increase is also included in the SK-IV detector simulation in this analysis.

\item[$S(\theta_i,\phi_i)$]
Correction for PMT coverage due to the fact that each PMT has a spherical surface, as a function for polar angle ($\theta_i$)  and azimuthal angle ($\phi_i$) of incident photons in the local coordinate of each PMT. 
This correction was evaluated with a \textsc{Geant4}~\cite{Agostinelli:2002hh, Allison:2006ve, Allison:2016lfl} based simulation of the detector geometry. 
This correction is improved from the previous analyses by introducing more realistic geometry to the simulation.
Details of improvements to this and the attenuation ($P_i$) correction are given in Appendix~\ref{sec:app_erec}. 

\item[$P_i$] 
Correction for light attenuation between the vertex and the $i$-th hit-PMT. 
In the previous analysis, a uniform light attenuation length was assumed, but this analysis accounts for variation of the optical properties of the water.

\item[$QE_i$] 
Correction for relative detection efficiency for the $i$-th hit-PMT. 
This factor includes all effects leading to a light sensitivity variation between individual PMTs.
This correction was re-evaluated using special data taken with a Ni-Cf source in February 2018. 
During this data taking, water convection in the SK-IV detector was intentionally evoked to realize uniform water quality across the detector volume.

\end{description}
Several changes are made to this analysis to improve energy reconstruction.  
The correction for PMT gain increase ($C_i$) is newly introduced for this analysis and improves the stability of the energy scale to be within $\pm 0.5\%$ over more than nine years of operation.
The corrections for PMT geometrical acceptance ($S(\theta_i,\phi_i)$) and light attenuation ($P_i$)  significantly improve the uniformity of the energy scale.
The correction for the dark rate contribution, $\varepsilon^{i}_{\mathrm{dark}}$, was also improved by incorporating PMT-by-PMT variation (in the previous analyses, an averaged value for all the PMTs was used.)
Figure~\ref{fig:neff_pos_dep} shows non-uniformity of $N_{\mathrm{eff}}$ for simulated electron events. The standard deviation of relative position dependence of $N_{\mathrm{eff}}$ was improved from $1.7\%$ to $0.5\%$. 
More details of the improved energy reconstruction method are given in Appendix~\ref{sec:app_erec}.

% <- Already described at an earl

\section{Energy calibration of the SK-IV detector} \label{sec_calib}

\subsection{LINAC calibration} \label{sec:linac}

\subsubsection{Overview}

The LINAC calibration system consists of an electron gun, a linear accelerator, beam pipes, collimators, magnets, and a beam trigger. 
The details of the system are described elsewhere~\cite{Nakahata:1998pz}. 
The last part of the beam pipe is inserted vertically into the SK detector and a single electron is injected at a time into the detector. 
The LINAC can inject electrons directly into the detector, which mimics the electrons produced by elastic scattering interactions of the $\mathrm{^{8}B}$ or $hep$ solar neutrinos. 

Mono-energetic single electrons are injected into the SK-IV detector at several positions. 
The energy of injected electron ranges from $4.4$~MeV to $18.9$~MeV in total energy. 

Since the previous report~\cite{Abe:2016nxk}, LINAC calibrations were conducted in in 2016 and 2017. 
We analyzed calibration data taken in SK-IV with the improved event reconstruction algorithm as described in Sec~\ref{sec_recon}. 
For the LINAC data analysis, a LINAC trigger signal from the trigger counter around the end of the beam pipe is required. 
To reject multiple electron events, separation by hit timing distribution after subtracting TOF~(defined in Sec.~\ref{sec_verdir}) is applied.
Since the timing of electrons occasionally overlaps, we rejected events whose energy is beyond $3\sigma$ from the mean of the energy distribution.

\subsubsection{Determination of the absolute energy scale in SK-IV}

\begin{figure*}[htp]
    \begin{center}
    \centering
        \includegraphics[width=0.8\linewidth]{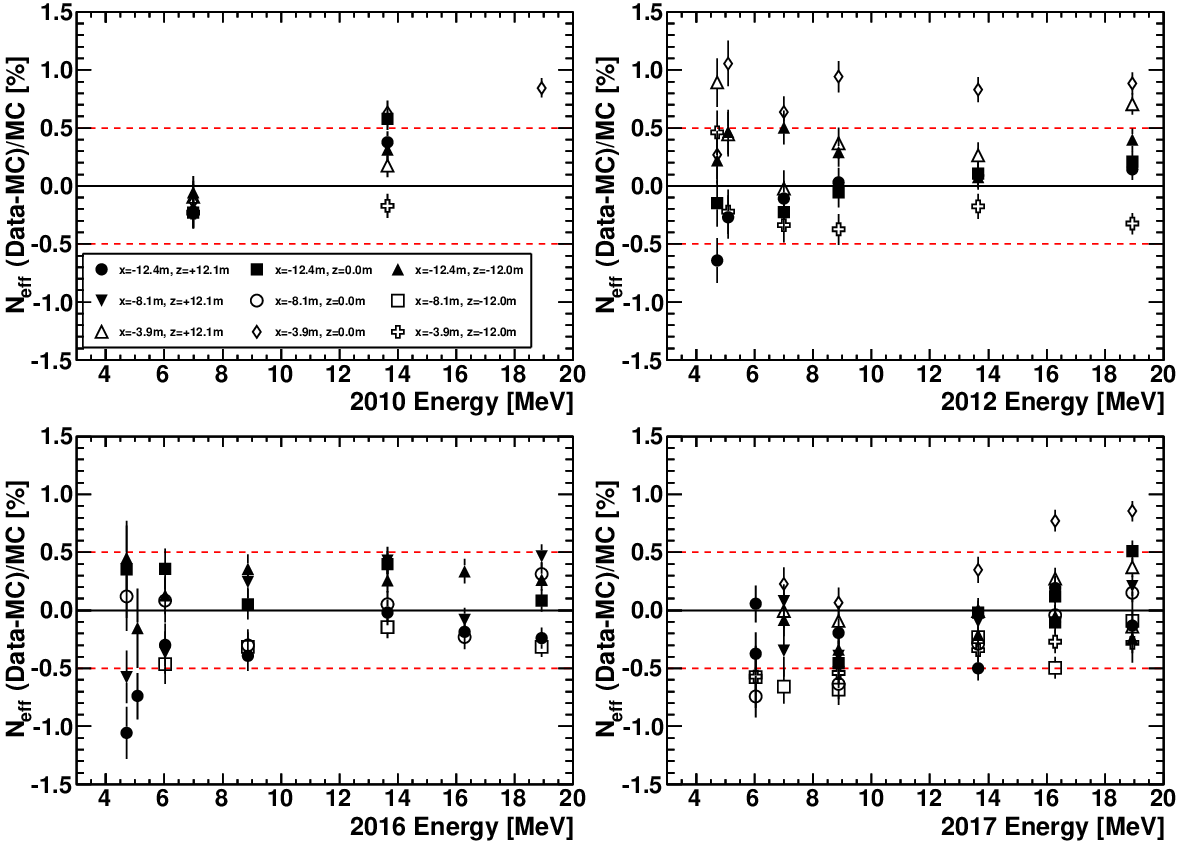}
    \caption{
    (Color online) The difference of the $N_{\mathrm{eff}}$ between the calibration data and the MC simulation.
    The markers show the position of the calibration. 
    The red (dashed) horizontal lines show $\pm0.5\%$. \label{fig:linac-neff-posi}}
    \end{center}
\end{figure*}

In this analysis, we use good-quality LINAC calibration data taken in 2010, 2012, 2016, and 2017 while the calibration data taken 
in 2009 is not used because of the lower quality of the water transparency at that time. 
The calibration data was taken at nine different positions in the SK detector. 
By comparing the peak of effective hits distribution~($N_{\mathrm{eff}}$ in Eq.~(\ref{eq:neff})) from these calibration data and simulated events, the absolute energy scale of the MC simulation code for SK-IV was tuned to match the LINAC data. 
In particular, the absolute correction factor for PMT quantum efficiency is determined by this analysis. 
Figure~\ref{fig:linac-neff-posi} shows the difference of the $N_{\mathrm{eff}}$ between the calibration data and the MC simulation 
after the tuning.

\begin{figure}[htp]
    \begin{center}
    \includegraphics[width=\linewidth]{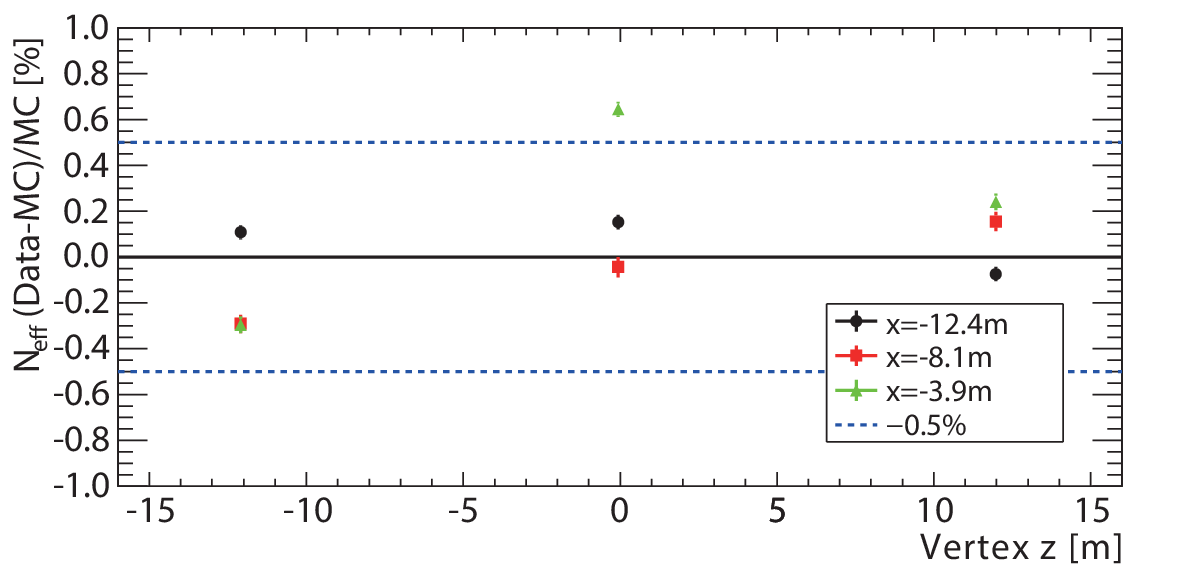}
    \caption{(Color online) The difference of the $N_{\mathrm{eff}}$ between the calibration data and the MC simulation after taking the weighted average. 
    The marker shows the position of the calibration. 
    The blue (dashed) horizontal lines show $\pm0.5\%$. \label{fig:linac-neff-posi-comb}}
    \end{center}
\end{figure}

To evaluate the position dependence of the energy scale during the SK-IV period, 
the weighted mean is obtained from the difference of $N_{\mathrm{eff}}$ between the data and MC simulation, 
where the injecting beam energies are $6.989$, $8.861$, $13.644$, and $18.938$~MeV. 
Figure~\ref{fig:linac-neff-posi-comb} shows the difference of data and MC simulation after taking the weighted mean for each position.
The remaining position-dependent energy scale uncertainty in SK-IV is calculated by taking the volume average 
for nine positions and 
the result is a difference of $0.40\%$. Therefore, $\pm0.40\%$ is used as the systematic uncertainty on the energy scale.

\begin{figure}[htp]
    \begin{center}
    \includegraphics[width=\linewidth]{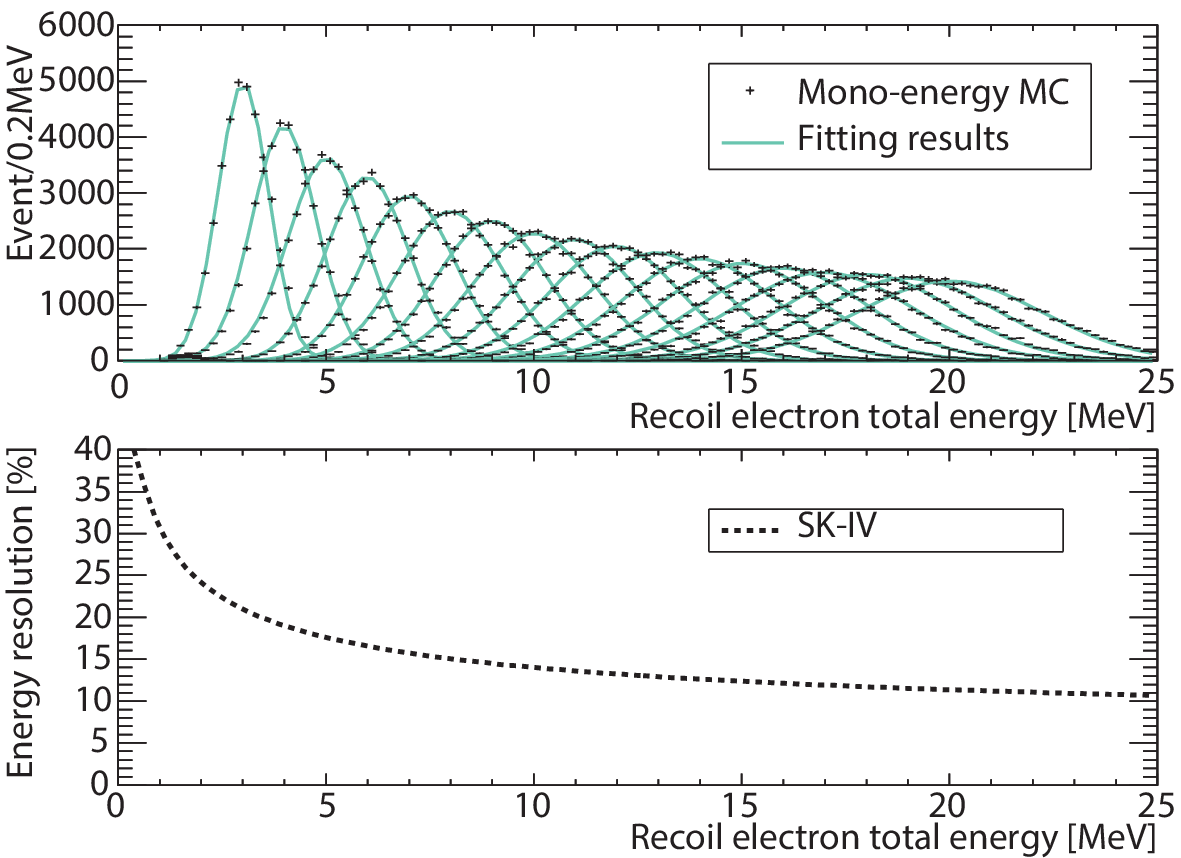}
    \caption{(Color online) (Top) The energy distributions of the mono-energy electrons, reconstructed in the standard fiducial volume. 
    (Bottom) The energy resolution~[$\%$] as a function of the electron's true total energy. \label{fig:ene_reso}}
    \end{center}
\end{figure}

After the MC simulation is tuned, it can be used to derive the conversion from $N_{\mathrm{eff}}$ to electron energy. 
For this purpose, mono-energetic electrons are generated with random vertex position and direction in the whole ID volume, 
and then the reconstructed energy distribution of events reconstructed in the standard fiducial volume is examined. 
Figure~\ref{fig:ene_reso} shows the reconstructed energy distribution of the mono-energetic electron simulations. 
Each distribution of reconstructed energy is fitted with a Gaussian function and the peak energy and the deviation are obtained. 
The fit results are used to create the conversion function, and determine the energy resolution function of the detector, which is found to be:
\begin{equation}
    \sigma(E)=-0.05525+0.3162\sqrt{E}+0.04572E
    \textrm{,}
\label{eq:res_fun}
\end{equation}
where $E$ is the reconstructed electron total energy in MeV.  
This is comparable to the energy resolution evaluated in the past publications: $\sigma(E)=-0.123+0.376\sqrt{E}+0.0349E$ in SK-III~\cite{Abe:2010hy}, 
and $\sigma(E)=-0.0839+0.349\sqrt{E}+0.0397E$ in SK-IV 1664 days~\cite{Abe:2016nxk}.
From here on, the energy scale tuned with this method will be used to estimate 
various performance metrics of the SK-IV detector.

\subsubsection{Energy resolution measurement} \label{sec:linac-reso}

\begin{figure*}[htp]
    \begin{center}
    \centering
        \includegraphics[width=0.8\linewidth]{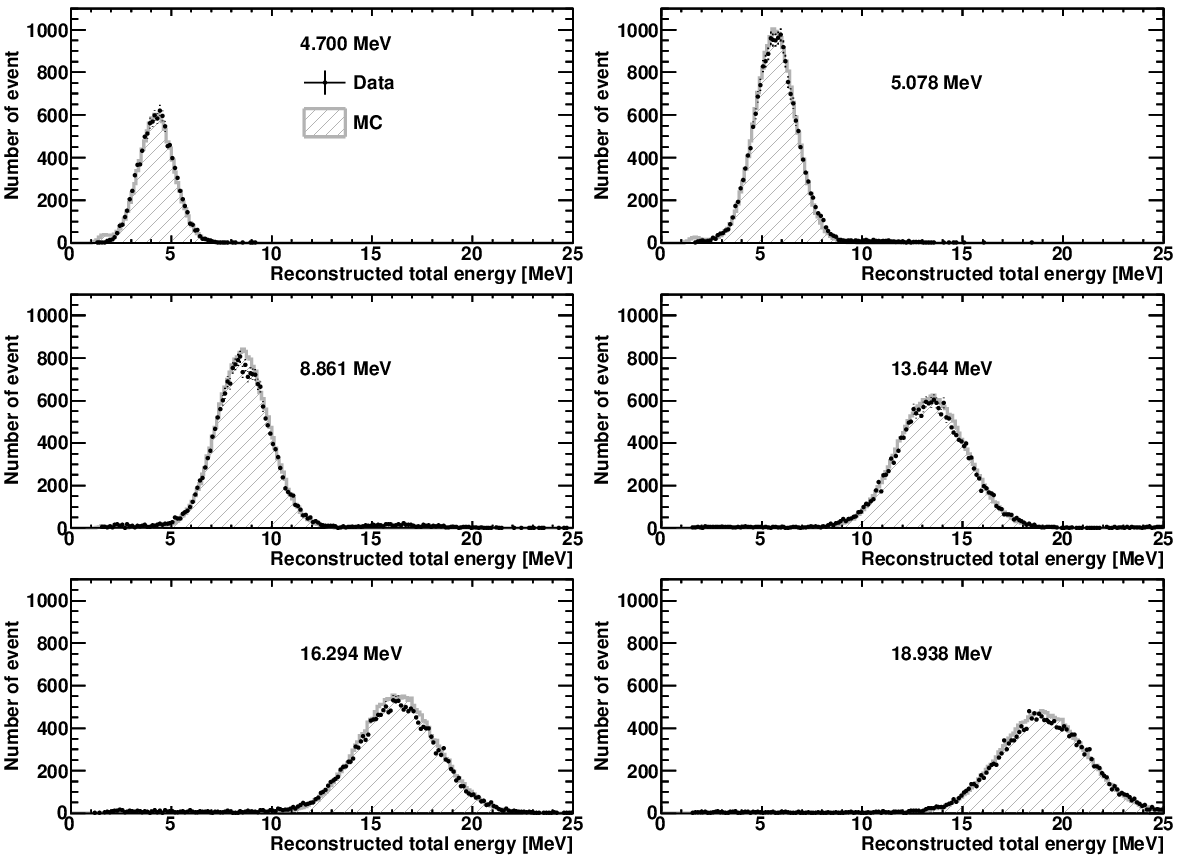}
    \caption{Typical energy distributions of the LINAC calibrations taken in 2016 at the position of $(X, Y, Z)=(-8.13, -0.707, -0.06)$~m. 
    The injected electron energies were measured by the Ge detector. \label{fig:linac-energy}}
    \end{center}
\end{figure*}

\begin{figure}[htp]
    \begin{center}
    \includegraphics[width=\linewidth,clip]{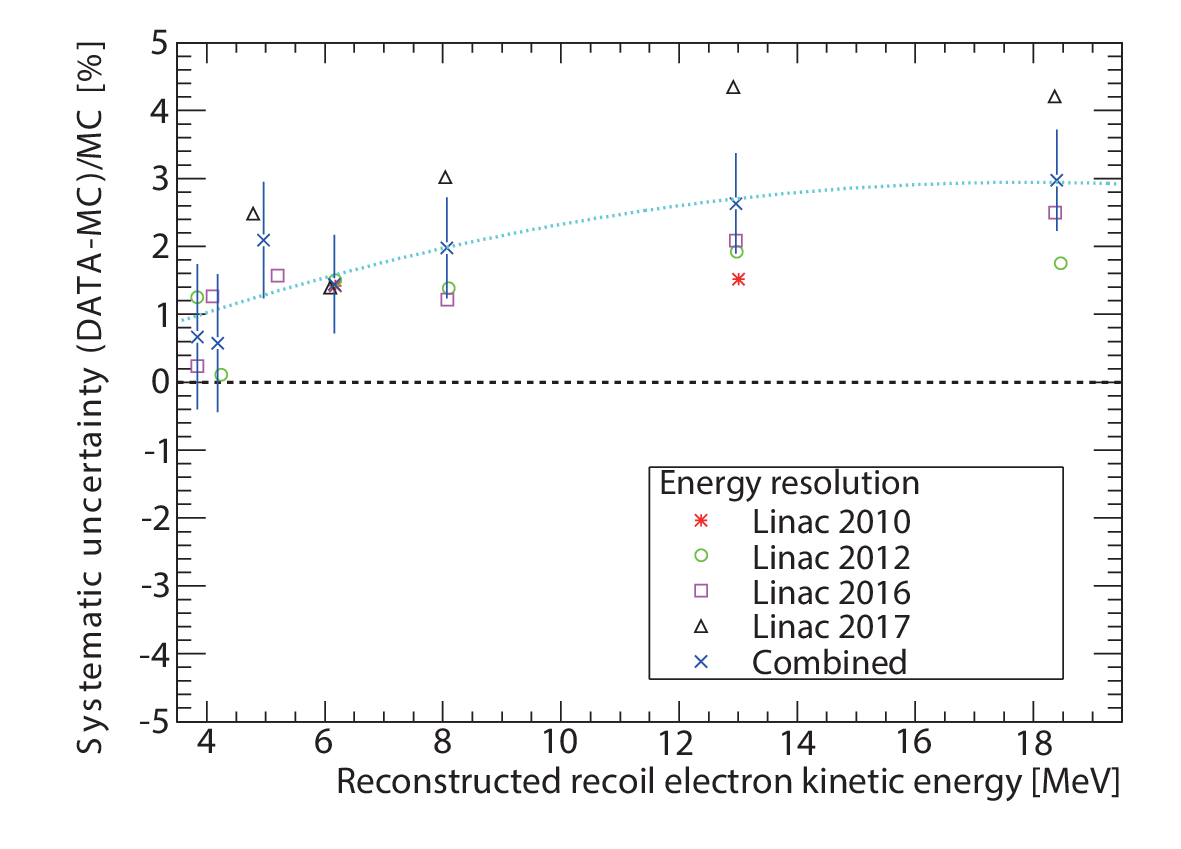}
    \caption{
    (Color online) The systematic uncertainties of the energy resolution are estimated by the LINAC calibration.
    The light-blue dashed line shows the fitting function for the combined values. \label{fig:linac_ene_reso}}
    \end{center}
\end{figure}

After determining the energy-scale function, the energy resolution of the LINAC calibration data and MC are compared.
Figure~\ref{fig:linac-energy} shows the typical energy distributions of the LINAC calibration.
The energy distributions are fitted with a Gaussian function and its peak value and deviation are obtained. 
Then, the energy resolution is calculated by dividing the deviation by the peak value. 
Figure~\ref{fig:linac_ene_reso} shows the systematic uncertainties of the energy resolution that comes from the difference between the data and MC.
As energy is higher, the uncertainty also increases. 
We used a polynomial function to describe the energy resolution as a function of the energy.
We estimated energy resolution systematic uncertainty in SK-IV is at most $3\%$ level in the range of the solar neutrinos.
This difference is taken into account in the analysis as an energy resolution systematic uncertainty. 

\subsubsection{Angular resolution measurement}

\begin{figure*}[htp]
    \begin{center}
    \centering
        \includegraphics[width=0.8\linewidth]{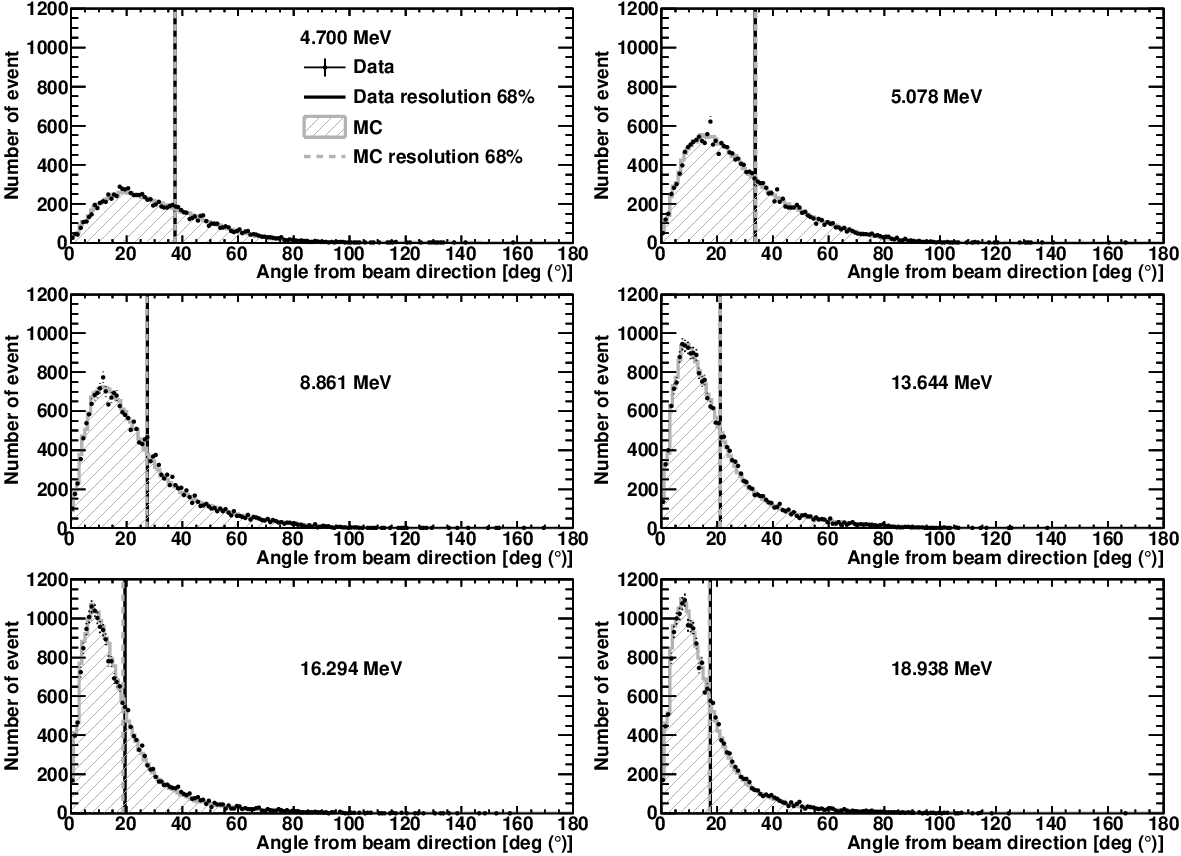}
    \caption{The typical opening angle distributions of the LINAC calibration data and the MC simulation in SK-IV. 
    The black solid~(gray dashed) line shows the angular resolution of data~(MC), which contains $68.3\%$ of events. 
    The data samples are the same as Fig.~\ref{fig:linac-energy}.
    \label{fig:linac-angle1}}
    \end{center}
\end{figure*}

Although the LINAC calibration is mainly used to determine the absolute energy scale of the detector, 
the electron beam makes it possible to evaluate the angular resolution for electrons in the water, since the direction of the LINAC beam is known. 
Figure~\ref{fig:linac-angle1} shows typical opening angle distributions between the direction of the beam injection and the reconstructed direction.
Because of the multiple scattering of an electron in water, the distribution of the opening angle 
spreads depending on the electron's energy.

\begin{figure}[htp]
    \begin{center}
\includegraphics[width=\linewidth]{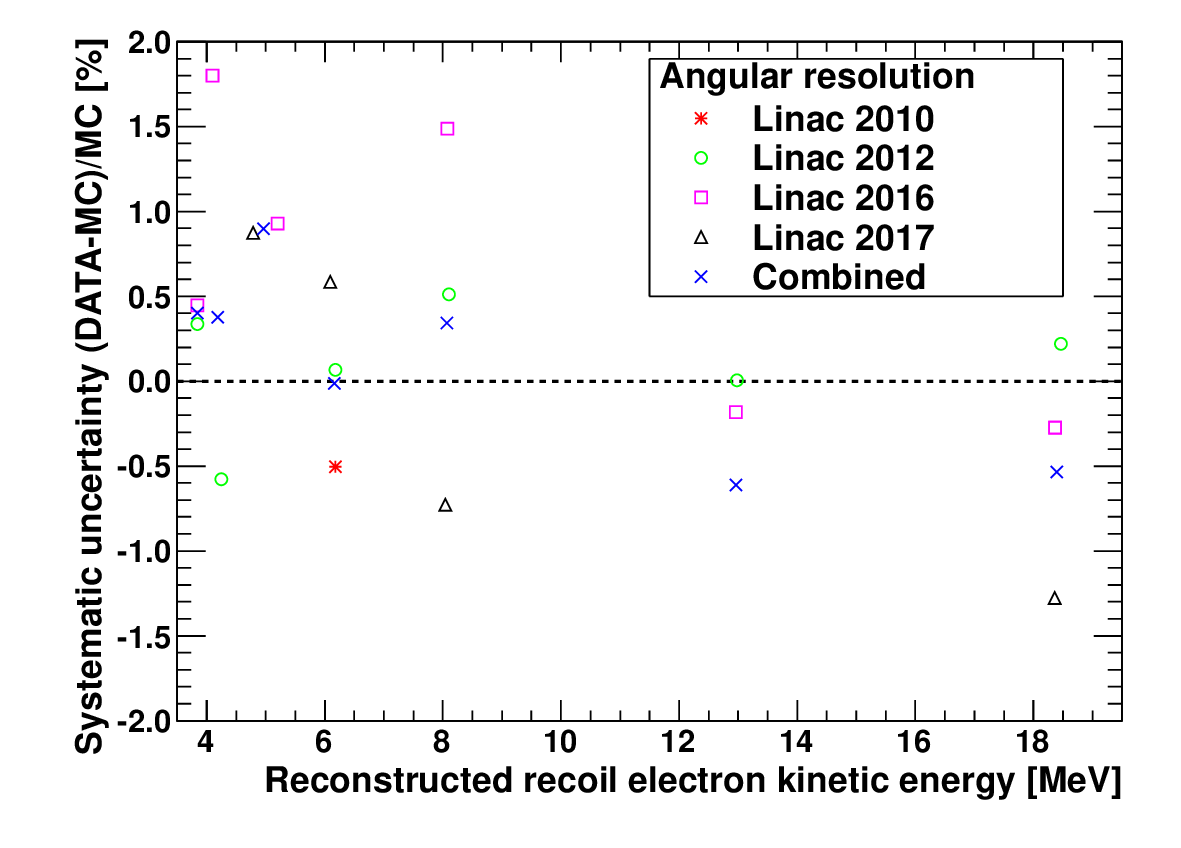}
    \caption{(Color online) The systematic uncertainties of the angular resolution estimated by the LINAC calibration. 
    The definitions of the markers are the same as Fig.~\ref{fig:linac_ene_reso}.
    \label{fig:sys_angle}}
    \end{center}
\end{figure}

Using the same definition of directional resolution as 
Sec.~\ref{sec_verdir}, the angular resolution of an electron is estimated. 
Figure~\ref{fig:sys_angle} compares the angular resolution obtained from LINAC data and from MC simulation.
This difference between the data and MC simulation in angular resolution leads to a difference of the fitted number of solar neutrino events of 
$0.1\% \sim 0.2\%$ in $3.49$--$19.49$ MeV energy region,
which is taken into account as a systematic uncertainty of the solar neutrino flux.

\subsection{DT calibration} \label{sec:calib_dt}

The LINAC uses a beam pipe inserted vertically through the calibration ports on the top of the SK detector. As such, electrons can only be injected in the downward vertical direction and so the LINAC system cannot be used to test the directional dependence of the energy scale.
Instead, the DT~(deuterium-tritium neutron generator) calibration source~\cite{Blaufuss:2000tp} is used to monitor the directional stability of the energy scale of the SK-IV detector. In addition, the DT device is designed to be portable and easy to operate and this allows us to calibrate the energy scale at many more positions than the LINAC calibration.  While the LINAC is used at $9$ positions, the DT calibrations are taken at $35$ different positions.

The DT device generates neutrons via the reaction of $\mathrm{^{2}H}+\mathrm{^{3}H} \to \mathrm{^{4}He}+n$. 
The neutrons are captured on $\mathrm{^{16}O}$ in the water and $\mathrm{^{16}N}$ is created via $(n,\alpha)$ reaction~\cite{David:1958}. 
The decay of $\mathrm{^{16}N}$, whose $Q$-value is $10.4$~MeV, mainly emits an electron with an energy of $4.3$~MeV and a single $\gamma$-ray with energy of $6.1$~MeV. These particles are emitted isotropically.
This allows probing directional systematic uncertainty in the SK energy reconstruction.

In the actual calibration procedure, the DT device is lifted about $100$~cm from the target position by the crane of the SK detector soon after its neutron generation. Although the $\mathrm{^{16}N}$ emits an electron and $\gamma$-ray isotropically, the upward events are affected by the shadow due to the DT device above the target position. In order to reduce such shadowing effect when calculating the average $N_{\mathrm{eff}}$, we removed the event whose direction is $0.8<\cos\theta_{\mathrm{DZ}}<1.0$, where the $\theta_{\mathrm{DZ}}$ is defined as the zenith angle respect to the vertical~($Z$) axis of the detector~(hereafter, we call this angle as the detector zenith angle).
Since the half life of $\mathrm{^{16}N}$ decay is $7.13$~s, we collected the data for $40$~s after the lifting of the DT device was completed. The former $20$~s is used for the signal region and the latter $20$~s is used for the background subtraction.

\begin{figure}[htp]
    \begin{center}
    \includegraphics[width=\linewidth,clip]{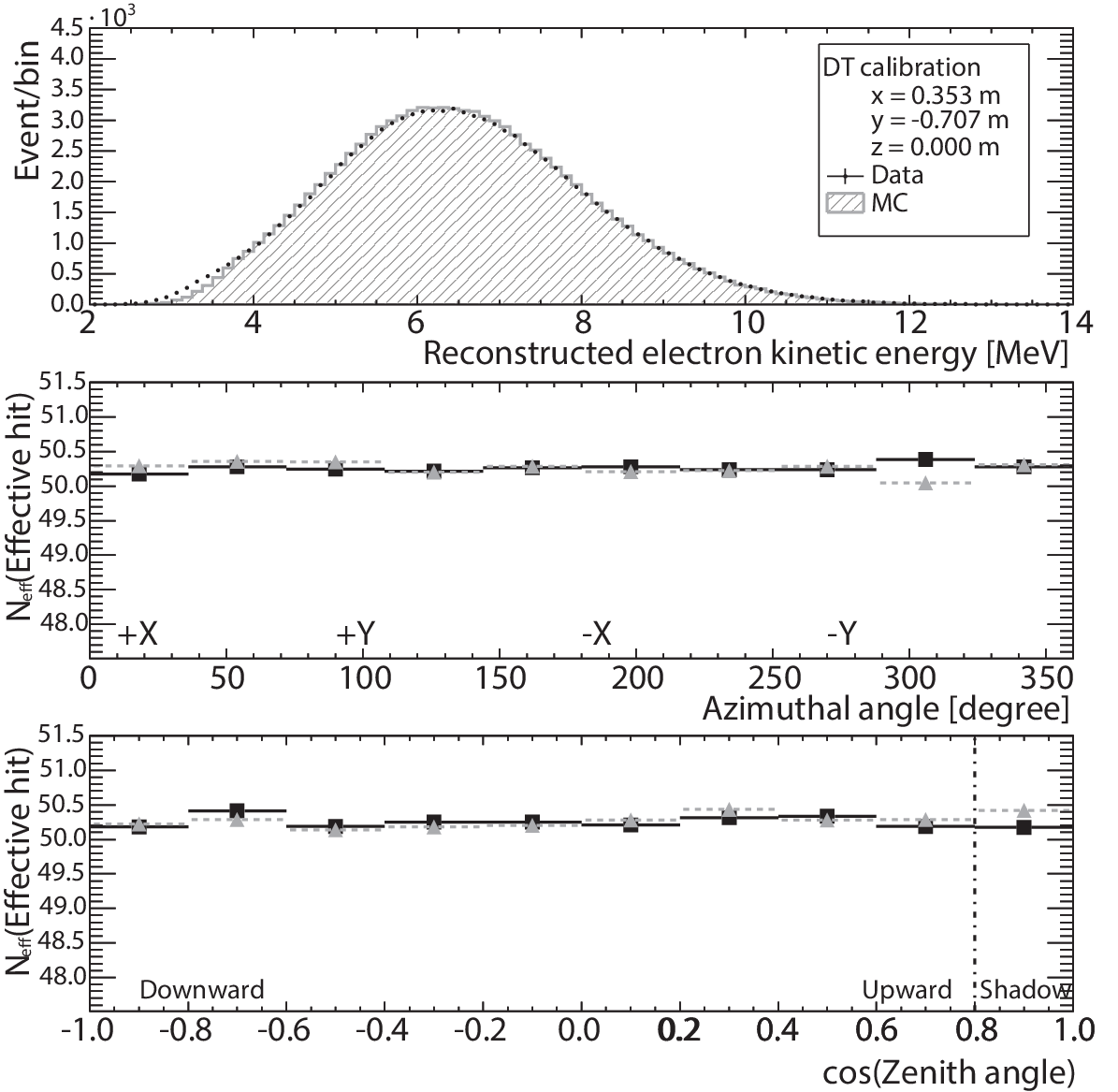}
    \caption{The typical distribution of the DT calibration at position $(X, Y, Z) = (+0.353, -0.707, 0.0)$~m taken on December 2016. The top panel shows the reconstructed kinetic energy of electrons, the middle panel shows the azimuthal angle dependence of the effective hits, and the bottom panel shows the zenith angle dependence of the effective hits. The effective hit is obtained by the peak position from the fitting with a Gaussian function. 
    \label{fig:dt-angle}}
    \end{center}
\end{figure}

Figure~\ref{fig:dt-angle} shows the typical distribution of the reconstructed energy and the directional dependence of the effective hits obtained from the DT calibration data by setting the calibration device at the central position of the SK detector together with the MC simulation. In this analysis, the detector azimuthal angle is defined as the angle in the $X$-$Y$ plane. 
The differences are sufficiently small for all azimuthal angles and cosine of the detector zenith angles. 

In order to check the stability of the energy scale throughout the SK-IV data set, we evaluate the time dependence of the energy scale by comparing the DT calibration data and the MC simulation. In this analysis, a position-weighted average is performed to obtain the one value from the deviation measured at various positions in the detector. For this purpose, we assigned a geometrical weight to each position where DT calibration data are taken. That weight is based on the fraction of the volume nearest each calibration position contributes to the total $22.5$~kt fiducial volume.
There is about $-0.5\%$ difference between the DT calibration data and the MC simulation, hereafter referred to as the offset.
Therefore, the DT calibration is used to evaluate relative variations from the offset, not absolute energy scale calibrations.
At the moment, we do not know the origin of the offset. 

\begin{figure}[htp]
    \begin{center}
    \includegraphics[width=\linewidth,clip]{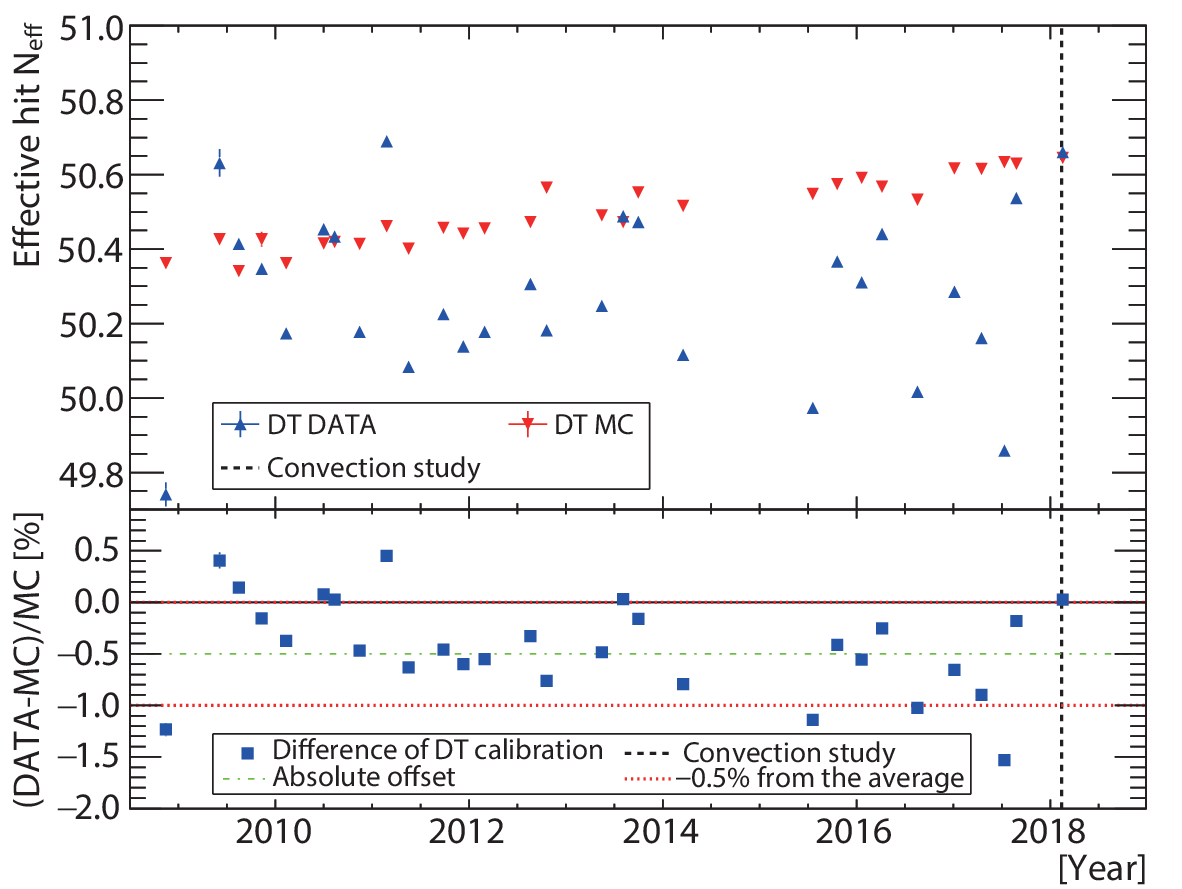}
    \caption{(Color online) The time variation of the effective hit~($N_{\mathrm{eff}}$) of the DT calibration data and MC simulation~(Top) together with their difference~(Bottom). The vertical dashed line shows the period of the convection study. The two horizontal red-dotted lines show $\pm0.5\%$ from the offset (green-dot-dashed line). 
    \label{fig:dt-time}}
    \end{center}
\end{figure}

Figure~\ref{fig:dt-time} shows the time variation of $N_{\mathrm{eff}}$ determined by the DT calibration.
The bottom panel of Fig.~\ref{fig:dt-time} also shows the difference between the data and the MC simulation. There may be a slight increase over the data-taking period, which probably originates from the modeling of the water transparency and the top-bottom asymmetry in the MC simulation described in Sec.~\ref{sec:water_sys}.  Although the difference between data and simulation fluctuates at the $\pm1\%$ level, the difference reduced to zero during the period of the convection study on February 2018 when the water quality throughout the detector is expected to be uniform, supporting the idea that the residual variation is related to non-uniformity of the water transparency.  

\begin{figure}[htp]
    \begin{center}
    \includegraphics[width=\linewidth]{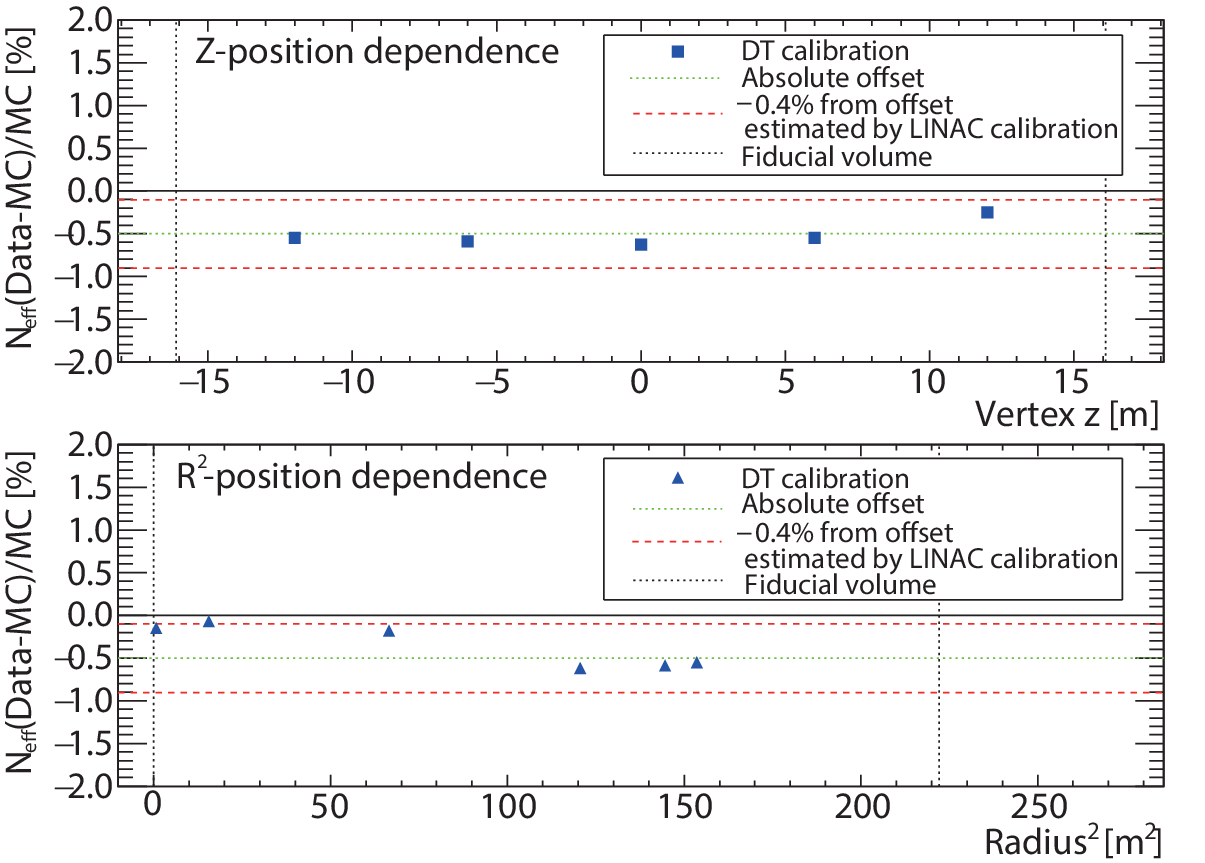}
    \caption{(Color online) The position dependence of the energy scale estimated from the DT calibration in SK-IV. The top~(bottom) panel shows the $Z$-position~($R^{2}$-position) dependence. 
    The two horizontal red-dashed lines show $\pm0.4\%$ from the offset (green-dotted line).
    There are $35$~calibration positions, i.e. seven calibration holes for the $X$-$Y$ plane, and five positions in height. For the $X$-$Y$ plane, $(X,Y)= (+0.353,-0.707)$, $(-3.889,-0.707)$, $(-8.131,-0.707)$, $(+10.958,-0.707)$, $(+0.353, \pm1201.9)$, and $(-12.370,-0.707)$~m. For the height, $Z=0$, $\pm6$, and $\pm12$~m. \label{fig:dt-position}}
    \end{center}
\end{figure}

Because the DT calibration data were taken at various positions in the SK detector, we can also evaluate the position dependence of the energy scale by comparing each calibration position. Figure~\ref{fig:dt-position} shows the variation of the energy scale with height and radius-squared.
Both of them are consistent within the $\pm0.4\%$ fluctuation from the offset,
which is the systematic uncertainty of the position dependence estimated from the LINAC calibrations in Sec.~\ref{sec:linac}.

\begin{figure}[htp]
    \begin{center}
    \includegraphics[width=\linewidth]{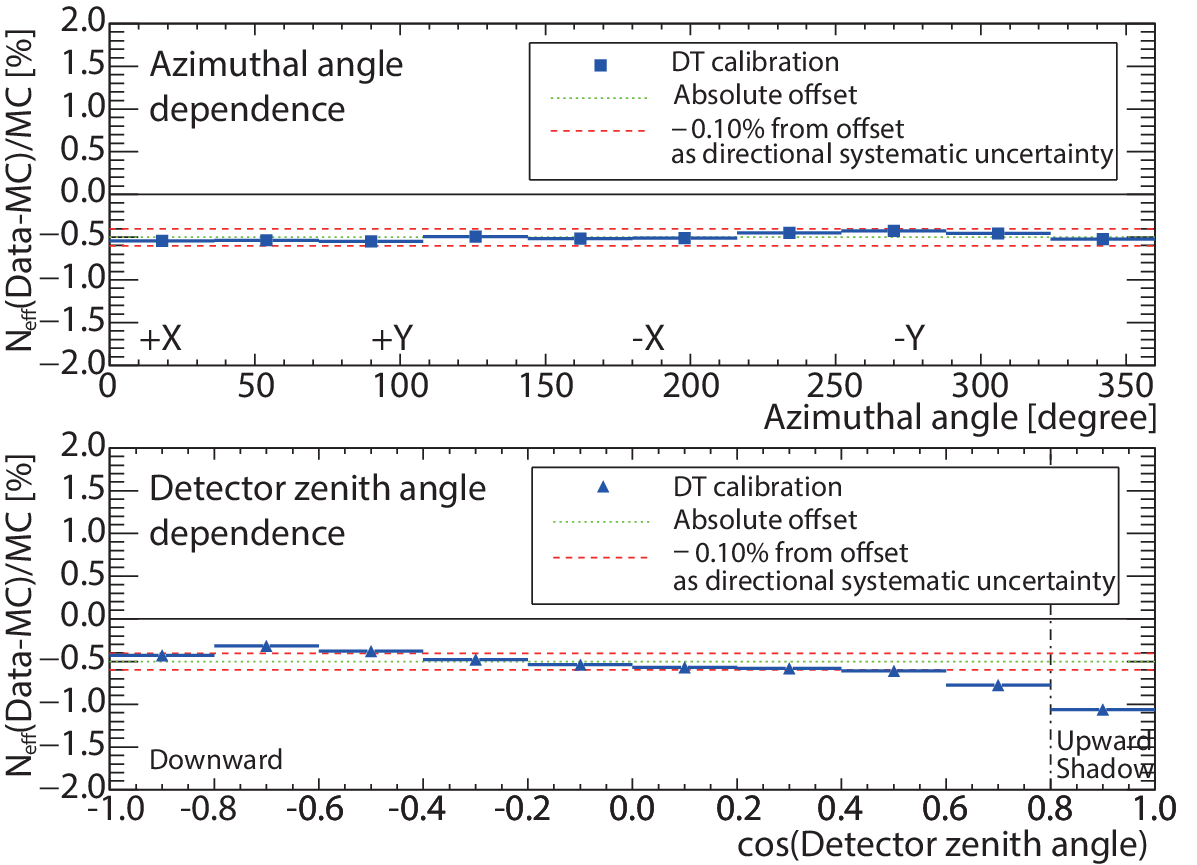}
    \caption{(Color online) The directional dependence of the difference of the effective hit between the calibration data and the MC simulation.  
    The value is obtained by considering the volume weight of the calibration position. 
    The two horizontal red-dashed lines show 
    $\pm0.1\%$
    from the offset~(green-dotted line). \label{fig:dt-direction}}
    \end{center}
\end{figure}

Figure~\ref{fig:dt-direction} shows the directional dependence of the energy scale on the detector azimuthal angle and detector zenith angle.  For the azimuthal angle, the directional systematic uncertainty is estimated by the variation from the offset. As seen in the top panel of Fig.~\ref{fig:dt-direction}, the fluctuation of the azimuthal angle dependence is 
less than $\pm0.1\%$
level throughout the detector.

For the detector zenith angle, the systematic uncertainty on the directional dependence of the energy scale is estimated using the same way as the azimuthal angle. 
However, the $N_{\mathrm{eff}}$ in the most upward bin~($0.8<\cos\theta_{\mathrm{DZ}}<1.0$) is noticeably reduced, an effect that was also seen in the previous analysis~\cite{Abe:2016nxk}. This is attributed to the increased shadowing of PMTs from the DT generator.  This is not implemented in the current DT simulation, so this bin is excluded when setting the anisotropy systematic.  We therefore assign a systematic uncertainty of $\pm0.10\%$ on the energy scale from the directional dependence.

\subsection{Summary of the systematic uncertainty of the energy scale in SK-IV}\label{sec_summary_energy_scale}

Table~\ref{tbl:linac_sys} shows a summary of the systematic uncertainties in the energy scale determination in SK-IV.
\begin{table}[htp]
    \begin{center}
\caption{Summary of the systematic uncertainties of the energy scale in SK-IV}
        \label{tbl:linac_sys}
        \begin{tabular}{lc}
            \hline
            \hline
            Source & Uncertainty~[$\%$]\\ \hline
            Position dependence from LINAC  & $\pm0.40$ \\
            Ge detector accuracy in LINAC   & $\pm0.21$ \\
            Water transparency during LINAC & $\pm0.11$ \\
            Directional dependence from DT  & $\pm0.10$ \\
            \hline 
            Total & $\pm0.48$ \\
\hline
            \hline
        \end{tabular}
    \end{center}
\end{table}
The position dependence of the energy scale is estimated to be $\pm0.40\%$ by taking the volume average of the difference of the effective hits in data and MC at different LINAC injection positions, as shown in Fig.~\ref{fig:linac-neff-posi-comb}. The accuracy of the LINAC calibration itself is set by the beam energy determination, which uses a  Ge detector and is estimated to be accurate to $\pm0.21\%$~\cite{Nakahata:1998pz}. Extrapolating this calibration to other time periods also has an uncertainty because of variations in the water transparency over time as shown in Fig.~\ref{fig:water_t}. These are continuously monitored with the decay electrons from stopping cosmic-ray muons, but the precision is limited by the number of stopping muons recorded during the reference period of LINAC operation. The energy scale uncertainty from this is estimated as $\pm 0.11\%$.  Finally, the directional dependence of the energy scale is evaluated by using the DT calibrations and the resulting uncertainty is 
$\pm0.10\%$ 
as described in Sec.~\ref{sec:calib_dt}. In total, we estimate an uncertainty of 
$\pm0.48\%$ on the absolute energy scale in SK-IV.

\section{Data Analysis} \label{sec_data_ana}
\subsection{Event selection}

Events used in the analysis must pass a series of selection criteria. 
A basic explanation of each selection step is given here, with more details to be found in the previous publication~\cite{Abe:2016nxk}. 
Using the full data sample of SK-IV, the selection criteria are optimized to maximize the significance $S/\sqrt{S+B}$, where $S$ and $B$ are defined as the number of signal and background events after the selection cut, respectively. 

\subsubsection{Run selection for solar neutrino analysis} \label{sec_runselection}

The typical run length in SK-IV is one day. 
For this solar neutrino analysis, a series of run selections are applied to select good-quality data.
As in previous analyses, short runs~(of less than five minutes), calibration runs, runs with hardware and/or software troubles, and 
runs with strange OD rates or strange numbers of bad PMT channels are rejected.
After the good run selection, the total live time for the solar analysis in SK-IV is 2970~days, from September 2008 to May 2018 as listed in Table~\ref{tb:phase}. 
This corresponds to an extra 1306 days more than the previous publication~(1664 days). 
Since May 2015, the experiment has run with a lower software trigger threshold as explained in the next section. In February and May 2018, convection studies were performed to obtain better estimates of systematic uncertainties; during these periods events below $5.49$~MeV are rejected since a large number of background events originating 
from the Rn in the detector are observed~\cite{nakano2023}.
Table~\ref{tb:dataset} summarises the data sets of solar neutrino analysis in SK-IV.
\begin{table*}[htp]
\begin{center}
\caption{Summary of the data set in SK-IV. During the two convection study periods, a higher reconstructed energy threshold of $5.49$~MeV is used.} 
\label{tb:dataset}
\begin{tabular}{l@{\quad}r@{\quad}rr}
\hline
\hline
Period &\multicolumn{1}{c}{Start} & \multicolumn{1}{c}{End} & Live time [day] \\ \hline
Threshold $34$ hits & 6th Oct. 2008 & 1st May 2015 & $2047$ \\
Threshold $31$ hits & 1st May 2015 & 30th May 2018 & $923$ \\
Total  & 6th Oct. 2008 & 30th May 2018 & $2970$  \\
\hline
Convection study 1 & 7th Feb. 2018 & 14th Apr. 2018 & $46$  \\
Convection study 2 & 9th May  2018 & 16th May  2018 &  $7$  \\
\hline
\hline
\end{tabular}
\end{center}
\end{table*}

\subsubsection{Trigger scheme}

The online software trigger in SK-IV is based on the number of coincident PMT signals. 
A lower threshold is desirable for the solar neutrino analysis, but causes the data rate to rise sharply, so events must pass a software trigger based on the number of hit-PMT within a $200\,\textrm{ns}$ window.
In the case of the solar neutrino events, the energy of recoil electrons is approximately proportional to the total number of the hit-PMTs
since most of the hit-PMTs receive one photon. 

\begin{figure}[htp]
    \begin{center}
\includegraphics[width=\linewidth]{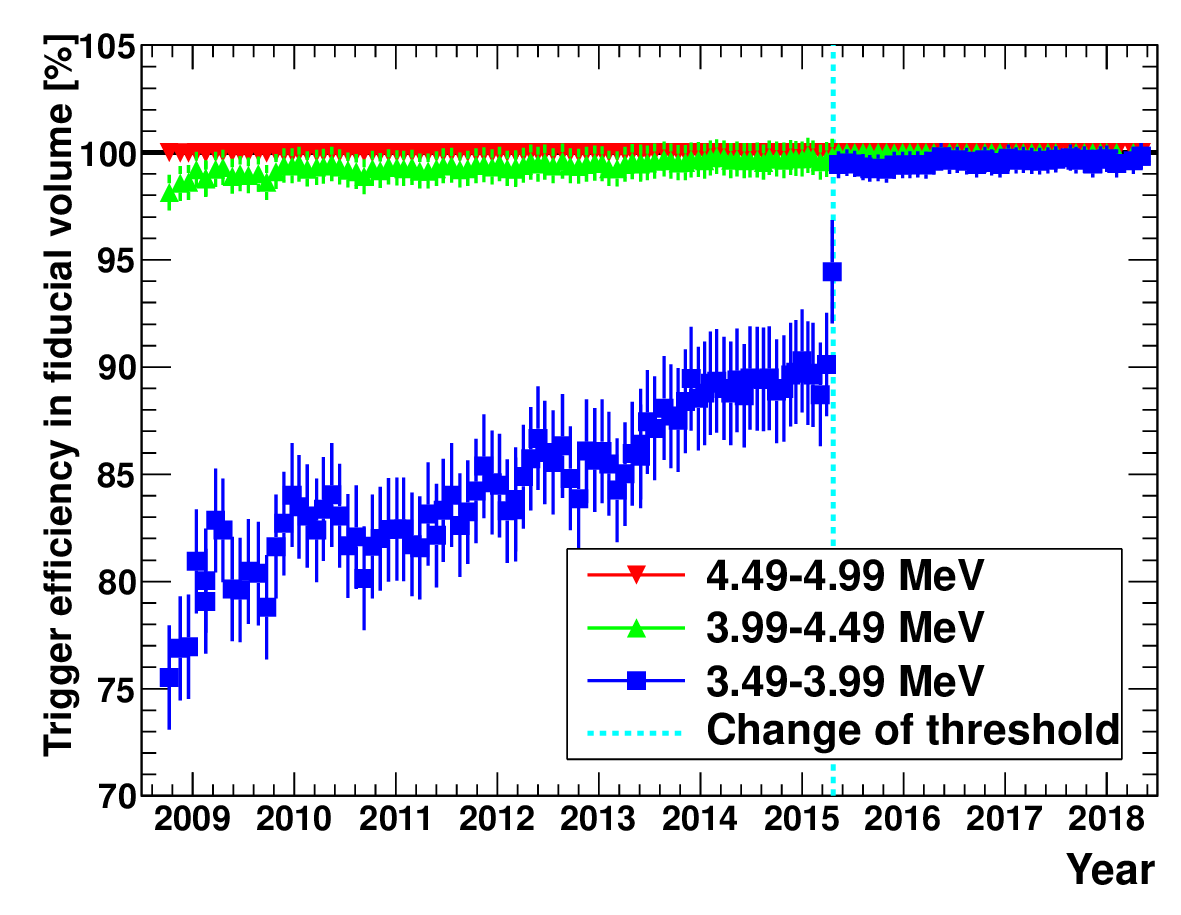}
    \end{center}
    \caption{(Color online) The time variation of the trigger efficiencies estimated by the MC simulation of $\mathrm{^{8}B}$ solar neutrinos. The trigger efficiency is highly correlated with the gain of PMTs~(Fig.~\ref{fig:gain}) and water transparency~(Fig.~\ref{fig:water_t}). 
    After changing the threshold of data acquisition~(vertical light-blue dashed line), the efficiency of $3.49$--$3.99$~MeV is improved significantly.
\label{fig:trig_mc}}
\end{figure}

In May 2015, the software trigger threshold was changed from 34~hit-PMTs to 31~hit-PMTs~\cite{Nishino:2009zu,Nakano:2015wdv}. 
Figure~\ref{fig:trig_mc} shows the time variation of the estimated trigger efficiencies near the trigger threshold using the MC simulation. For events above $4.99$~MeV the trigger efficiency is basically $100\%$.  For events between $3.49$ and $3.99$~MeV, there was a noticeable inefficiency due to the 34-hit threshold, this is significantly reduced by moving to the lower 31-hit threshold.

\subsubsection{1st reduction}

The first step of the data reduction is the elimination of obvious background events and noise originating from hardware in the detector.   The method of selection is the same as in the previous analysis. 

\begin{itemize}
\item 
Since the PMTs are operated at high voltage, an arc discharge can occasionally occur between the dynodes during data taking. When this happens, other PMTs also register optical signals.
If the maximum charge of a single PMT is larger than $50$~p.e. and there are more than three hits on the surrounding $24$~PMTs, the event is identified as a `flasher' event, and removed.
\item
Events that occur less than 50~$\mu$s after a previous event are rejected to remove decay electrons from stopping cosmic-ray muons,
and instrumental noise from PMT after pulses. 
\item
Several calibration sources remain inside the detector during data taking. Events that are triggered by external calibration triggers and scheduled calibration events are obvious and rejected.  In addition, the calibration sources inevitably have higher radioactivity than the water they displace.  Therefore events below $4.99$~MeV are excluded if their vertex position is less than $200$~cm from the calibration sources or $100$~cm from the cables supporting the sources (all cables run parallel to the $Z$-axis from the top of the detector to the source position).
\end{itemize}

\subsubsection{Spallation cut} \label{sec:spallation}

Cosmic-ray muons pass through SK at ${\sim}2$~Hz, and will occasionally shower; some of these showers produce hadrons such as pions or neutrons (hadronic showers)~\cite{Li:2014sea, Li:2015kpa, Li:2015lxa}. The showers can interact with or break apart $\mathrm{^{16}O}$ nuclei, generating radioactive isotopes~(spallation).  Those isotopes that undergo $\beta$s and/or $\gamma$ decays have similar reconstructed energies as solar $\mathrm{^{8}B}$ and $hep$ neutrino interactions. Their half-lives range from milliseconds to tens of seconds, with the most abundantly produced isotope, $\mathrm{^{16}N}$, having a half-life of $7.3$~s. This ``spallation background'' dominates all others above $5.5$~MeV and motivates a number of selection cuts to reduce its impact.

Previous solar neutrino analyses of SK data reduced this background using three variables calculated and checked against any cosmic-ray muons in the previous $100$~s:  the closest distance between the muon track and the candidate; the time difference between the muon and the candidate and the excess light from the muon compared to that expected from minimum ionization. A likelihood was formed from these three variables, and a cut was placed on this likelihood. The cut was tuned to have $20\%$ loss of solar neutrinos (effectively the same as a dead time), and was $90\%$ effective at removing spallation. The details of
the previous method can be found in Ref.~\cite{Hosaka:2005um,Abe:2016nxk}.
In this analysis, the ``spallation cut'' is complemented and improved. The improvements are summarised in this section, and full details will be provided in the recent publication~\cite{scott2021}.

Firstly, a completely new method directly tags hadronic showers responsible for isotope production. Such showers typically produce a large number of neutrons, and in pure water, these neutrons are captured on hydrogen via $n + \mathrm{^{1}H} \rightarrow \mathrm{^{2}H} + \gamma~(2.2~\mathrm{MeV})$, and this can be used to tag neutrons.  While the total light produced by a $2.2$~MeV $\gamma$ is small (typically only seven detected photo-electrons) and therefore difficult to see, the large neutron multiplicity allows this method to remove $54\%$~of spallation-induced events with only a $1.4\%$ loss of signal efficiency.  The updated cut depends on the location of the 
solar neutrino candidate relative to a cluster of tagged neutrons, the multiplicity of the neutron cluster, and the time between the candidate and the muon responsible for the shower. Due to the difficulty of triggering on such a small signal, the data acquired through a special Wide-band Intelligent Trigger~(WIT)~\cite{Carminati:2015jna, Elnimr:2017nzi} 
must be used; its trigger efficiency for a $2.2$~MeV $\gamma$ is about $13\%$. The WIT was only available for $388$~days, towards the end of the SK-IV period.

Secondly, a new method uses the spallation decays to veto themselves. 
Events within the standard fiducial volume that pass most
event quality cuts, 
and above $4.99$~MeV are used to create a sample of ``veto events''. 
Solar neutrino candidates within $4$~m and $60$~s of the veto events are removed. 
This removes $47\%$ of spallation events with an effective dead time of $1.3\%$. 

\begin{figure}[htp]
    \begin{center}
        \includegraphics[width=\linewidth]{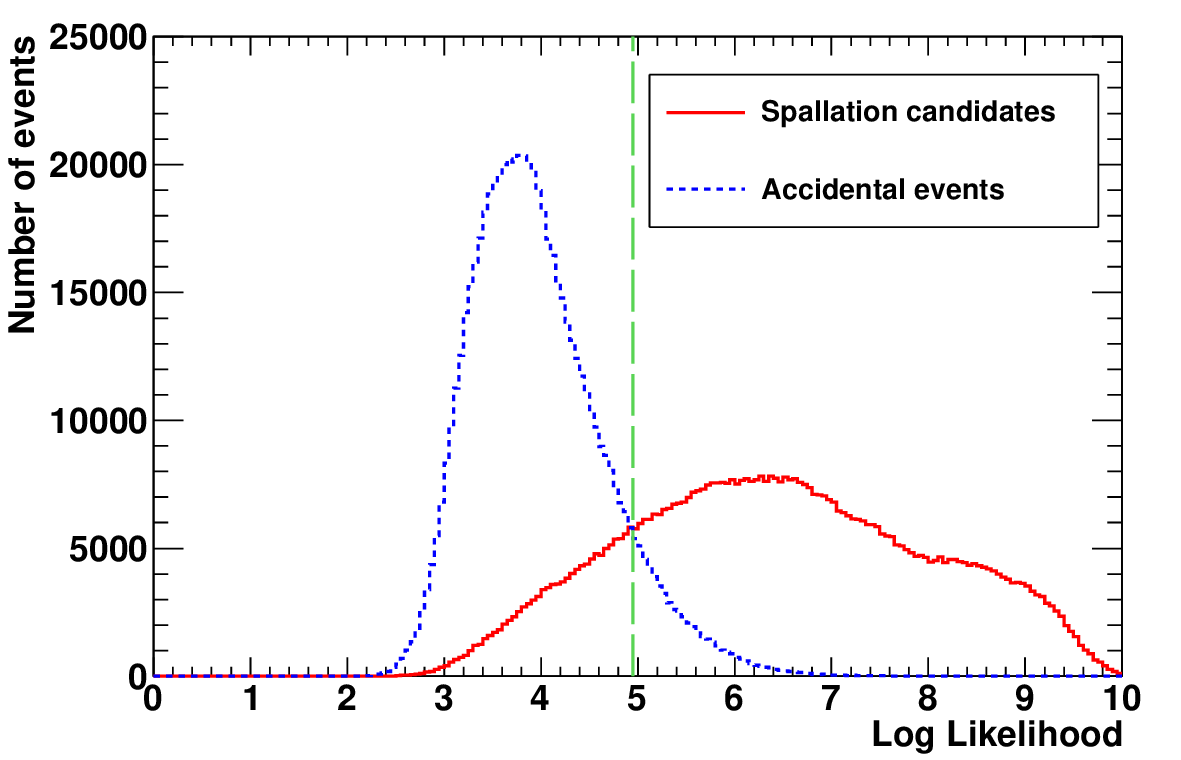}
    \end{center}
    \caption{(Color online) Comparison of the log-likelihood for the signal and background of the non-neutron data period, without the WIT system. 
    The spallation signal is shown by the red solid line, accidentals by the blue dotted line, and the tuned cut value is shown in the vertical green dashed line. 
    The multiple spallation cut is already applied, so only $82\%$ of remaining spallation needs to be removed to achieve $90\%$ overall spallation removal effectiveness.}
    \label{fig:spa_loglike}
\end{figure}

Thirdly, the existing spallation cut was updated and re-tuned to more effectively tag spallation remaining after the two new cuts were applied. The updates included: a better muon fitter; the removal of artificial saturation of PMTs; updated probability density functions~(PDFs) for the three input variables and the addition of a fourth variable. This new input variable utilizes the distance between the solar neutrino candidate and the peak of the muon energy loss along the track. This peak typically corresponds to the position of electromagnetic showers.  The resulting likelihood cut value was tuned separately for cases with and without neutron cluster data, to achieve the same ($90\%$) reduction in spallation-induced events as used previously.
Figure~\ref{fig:spa_loglike} shows a comparison of the likelihood of the signal and background of the non-neutron data period.

\begin{figure}[htp]
    \begin{center}
        \includegraphics[width=\linewidth]{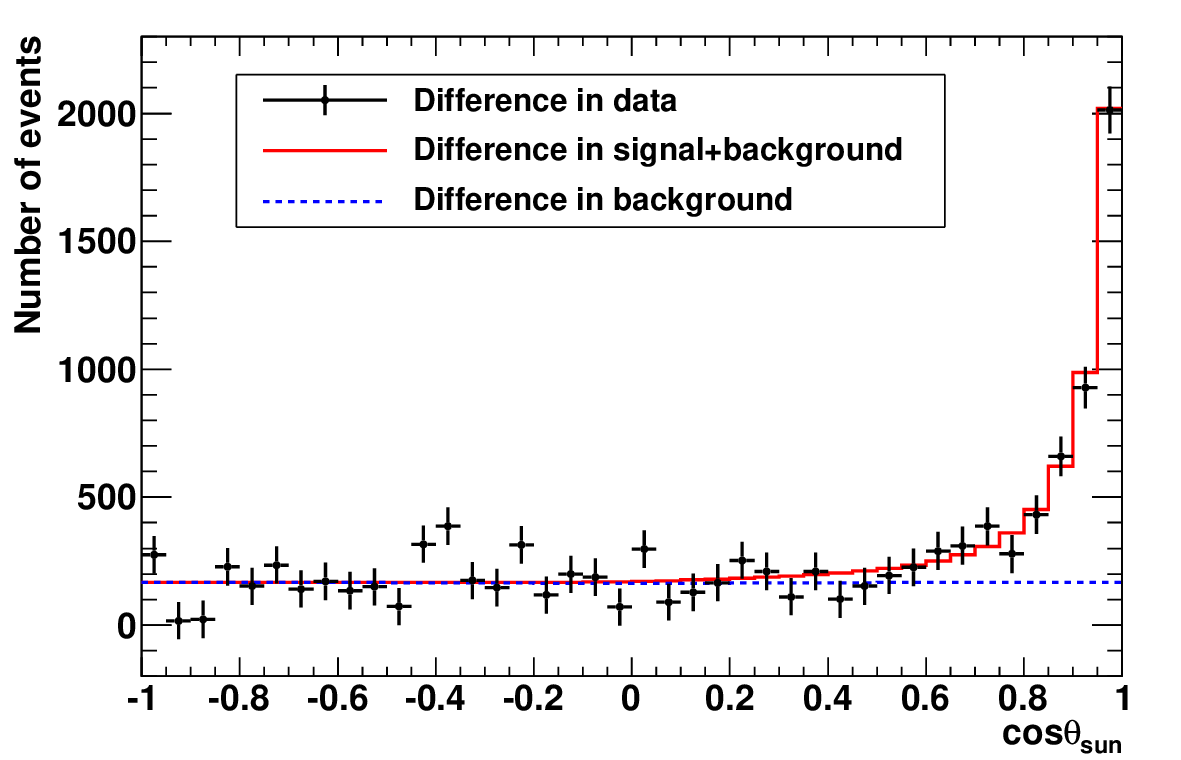}
    \end{center}
    \caption{(Color online) Difference in solar neutrino signal events between $5.99$--$19.49$~MeV using the updated spallation cuts compared to the previous method. 
    The red solid~(blue dashed) line is the difference in total~(background), respectively. 
    The increase in background is from the change of cut tuning unrelated to spallation.}
    \label{fig:spa_diff}
\end{figure}

The improvement is quantified in the reduction of the effective dead time: $8.6\%$~($10.5\%$) for the dataset with~(without) neutron clusters. 
The total SK-IV effective dead time from these cuts is $10.2\%$. 
This leads to an increase of $12\%$ in the number of solar neutrino signal events~(${\sim}7000$~events) in the 2970-day sample.
Figure~\ref{fig:spa_diff} shows the difference in the number of events accepted using the updated spallation cuts, which includes a large contribution to the solar neutrino signal sample.

\subsubsection{Summary of the event selection}

After the spallation cut, we applied the same event selections as the previous analysis~\cite{Abe:2016nxk}, with cut points re-tuned, to obtain the final data sample.

\begin{figure}[htp]
    \begin{center}
	    \includegraphics[width=\linewidth]{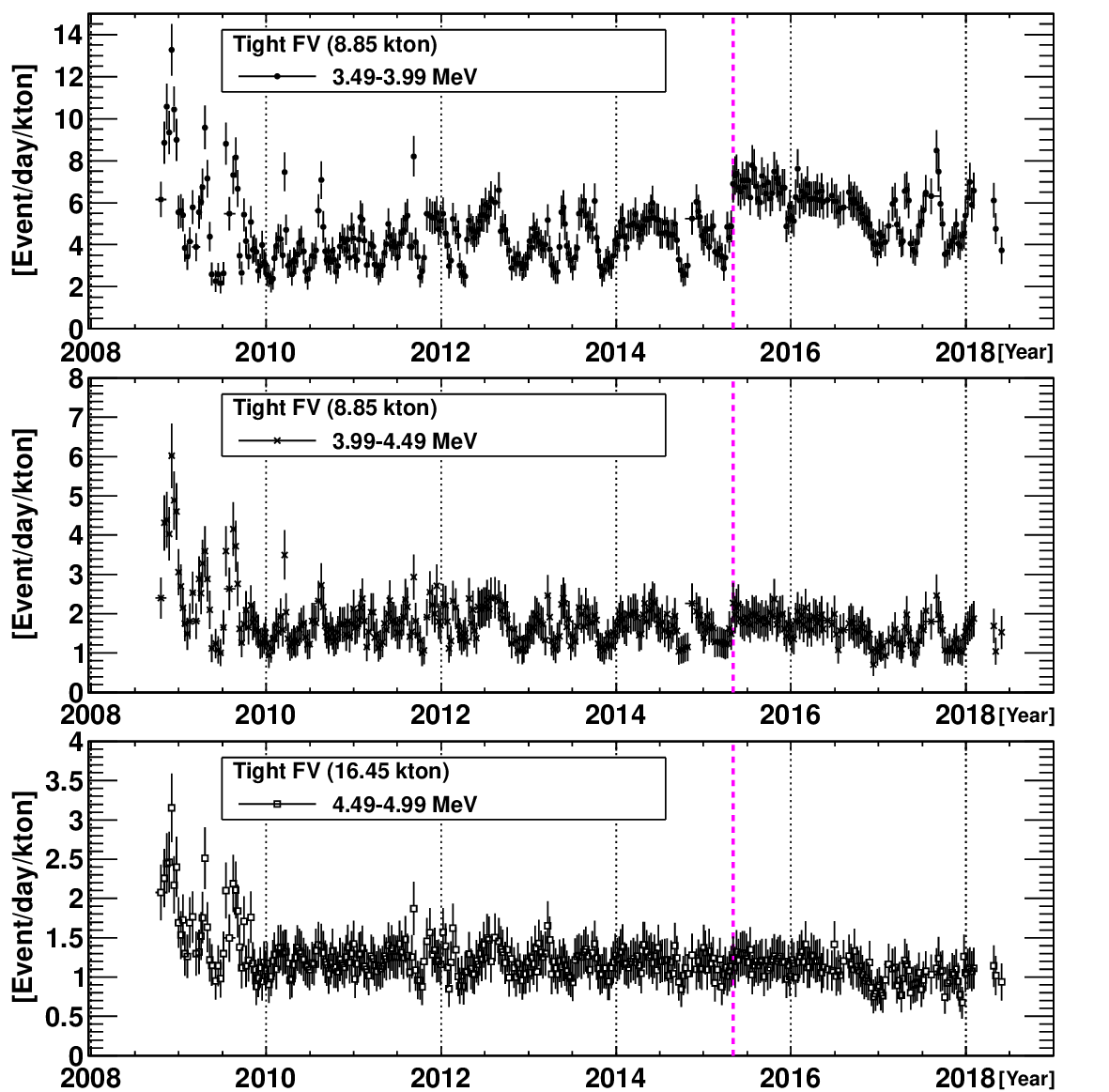}
	    \includegraphics[width=\linewidth]{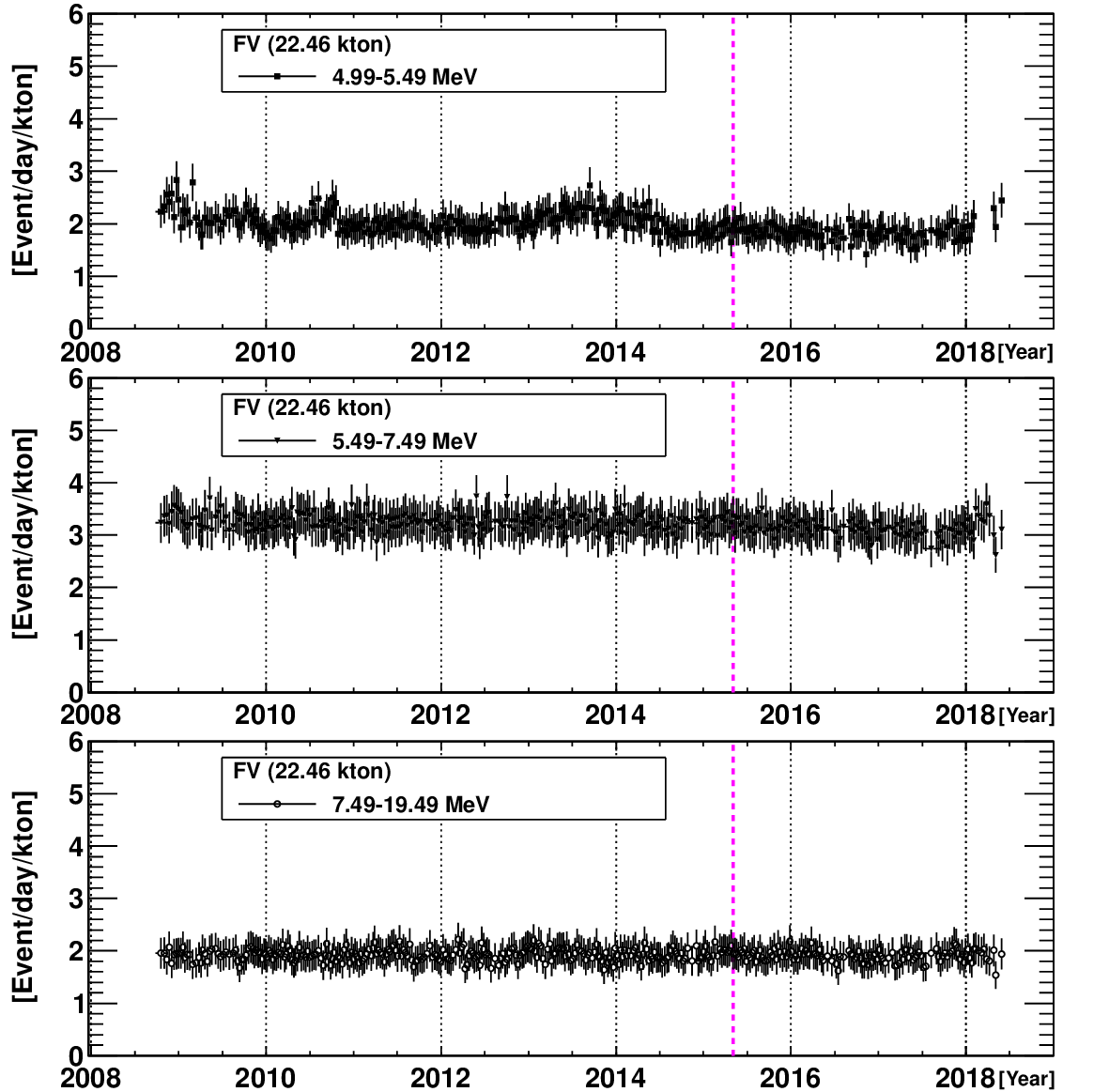}
	\end{center}
	\caption{The time variation of event rates in the 2970-day final data sample. 
	The vertical dashed line shows the threshold change from $34$~hits to $31$~hits. 
	The events below $5.49$ MeV during convection periods in 2018 are removed. \label{fig:sk4_evrate}}
\end{figure}

Figure~\ref{fig:sk4_evrate} shows the event rate of the SK-IV 2970-day final data sample.
The event rate in $3.49$--$3.99$~MeV increased after the software trigger threshold was lowered in May 2015. 
Other fluctuations in the event rate below $4.49$~MeV are strongly correlated with the variation of water attenuation length shown in Fig.~\ref{fig:water_t}. 
The lower energy ranges are most affected; event rates in higher energy ranges are relatively stable throughout the data set.

\begin{figure}[htp]
    \begin{center}
        \includegraphics[width=\linewidth]{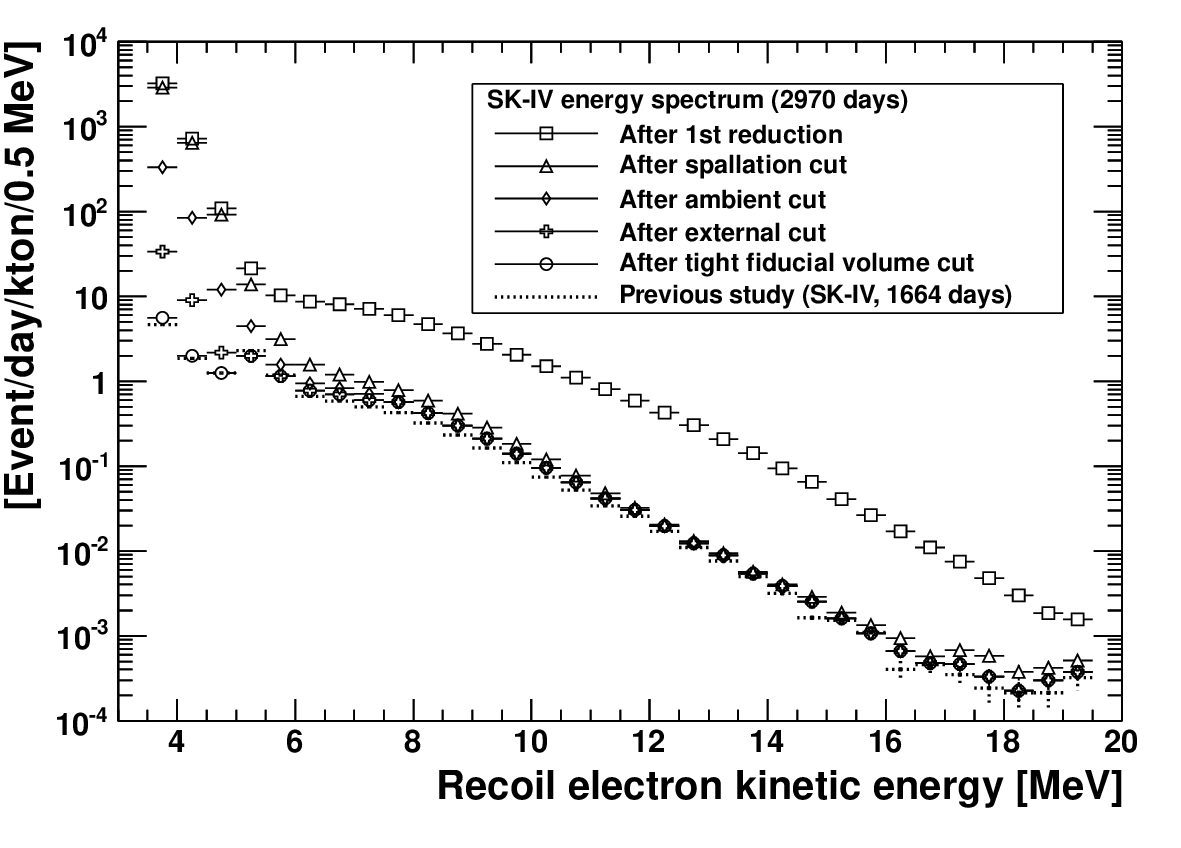}
    \end{center}
    \caption{Reconstructed electron kinetic energy distribution of the observed data after each reduction step.
    The square, triangle, rhombus, cross, and circle correspond 
    to after 1st reduction, spallation cut, ambient cut, external cut, and tight fiducial volume cut, respectively.
    The dotted lines correspond to the SK-IV 1664-day final data sample in the previous analysis~\cite{Abe:2016nxk}
    and can be compared to the circle markers.
    \label{fig:data_reduc}}
\end{figure}

Figure~\ref{fig:data_reduc} shows the energy distribution after each reduction step of the SK-IV 2970-day data set. 
Background events above $5.5$~MeV are mostly removed by the spallation cut. 
On the other hand, background events below $5.5$~MeV are mostly removed by the ambient cut, 
external cut, and tight fiducial volume cut~\cite{Abe:2016nxk}.

\begin{figure}[htp]
    \begin{center}
    \includegraphics[width=\linewidth]{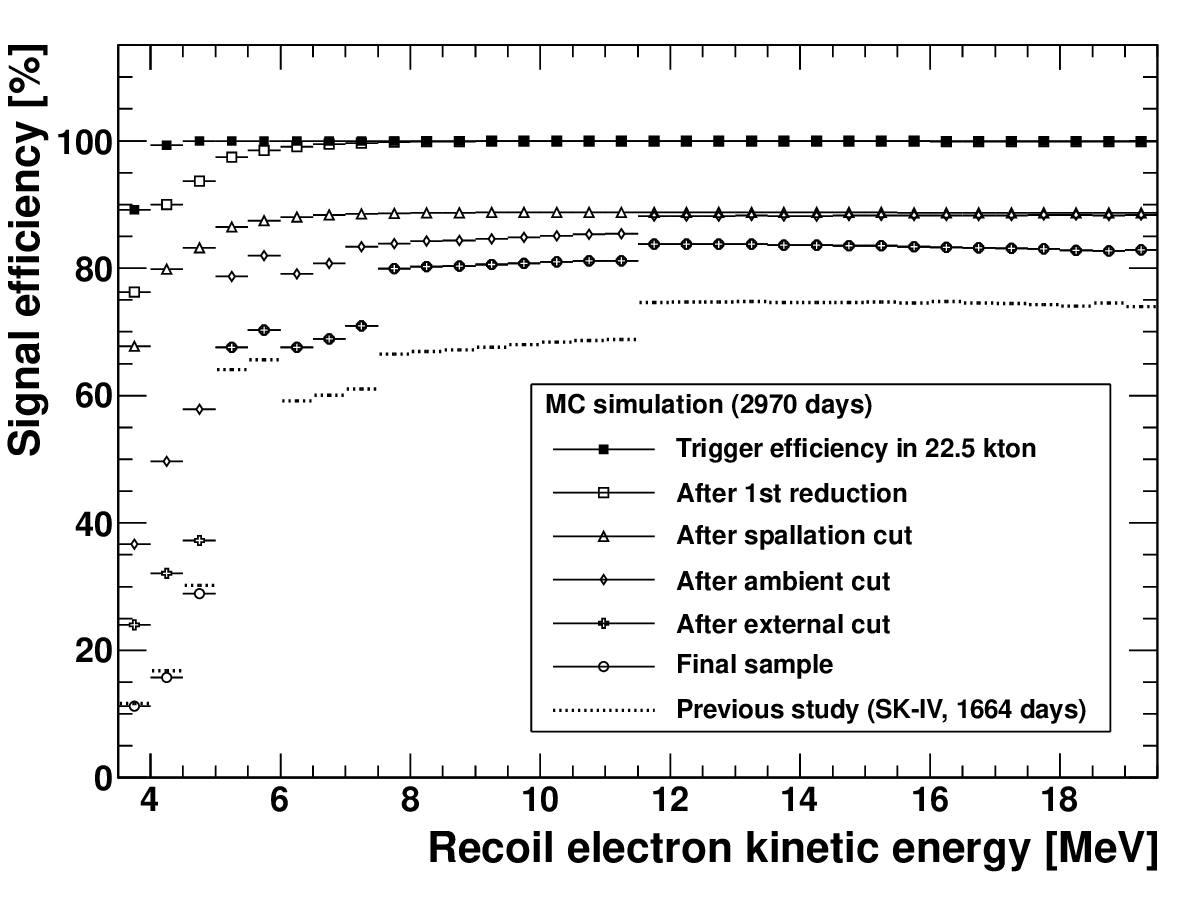}
 \end{center}
    \caption{Signal efficiency after each reduction step as a function of the reconstructed electron kinetic energy. 
    The filled-square corresponds to the number of events after the trigger cut in the $22.5$~kt standard fiducial volume. 
    Other markers are the same as Fig.~\ref{fig:data_reduc}. 
      \label{fig:mc_reduc_eff}}
\end{figure}

Figure~\ref{fig:mc_reduc_eff} shows the signal efficiency after each reduction step as a function of reconstructed recoil electron kinetic energy.
The improvement of the signal efficiency above about 6~MeV is due to the new spallation cuts, as described in Sec.~\ref{sec:spallation}.
In the lower energy regions, tighter cuts are applied to remove more background events, which was optimized by the cut point tuning in this analysis. 

\subsection{Signal extraction}\label{sec:solfit}

The observed number of solar neutrino signal events are extracted from the final data sample
using an extended maximum likelihood function fit.
The fit is made to the angle, $\theta_{\rm{Sun}}$, between the reconstructed event direction and the Sun at the time of detection.  This distribution is used since neutrino-electron elastic scattering is peaked in the forward direction, so the solar neutrino signal corresponds to a peak on top of a uniform background.
This extraction method is similar to previous analysis~\cite{Abe:2016nxk} but in the SK-IV 2970-day analysis 
the binning with multiple scattering goodness~(MSG) parameter, defined in~\cite{Abe:2016nxk}, is newly considered.

The likelihood function is defined as
\begin{multline}
        \mathcal{L_{\mbox{\tiny signal}}}  =  \mathrm{e}^{- \left( \sum_{ij} B_{ij} + S\right) }
        \prod_{i=1}^{N_{\rm{energy}}}
        \prod_{j=1}^{N_{\rm{MSG}_{\it{i}}}} \prod_{k=1}^{n_{ij}}
        \\
        \big( B_{ij} \cdot b_{ij}(E_{ijk},\theta_{\rm{Sun, ijk}}) 
         + S \cdot Y_{ij} \cdot s_{ij}(E_{ijk},\theta_{\rm{Sun, ijk}}) \big),
    \label{eq:solfit}
    \end{multline}
where
$N_{\rm{energy}}$ is the total number of the energy bins,
$N_{\mathrm{MSG}_i}$ is the total number of the MSG bins,
$n_{ij}$ is the total number of events in the  $\{ij\}$ bin,
$b_{ij}$ ($s_{ij}$) are the background (signal) probability density function,
$Y_{ij}$ is the fraction of the total signal events ($S$) in the $\{ij\}$ bin,
$B_{ij}$ are free parameters corresponding to the number of background events in the  $\{ij\}$ bin,
$S$ is a free parameter corresponding to the total number of solar neutrino events in all energy and MSG bins.
The fitting parameters, $S$ and $B_{ij}$, represent the number of signal and background events, respectively. 
These parameters are obtained by maximizing the likelihood function.
The uncertainty of this extraction method is evaluated with dummy solar angle distributions produced with solar MC events. The difference between generated and extracted solar neutrino events is taken into account as the signal extraction systematic uncertainty.
Details are explained in Appendix~\ref{sec:app_solfit}.

\subsection{Systematic uncertainty}

The uncertainties on solar neutrino measurements are estimated using simulation programs and calibration data.
For the efficiency of the reduction steps, LINAC data and MC events are compared.
For the vertex shift, Ni-Cf calibration data and MC are used in the same way as previous analyses.

Table~\ref{tb:sys_table} shows a summary of the systematic uncertainty on the observed solar neutrino event rate in SK across the whole energy range. 
The total systematic uncertainty on the solar neutrino flux in SK-IV is estimated as $\pm 1.4$\%
in the $3.49$--$19.49$~MeV region.
\begin{table}[htp]
\caption{Summary of the systematic uncertainty on the observed solar neutrino event rate for SK-IV and compared to SK-I~\cite{Hosaka:2005um}, SK-II~\cite{Cravens:2008aa}, and SK-III~\cite{Abe:2010hy}.
\label{tb:sys_table}}
\centerline{\begin{tabular}{lcccc}
\hline\hline
 & SK-I & SK-II & SK-III & SK-IV \\
Threshold~[MeV] & $4.49$ & $6.49$ & $3.99$ & $3.49$ \\
\hline
Trigger efficiency & $\pm0.4\%$ & $\pm0.5\%$ & $\pm0.5\%$ & $\pm0.1\%$ \\
Angular resolution & $\pm1.2\%$ & $\pm3.0\%$ & $\pm0.7\%$ & $\pm0.1\%$ \\
Reconstruction goodness & $^{+1.9}_{-1.3}\%$ & $\pm3.0\%$ & $\pm0.4\%$ & $\pm0.5\%$ \\
Hit pattern & $\pm0.8\%$ & -- & $\pm0.3\%$ & $\pm0.4\%$ \\
Small hit cluster & -- & -- & $\pm0.5\%$ & $\pm0.1\%$ \\
External event cut & $\pm0.5\%$ & $\pm1.0\%$ & $\pm0.3\%$ & $\pm0.1\%$ \\
Vertex shift & $\pm1.3\%$ & $\pm1.1\%$ & $\pm0.5\%$ & $\pm0.2\%$ \\
Second vertex fit & $\pm0.5\%$ & $\pm1.0\%$ & $\pm0.5\%$ & -- \\
Background shape & $\pm0.1\%$ & $\pm0.4\%$ & $\pm0.1\%$ & $\pm0.1\%$ \\
Multiple scattering goodness & -- & $\pm0.4\%$ & $\pm0.4\%$ & $\pm0.4\%$ \\
Live time & $\pm0.1\%$ & $\pm0.1\%$ & $\pm0.1\%$ & $\pm0.1\%$ \\
Spallation cut & $\pm0.2\%$ & $\pm0.4\%$ & $\pm0.2\%$ & $\pm0.2\%$ \\
Signal extraction & $\pm0.7\%$ & $\pm0.7\%$ & $\pm0.7\%$ & $\pm0.7\%$ \\
Cross section & $\pm0.5\%$ & $\pm0.5\%$ & $\pm0.5\%$ & $\pm0.5\%$ \\
\hline
Subtotal & $\pm2.8\%$ & $\pm4.8\%$ & $\pm1.6\%$ & $\pm1.1\%$ \\
\hline
Energy scale & $\pm1.6\%$ & $^{+4.2}_{-3.9}\%$ & $\pm1.2\%$ & $\pm0.8\%$ \\
Energy resolution & $\pm0.3\%$ & $\pm0.3\%$ & $\pm0.2\%$ & $\pm0.1\%$ \\
$\mathrm{^{8}B}$ spectrum & $^{+1.1}_{-1.0}\%$ & $\pm1.9\%$  & $^{+0.3}_{-0.4}\%$ & $^{+0.3}_{-0.4}\%$ \\
\hline
Total & $^{+3.5}_{-3.2}\%$ & $^{+6.7}_{-6.4}\%$& $\pm2.2\%$& $\pm1.4\%$ \\
\hline\hline
\end{tabular}}
\end{table}
Table~\ref{tb:syst_spect} shows the uncertainty in each energy range in SK-IV.
\begin{table*}[htb]
\begin{center}
\caption{Energy-uncorrelated systematic uncertainty in each energy region in SK-IV.\label{tb:syst_spect}}
\begin{tabular}{lccccccccc} 
\hline\hline
Energy~[MeV] & $3.49$--$3.99$ & $3.99$--$4.49$ & $4.49$--$4.99$ & $4.99$--$5.49$ & $5.49$--$5.99$ & $5.99$--$6.49$ & $6.49$--$6.99$ & $6.99$--$7.49$ & $7.49$--$19.49$ \\
\hline
Trigger efficiency & $^{+3.5}_{-3.2}\%$ & $ \pm 0.7\%$ & -- & -- & -- & -- & -- & -- & -- \\
Angular resolution & $\pm0.2\%$ & $\pm0.2\%$ & $\pm0.2\%$ & $\pm0.1\%$ & $\pm0.1\%$ & $\pm0.1\%$ & $\pm0.1\%$ & $\pm0.1\%$ & $\pm0.1\%$ \\
Reconstruction goodness &$\pm0.1\%$& $\pm0.2\%$ & $\pm0.1\%$ & $\pm 0.1\%$ & $\pm0.1\%$ & $\pm 0.3\%$ & $\pm 0.5\%$ & $\pm 0.7\%$ & $\pm 0.4\%$ \\
Hit pattern & -- & -- & -- & -- & -- & $\pm0.5\%$ & $\pm 0.5\%$ & $\pm 0.4\%$ & $\pm 0.4\%$ \\
Small hit cluster &$\pm0.1\%$ & $ \pm 0.1\%$& $\pm0.1\%$ & -- & -- & -- & -- & -- & -- \\
External event cut & $\pm 0.1\%$ & $\pm 0.1\%$ & $\pm 0.1\%$ & $\pm 0.1\%$ & $\pm 0.1\%$ & $\pm 0.1\%$ & $\pm 0.1\%$ & $\pm 0.1\%$ & $\pm 0.2\%$ \\
Vertex shift & $\pm 0.4\%$ & $\pm 0.4\%$ & $\pm 0.4\%$ & $\pm 0.7\%$ & $\pm 0.4\%$ & $\pm 0.4\%$ & $\pm 0.4\%$ & $\pm 0.4\%$ & $\pm 0.1\%$ \\
Background shape & $\pm 2.7\%$ & $\pm 0.6\%$ & $\pm 0.6\%$ & $\pm 0.2\%$ & $\pm 0.2\%$ & $\pm 0.2\%$ & $\pm 0.2\%$ & $\pm 0.2\%$ & $\pm 0.1\%$ \\
Signal extraction & $\pm 2.1\%$ & $\pm 2.1\%$ & $\pm 2.1\%$ & $\pm 0.7\%$ & $\pm 0.7\%$ & $\pm 0.7\%$ & $ \pm 0.7\%$ & $\pm 0.7\%$ & $\pm0.7\%$ \\
Cross section &$\pm 0.2\%$ & $\pm 0.2\%$ & $\pm 0.2\%$ & $\pm 0.2\%$ & $\pm 0.2\%$ & $\pm 0.2\%$ & $\pm 0.2\%$ & $\pm 0.2\%$ & $\pm0.2\%$ \\
Multiple scattering goodness & $\pm 0.4\%$ & $\pm 0.2\%$ & $\pm 0.3\%$ & $\pm 0.3\%$ & $\pm 0.3\%$ & $\pm 0.6\%$ & $\pm 1.3\%$ & $\pm 1.3\%$ & -- \\ \hline
Total &$^{+4.9}_{-4.8}\%$ & $\pm2.4\%$ & $\pm2.3\%$ & $\pm 1.1\%$ & $\pm0.9\%$ & $\pm 1.2\%$ & $\pm 1.7\%$ & $^{+1.8}_{-1.7}\%$ & $\pm 0.9\%$ \\
\hline\hline
\end{tabular}
\end{center}
\end{table*}
These are treated as energy-uncorrelated uncertainties in the oscillation analysis.

\begin{figure}[htb]
    \begin{center}
        \includegraphics[width=\linewidth]{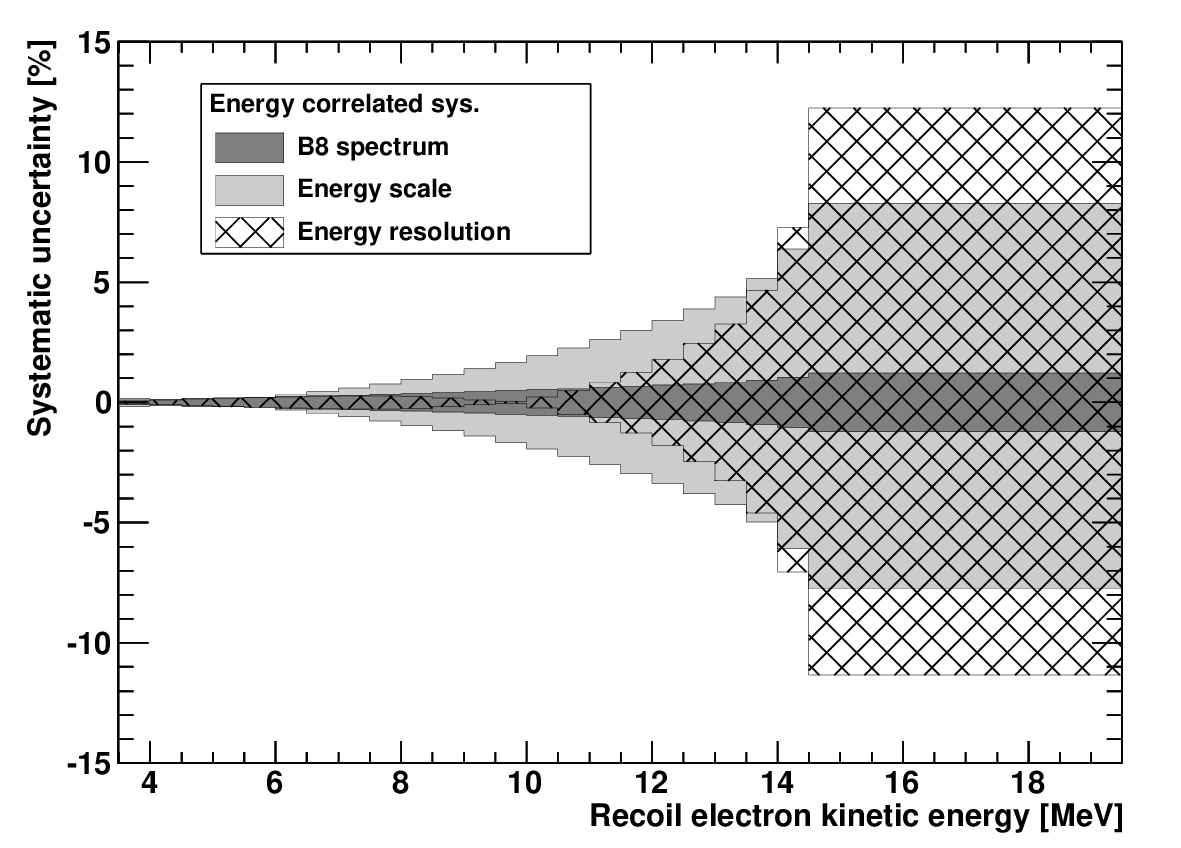}
    \end{center}
    \caption{The energy-correlated systematic uncertainties in SK-IV 2970-day data set. 
    The gray, light-gray, and cross-pattern show the systematic uncertainty of the $\mathrm{^{8}B}$ spectrum shape, energy scale, and energy resolution uncertainties, respectively. \label{fig:ene_corr}}
\end{figure}

Energy-correlated uncertainties, like energy scale, resolution, and expected solar neutrino spectrum shape, 
are estimated with the simulation by artificially shifting the relevant parameters.
To evaluate the systematic uncertainties related to the energy scale determination, 
the energy scale in the simulation program is artificially changed according to the estimations in Sec.~\ref{sec_summary_energy_scale}.
Using this modified simulation program, the solar neutrino events are generated and then applied same reduction process. Finally, we investigate the difference of the extracted solar neutrino signal events.
For the energy resolution, the same treatment is done with the uncertainty estimated in Sec.~\ref{sec:linac-reso}.
In addition, the theoretical uncertainty of the $\mathrm{^{8}B}$ spectrum shape is also considered~\cite{Bahcall:1995mm}. 
These uncertainties are correlated in the solar neutrino energy spectrum.
Figure~\ref{fig:ene_corr} shows the energy-correlated systematic uncertainties in the SK-IV 2970-day data set.
These correlations are considered in the oscillation analysis.
The effects in $3.49$--$19.49$~MeV region are shown in Table~\ref{tb:sys_table}.

\section{Flux and energy spectrum measurements} \label{sec_result}

In this section, the results using the SK-IV 2970-day data set as well as the combined results with other phases are described.
The basic conditions for the solar neutrino simulation events used in this analysis are given in  Sec.~\ref{sec:simulation}. 
The analysis in this section does not take into account the effects of neutrino oscillations. 
For this reason, it is explicitly indicated as "MC(Unoscillated)" in figures.

\subsection{Total $\mathrm{^{8}B}$ neutrino flux}

\begin{figure}[htp]
    \begin{center}
        \includegraphics[width=\linewidth]{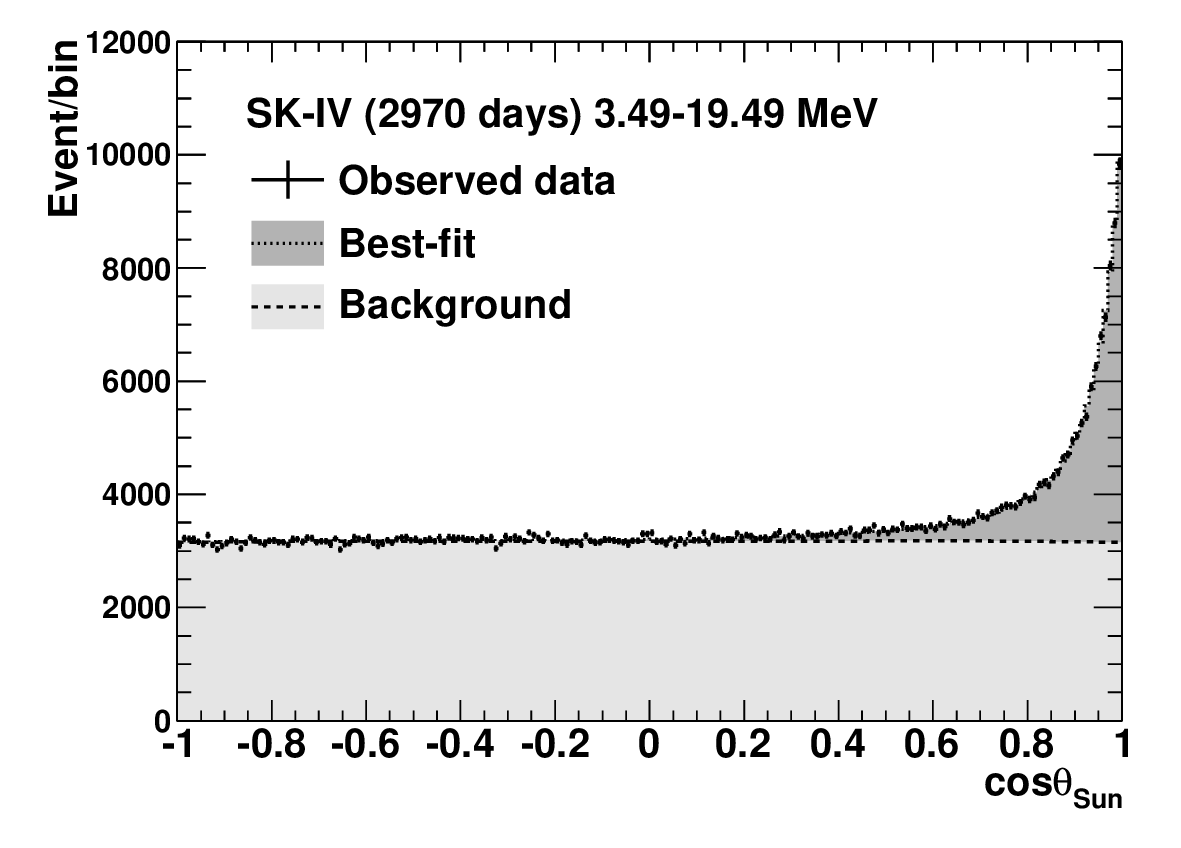}
    \end{center}
    \caption{Solar angle distribution for the energy range of $3.49$--$19.49$~MeV using SK-IV 2970-day data set.
    The black points show the observed data, the dark and light gray shades show the background and the best-fit signal, respectively.  
    \label{fig:solar_dist}}
\end{figure}

\begin{figure}[htp]
    \begin{center}
\includegraphics[width=\linewidth]{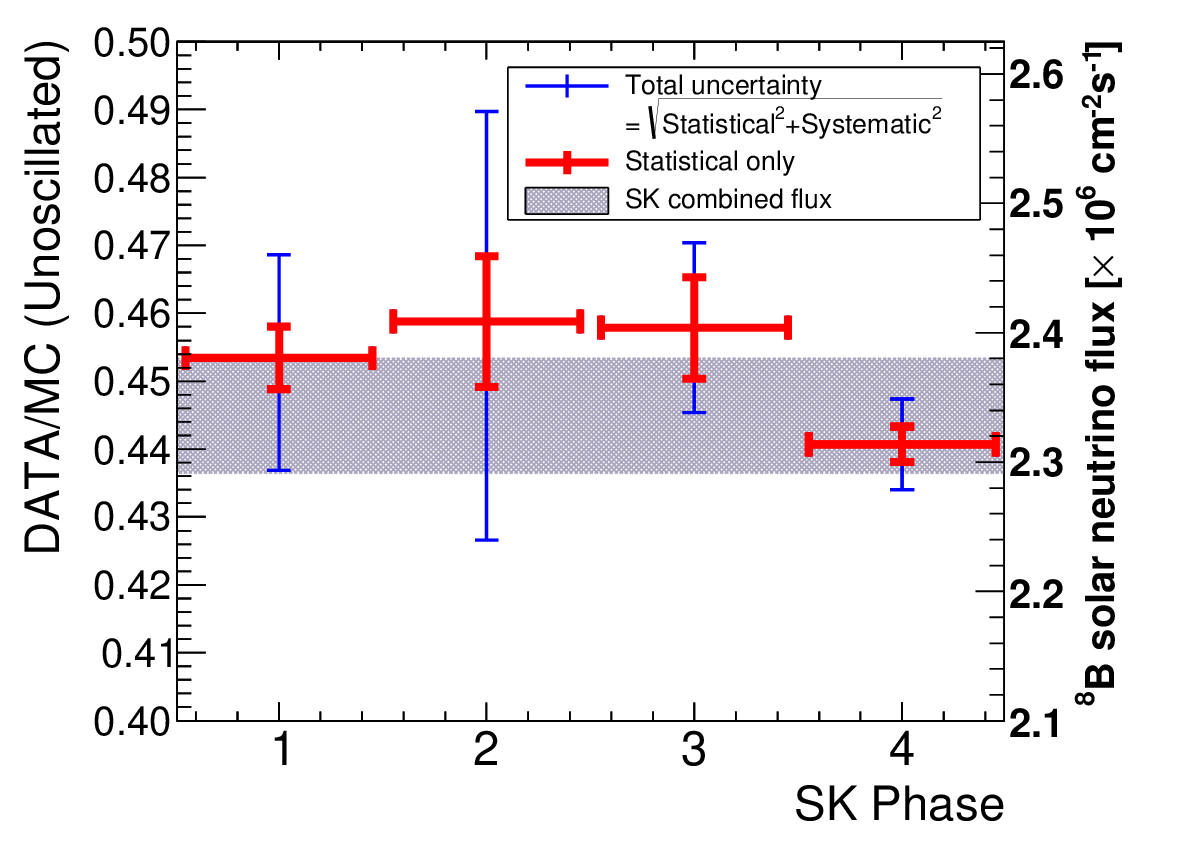} 
    \end{center}
    \caption{(Color online) Comparison of $\mathrm{^{8}B}$ solar neutrino flux among different SK phases. 
    The thick red~(thin blue) bars illustrate the statistical uncertainty~(total uncertainty).
    The gray band shows the combined value. 
    \label{fig:sk_flux}}
\end{figure}

The direction of a recoil electron from elastic scattering is strongly correlated to that of the incident solar neutrino.
Figure~\ref{fig:solar_dist} shows the solar angle distribution of the final data sample from $3.49$ to $19.49$~MeV obtained in this analysis. 
The number of extracted solar neutrino events is $65,443^{+390}_{-388}\,(\mathrm{stat.})\pm925\,(\mathrm{syst.})$.
This number corresponds to a $\mathrm{^{8}B}$ solar neutrino flux of
\begin{equation}
    \Phi_{\mathrm{^{8}B},\mbox{\tiny SK-IV}}=(2.314\pm0.014\pm0.040)~\mathrm{\times10^{6}~cm^{-2}\,s^{-1}}
\end{equation}

assuming a pure electron neutrino flavor component, without neutrino oscillation. 
The ratio of the extracted events to that expected by assuming SNO's neutral current flux~\cite{Aharmim:2011vm} is $0.441\pm0.003\,(\mathrm{stat.})\pm0.007\,(\mathrm{syst.})$. The $\mathrm{^{8}B}$ solar neutrino flux measured in SK-IV is therefore consistent with those from previous phases within their total uncertainties. Figure~\ref{fig:sk_flux} shows a summary of the measured flux ratios in SK. 

\begin{figure}[htp]
    \begin{center}
        \includegraphics[width=\linewidth]{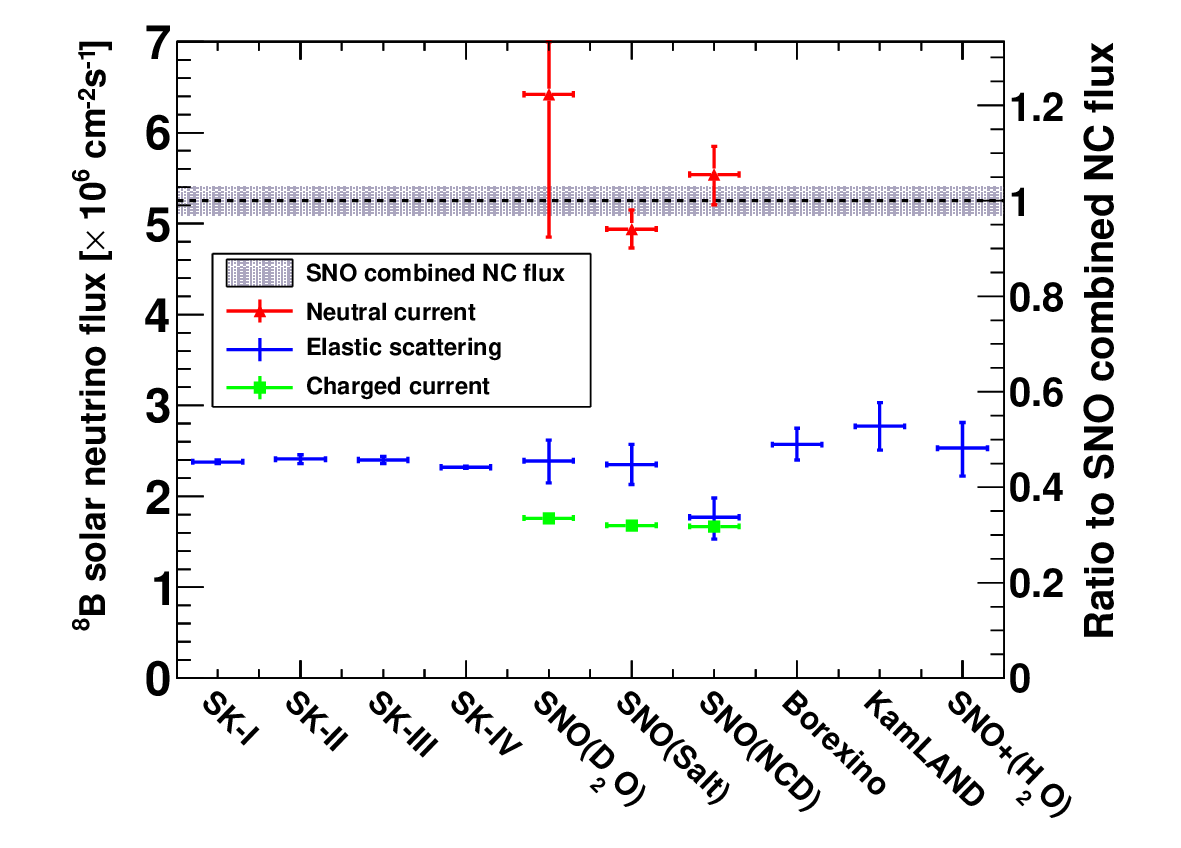} 
    \end{center}
    \caption{(Color online) Comparison of $\mathrm{^{8}B}$ solar neutrino flux among solar neutrino experiments. The measured flux of other experiments are obtained from the three phases of SNO~\cite{Aharmim:2005gt,Aharmim:2007nv,Aharmim:2008kc,Aharmim:2011yq}, Borexino~\cite{Agostini:2017cav}, KamLAND~\cite{Abe:2011em}, and SNO{+}~(pure water)~\cite{Anderson:2018ukb}. The left~(right) vertical axis shows the measured flux without neutrino oscillation~(the ratio of the measured flux to the SNO's neutral current flux~\cite{Aharmim:2011vm}).
    \label{fig:flux_comp}}
\end{figure}

The combined flux ratio from all the SK phases is $0.445\pm0.002\,(\mathrm{stat.})\pm0.008\,(\mathrm{syst.})$, which corresponds to 
\begin{equation} 
\Phi_{\mathrm{^{8}B},\mbox{\tiny SK}} = (2.336\pm0.011\pm0.043)\mathrm{\times10^{6}~cm^{-2}\,s^{-1}} \textrm{.}
\end{equation}
Figure~\ref{fig:flux_comp} shows the measured $\mathrm{^{8}B}$ solar neutrino flux among 
real-time solar neutrino experiments.
A comparison between different interaction channels is a clear demonstration of the existence of non-electron neutrino components in $\mathrm{^{8}B}$ solar neutrinos.

\begin{figure}[htp]
    \begin{center}
        \includegraphics[width=\linewidth,clip]{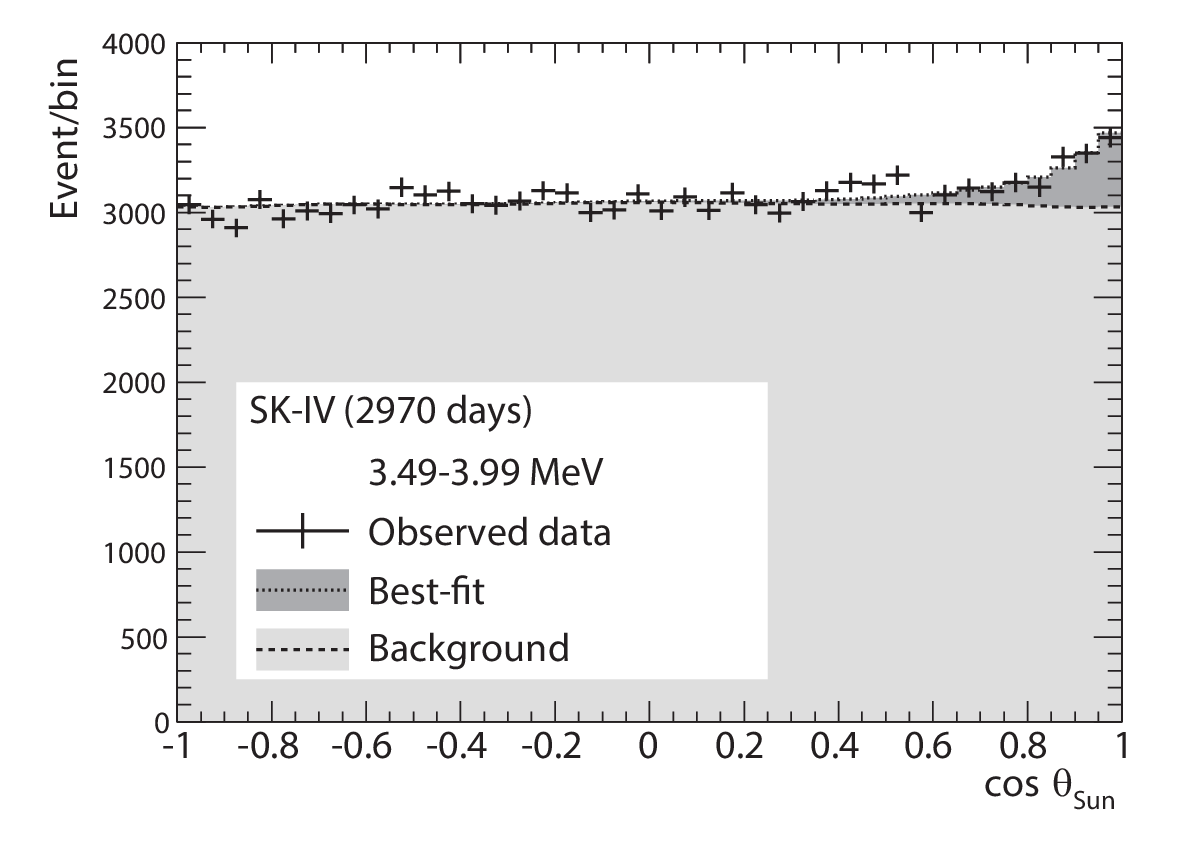}
    \end{center}
    \caption{Solar angle distribution for the energy range of $3.49$--$3.99$~MeV using SK-IV $2970$~days data sample. 
    The definitions are the same as in Fig.~\ref{fig:solar_dist}.     
    \label{fig:solar_dist_low}}
\end{figure}

At lower neutrino energies, the MSW effect in the Sun is weaker and neutrino flavor change reverts to a vacuum-like mechanism. For solar neutrinos this transition is expected to occur around $3$~MeV, so SK's lowest-energy events are of particular interest to confirm our understanding of the MSW effect.  
Figure~\ref{fig:solar_dist_low} shows the solar angle distribution from $3.49$ to $3.99$~MeV. 
Although the signal-to-noise ratio is small because of the contamination from the background events, 
the peak of the solar neutrino signal is clearly observed over the background rate.
The number of extracted events in the energy range of $3.49$--$3.99$~MeV is
\begin{equation}
S_{3.49\mathchar`-3.99} = 1871\,^{+167}_{-165}\,(\mathrm{stat.})\,^{+92}_{-90}\,(\mathrm{syst.}).
\end{equation}
The statistical significance is about 10 sigma.
This is achieved by reducing the background events, mainly from Rn daughters, in the central region of the SK-IV detector~\cite{Nakano:2019bnr}.

\subsection{Yearly flux}

The solar activity cycle is a roughly 11-year
periodic change of sunspot numbers and reconfiguration of the magnetic field at the surface of the Sun.
Although the standard solar model predicts the production rate of solar neutrinos is constant on this timescale, 
it does not consider the periodical activities of the Sun, such as the rotation inside the Sun or the variation of the sunspot numbers.
The combined SK data sample contains the period from 1996 to 2018 and this long-term observation covers nearly two solar activity cycles, cycle $23$ and $24$, so it can be used to check for any correlation.

\begin{figure}[htp]
    \begin{center}
\includegraphics[width=\linewidth]{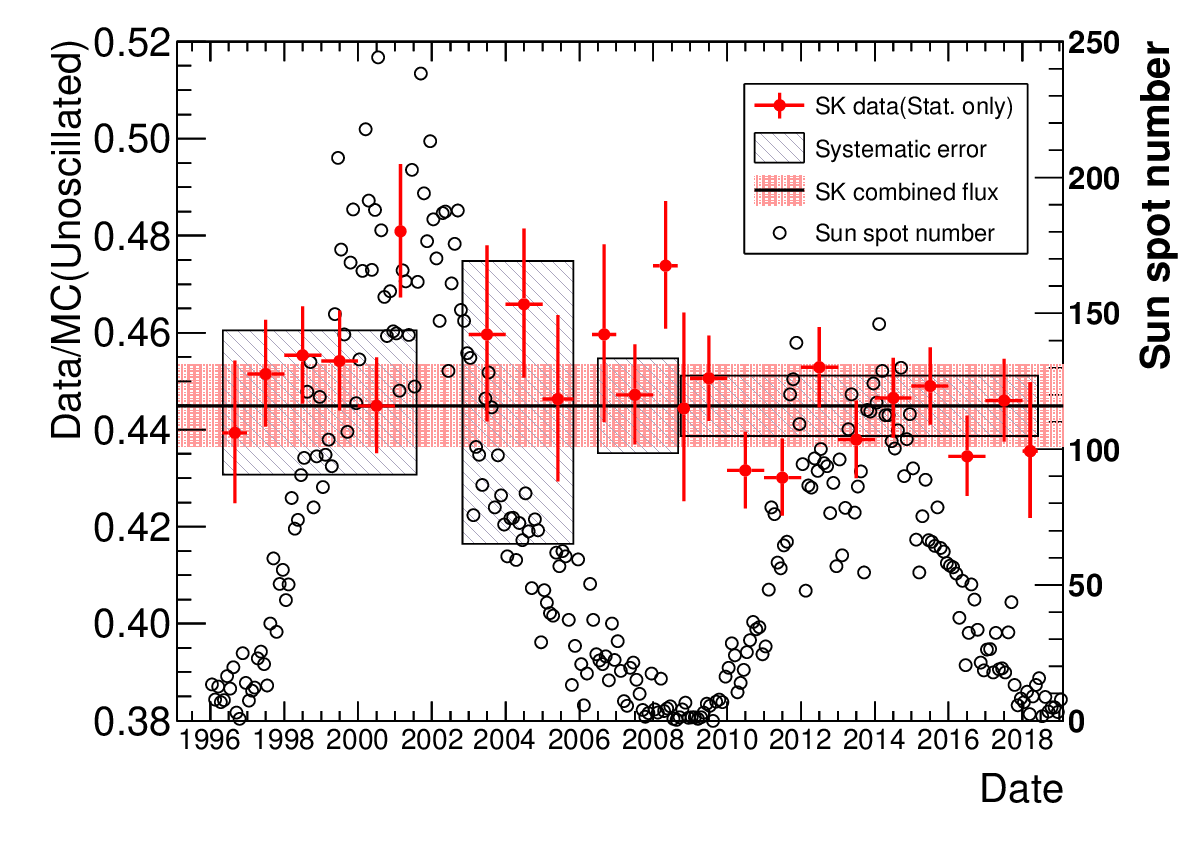}
    \end{center}
    \caption{(Color online) Yearly solar neutrino flux measured by SK. 
    The red-filled circle points show the SK data with statistical uncertainty and the gray striped area shows the systematic uncertainty for each phase. 
    The horizontal black solid line~(red shaded area) shows the combined value of measured flux~(its combined uncertainty). The black-blank circle points show the sunspot numbers from 1996 to 2018. The sunspot numbers are taken from Ref.~\cite{sun_spot}. \label{fig:flux_year}}
\end{figure}

Figure~\ref{fig:flux_year} shows the yearly averaged fluxes observed in the different phases 
of the SK detector together with the corresponding sunspot numbers.
To test the correlation between the observed yearly fluxes and the sunspot numbers, the $\chi^{2}$ between the average flux value and the yearly flux values is defined as:
\begin{equation}
\chi^{2}(r) = \mathrm{min}\left[ 
\sum_{p=1}^{4} \sum_{t=1}^{n_{p}} \left(\frac{r-(d_{p,t}+\alpha_{p})}{\sigma_{p,t}}\right)^{2}+\left(\frac{\alpha_{p}}{\tau_{p}}\right)^{2}
\right]
\end{equation}
\noindent where the summations are over~$p$, the SK phase~(I to IV); $n_p$ is the number of operating years in each phase; and $t$ is year within a SK phase. The combined SK average flux~(with no annual variation) is~$r$, while $d_{p,t}$ is the observed yearly flux in year~$t$ of SK phase~$p$.  The statistical uncertainty on each~$d_{p,t}$ is~$\sigma_{p,t}$. Systematic variations of SK phases are described by the nuisance parameters~$\alpha_{p}$, where $\tau_{p}$ is the systematic uncertainty for each SK phase~$p$.
Using the observed data in the energy range from $4.49$~MeV~(exceptions: $6.49$~MeV for SK-II and $5.49$~MeV for the 2018 data point) 
to $19.49$~MeV, the minimum $\chi^{2}$ for a steady flux (parameter to be varied) is calculated with the total experimental uncertainties as:
\begin{equation}
\chi^{2}/N_\mathrm{d.o.f.} = 19.94/22, \label{eq:chi2_yearly} \end{equation}
\noindent where $N_{\mathrm{d.o.f}}~(=22)$ is the degree of freedom for $\chi^{2}$~\footnote{We exclude the data taken in 2002 after the reconstruction toward SK-II because of the short live time from October to December. This data is not included in the calculation of $\chi^{2}$. The energy threshold of the data in 2018 is changed from $4.49$~MeV to $5.49$~MeV since the high background rate below $5.49$~MeV was observed due to radioactive impurities during the convection study.}. This corresponds to a probability of $58.9\%$. The solar neutrino rate measurements in SK are fully consistent with a constant solar neutrino flux emitted by the Sun.

\subsection{Flux Measurement for Day and Night} \label{sec_dn}

Due to flavor-specific interactions of solar neutrinos with the Earth's matter, the solar neutrino interaction rate measured by SK depends on the solar zenith angle~$(\theta_{\mathrm{z,solar}})$, which is the angle between the vertical~($Z$) direction of the SK detector 
and the neutrino direction from the Sun when an event occurs~(i.e. the time of day). In most cases, the Earth matter effects lead to a ``regeneration" of electron flavor, i.e. the electron flavor survival probability $P_{ee}$ is larger during the night compared to the day.
The apparent day-time flux and the night-time flux of solar neutrinos in SK are measured separately in the SK-IV 2970-day data set to test this Earth matter effect:
\begin{equation*}
    \begin{cases}
        \Phi_{\mathrm{^{8}B},\mbox{\tiny SK-IV}}^{\mathrm{day}}  = & (2.284\pm0.020\pm0.032)~\mathrm{\times10^{6}~cm^{-2}\,s^{-1}}, \\
        \Phi_{\mathrm{^{8}B},\mbox{\tiny SK-IV}}^{\mathrm{night}}  = &  (2.341\pm0.019\pm0.033)~\mathrm{\times10^{6}~cm^{-2}\,s^{-1}},
    \end{cases}
\end{equation*}
The live time of the day (night) is $1434$~days ($1536$~days).
The day/night asymmetry parameter
\begin{align}
    A_{\mathrm{D/N}}  & =  \frac{\Phi_{\mathrm{^{8}B}}^{\mathrm{day}}-\Phi_{\mathrm{^{8}B}}^{\mathrm{night}}}{\frac{1}{2}(\Phi_{\mathrm{^{8}B}}^{\mathrm{day}}+\Phi_{\mathrm{^{8}B}}^{\mathrm{night}})}. 
    \label{eq:day-night}
\end{align}
is then calculated as
\begin{align}
    A_{\mathrm{D/N}}^{\mbox{\tiny SK-IV, calc}} 
& = -0.025\pm0.012\,(\mathrm{stat.})\pm0.014\,(\mathrm{syst.}).
\end{align}

\begin{figure}[htp]
    \begin{center}
\includegraphics[width=\linewidth]{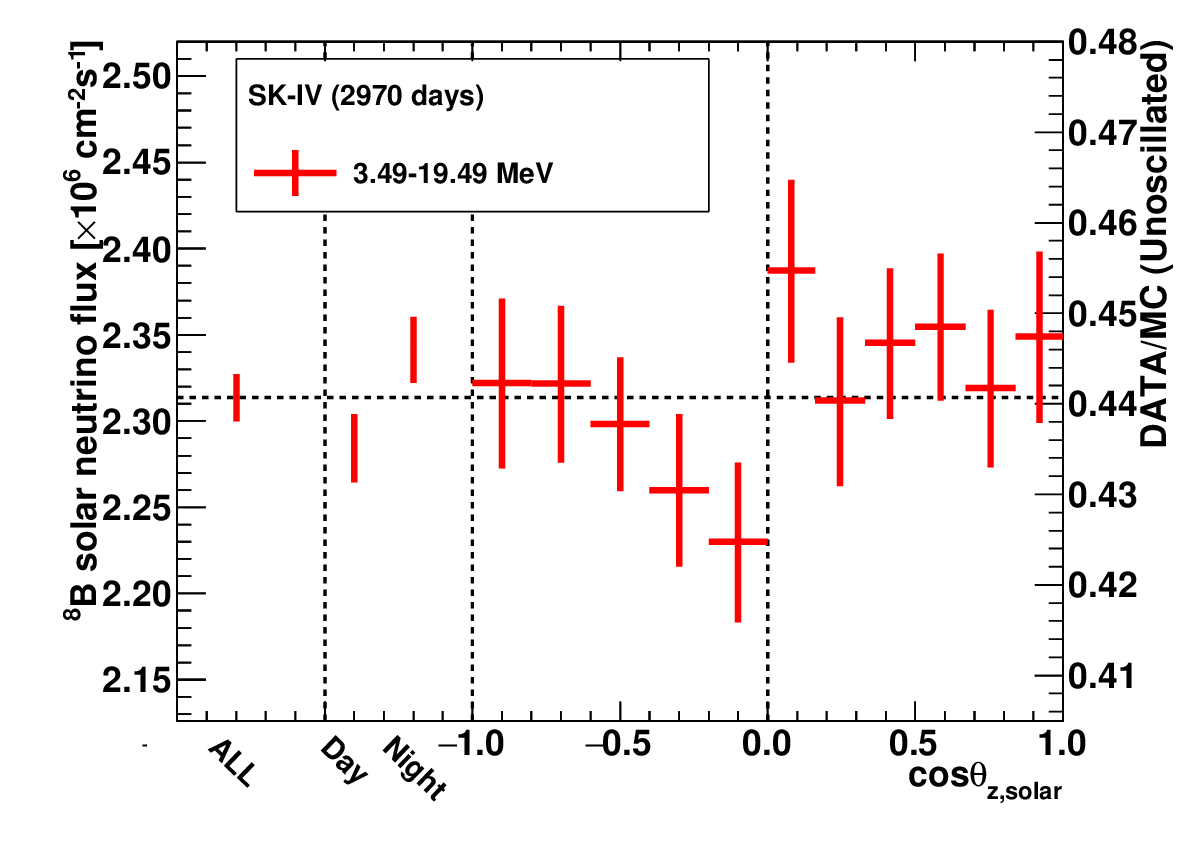}
    \end{center}
    \caption{(Color online) Solar zenith angle dependence of the measured $\mathrm{^{8}B}$ solar neutrino flux in SK-IV. The vertical bars show the statistical uncertainty. 
    \label{fig:flux_zenith_sk4}}
\end{figure}

In addition, we fit the amplitude of the expected zenith angle variation to the observed data. 
Here, we use the convention that during the day-time $\cos\theta_{\mathrm{z,solar}} \leq 0$ and during night-time
$\cos\theta_{\mathrm{z,solar}}>0$.
The day/night asymmetry parameter extracted from the best-fit amplitude to the solar zenith angle variation (see Sec.~\ref{sec:osc_dn}) is
\begin{align}
    A_{\mathrm{D/N}}^{\mbox{\tiny SK-IV, fit}} 
& = -0.0262\pm0.0107\,(\mathrm{stat.})\pm0.0030\,(\mathrm{syst.}).
\end{align}
Unlike $A_{\mathrm{D/N}}^{\mbox{\tiny SK-IV, calc}}$, this parameter depends on the oscillation parameters. The expected asymmetry depends also on the oscillation parameters as well as the energy range in which the flux is measured. With current oscillation parameters, the day/night asymmetry is expected to be at the few-percent level above the MeV region while no asymmetry is expected in the keV region. 
In a previous publication~\cite{Renshaw:2013dzu}, an indication for a non-zero day/night flux asymmetry was found in SK; the best fit was at the few-percent level. 
Borexino also measured the day/night flux asymmetry of solar $\mathrm{^{7}Be}$ neutrinos ~\cite{Bellini:2011yj} and the difference is 
consistent with zero. The SNO collaboration~\cite{Aharmim:2011vm} measured a day/night asymmetry consistent with zero, and consistent with the SK result.
All three results favor the MSW-LMA solutions. 
Figure~\ref{fig:flux_zenith_sk4} shows the solar zenith angle distribution of the observed flux divided into five-day and six-night bins. 

\subsection{Spectrum results}

\begin{figure}[htp]
    \begin{center}
\includegraphics[width=\linewidth,clip]{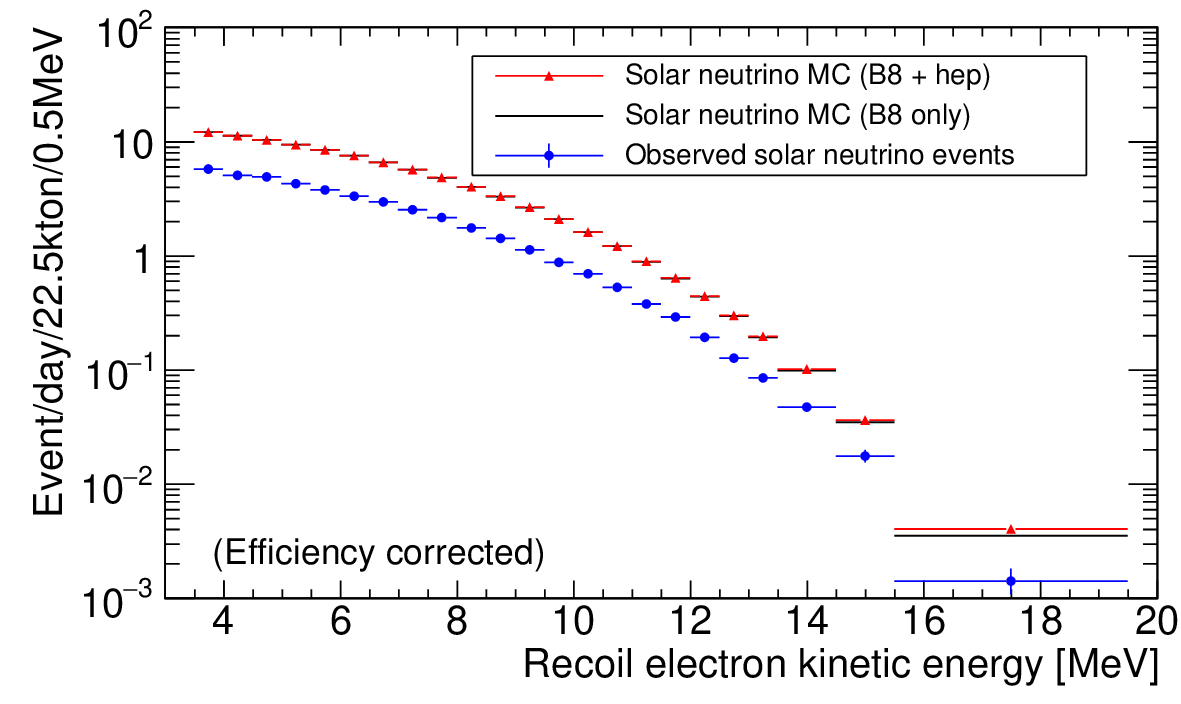}    
    \end{center}
    \caption{(Color online) Energy spectrum of the solar neutrino signal using the SK-IV 2970-day data set.
    The blue circle markers show the observed event rate as a function of the recoil electron energy.
    The red triangle markers show the expected rate from the $\mathrm{^{8}B}+hep$ MC simulation without the oscillation effect. 
    The signal efficiency shown in Fig.~\ref{fig:mc_reduc_eff} 
    is corrected.
    \label{fig:observed-rate}}
\end{figure}

Figure~\ref{fig:observed-rate} shows the observed signal spectrum as well as the expected spectrum calculated assuming the SNO NC flux and the $hep$ flux of the standard solar model~\cite{Aharmim:2011yq, Bahcall:2004fg}. 
In the calculation of the expected spectrum, no neutrino oscillation is assumed and the signal efficiency in Fig.~\ref{fig:mc_reduc_eff} is corrected. 
To test the contribution of $hep$ neutrinos, the expected spectrum without $hep$ neutrinos is also shown.
Although the total $hep$ neutrinos flux is $3$~orders of magnitude smaller than the $\mathrm{^{8}B}$ neutrinos, its contribution can be seen in the the highest energy bins of the recoil electron spectrum since the end-point of the $hep$ neutrino energy spectrum~($18.8$~MeV) is slightly higher than that of $\mathrm{^{8}B}$ neutrinos~($16$~MeV)~\cite{Bahcall:1998se}.

\begin{figure}[htp]
    \begin{center}
\includegraphics[width=\linewidth]{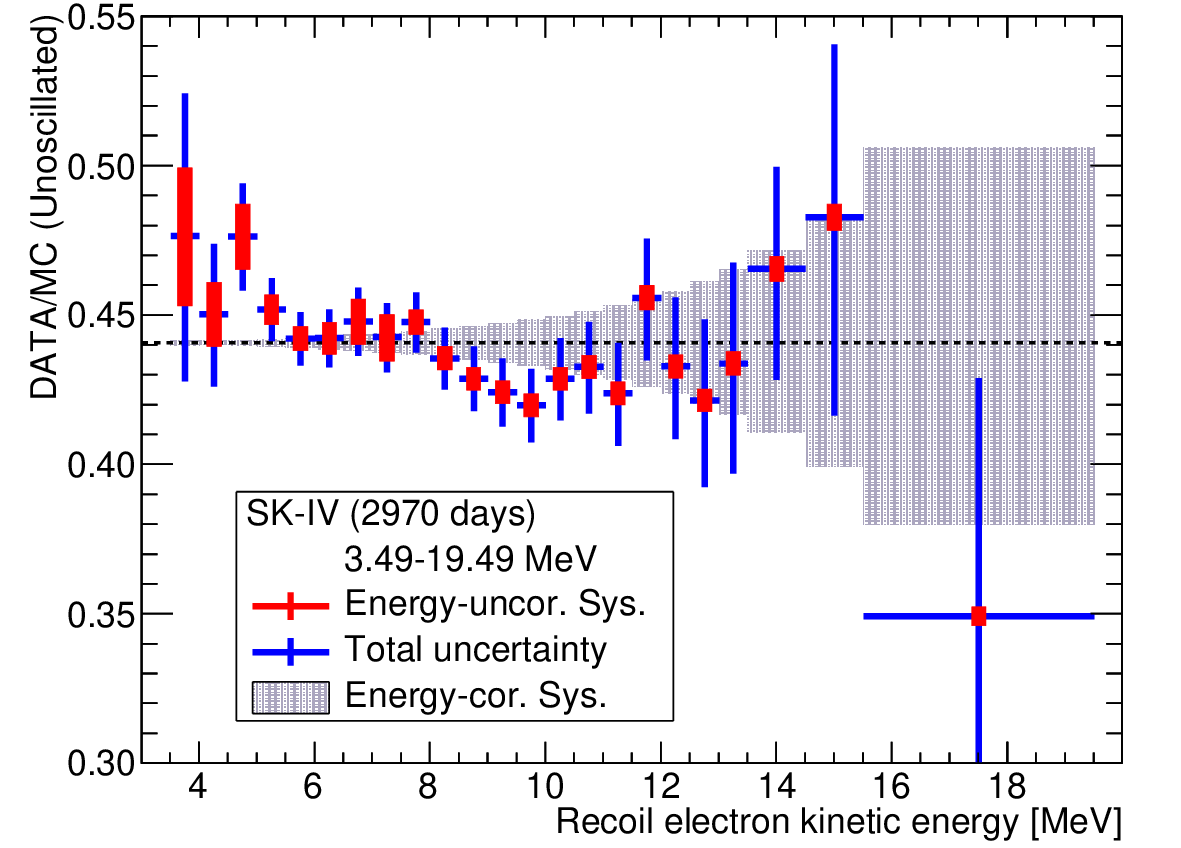}
    \end{center}
    \caption{(Color online) The measured energy spectrum in SK-IV. 
    The blue points and bars show the observed rate divided by the expected event rate assuming no neutrino oscillation,
    with statistical and energy-uncorrelated uncertainties.
    The red bars and gray bands show the energy-uncorrelated and energy-correlated systematic uncertainties.  
    \label{fig:spect_sk4_msg}}
\end{figure}

\begin{figure}[htp]
    \begin{center}
\includegraphics[width=\linewidth]{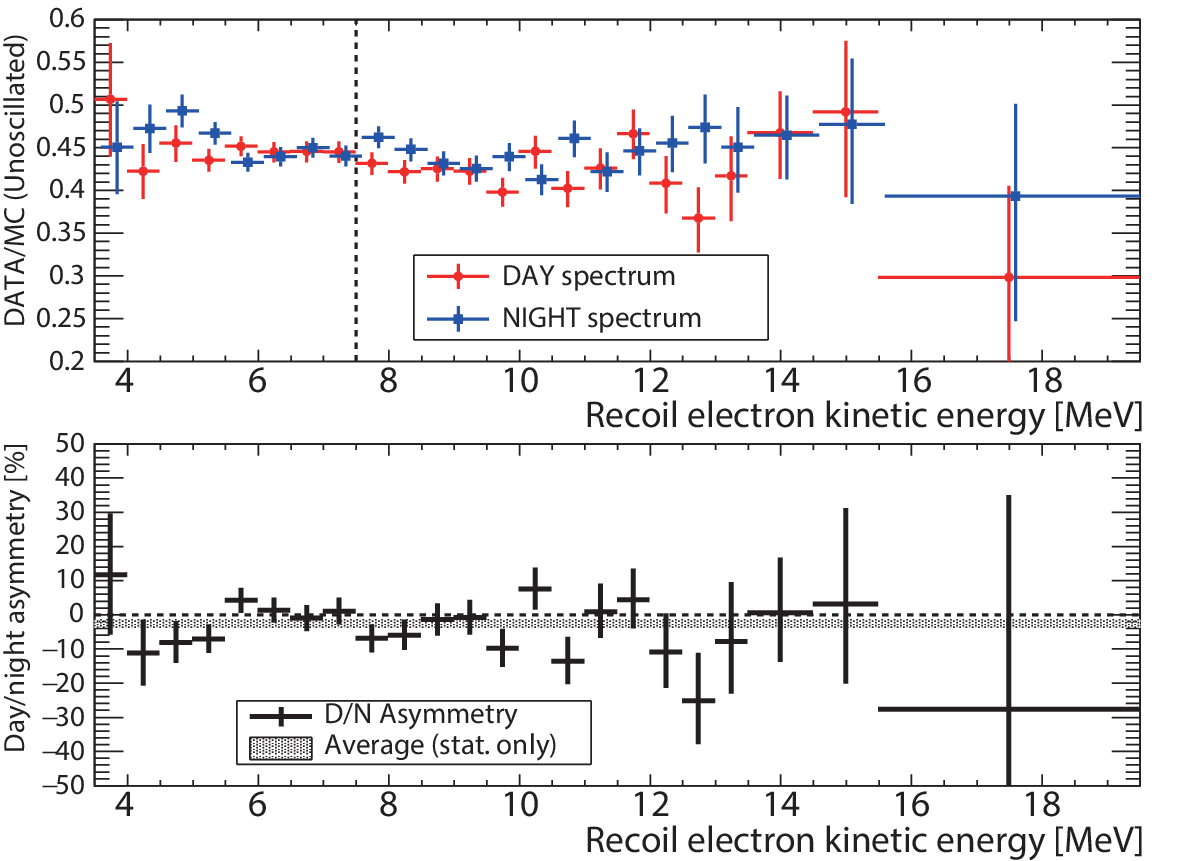}
    \end{center}
    \caption{(Color online) (Top) The recoil electron energy spectrum taken during day~(night) time in SK-IV. The red filled circle (blue filled square) points show the day spectrum~(night spectrum). To improve visibility, the night spectrum is shifted by $+0.1$~MeV.
    (Bottom) The straight day/night asymmetry~(black cross) as a function of the recoil electron energy. 
    The red shaded area shows the statistical average of the day/night asymmetry, which is shown in Sect.~\ref{sec_dn}. \label{fig:spect_sk4_dn}}
\end{figure}

Figure~\ref{fig:spect_sk4_msg} shows the energy spectrum of solar neutrinos measured in the SK-IV 2970-day data set. 
The top panel of Fig.~\ref{fig:spect_sk4_dn} shows the recoil electron spectrum taken in day or night time in SK-IV. 
Based on the spectrum results between day-time and night-time, the spectral straight day/night asymmetry is obtained as shown in the bottom panel of Fig.~\ref{fig:spect_sk4_dn}.

\section{Oscillation analysis} \label{sec_osc}

\subsection{SK-IV Analysis}
\label{sec:osc_method}
The solar neutrino oscillation analysis is based on the same methods as~\cite{Abe:2016nxk} which were first described in Ref.~\cite{Smy:2003jf, Hosaka:2005um}.
The survival probability is calculated in a two-neutrino framework: $P_{ee}^\mathrm{2\nu} (A_\mathrm{mat},\theta_{12}, \Delta m^2_{21})$, where $\theta_{12}$ and $\Delta m^2_{21}$ are the vacuum oscillation parameters, and $A_{\mathrm{mat}}$ is the potential caused by non-zero electron density when the neutrino travels in matter.

To account for three-neutrino effects---specifically a non-zero value of $\theta_{13}$---this is modified, following~\cite{Fogli:2000bk}: \begin{multline}
    P_{ee}^{\mathrm{3\nu}}(\theta_{12}\,\theta_{13}\,\Delta m^2_{21})=  \sin^{4}\theta_{13}\\
    +\cos^{4}\theta_{13}
    P_{ee}^{\mathrm{2\nu}}\left(\cos^{2}\theta_{13} A_{\mathrm{mat}},\, \theta_{12},\, \Delta m^{2}_{21}\right) 
\end{multline}
In fitting neutrino oscillations, we use a constraint on $\theta_{13}$ derived from reactor neutrino experiments~\cite{Zyla:2020zbs}: \begin{equation}
    \sin^{2}\theta_{13} = 0.0218\pm0.0007.
\end{equation}
Since solar neutrino measurements are not sensitive to changes in $\sin^2\theta_{13}$ less than $0.005$, the closest calculation point to this value~$(0.020)$ is effectively used.

The SK analysis constrains neutrino flavor oscillation by measuring the rate of elastic scattering of $^{8}$B and {\it hep} neutrinos with electrons, the spectrum of the recoiling electrons, and the time variation of the interaction rate. Such variations of rate occur due to matter effects on neutrino oscillations in the Earth, and therefore depend on the time of day. The most stringent constraints are obtained when combining all SK operating phases, however, SK-IV dominates in the combined fit: SK-IV observed the largest number of solar neutrinos, it has the lowest energy threshold, and the smallest systematic uncertainties.
An unbinned likelihood $\mathcal{L}$ is used to fit the number of solar neutrino interactions to the angular distribution (see Eq.~(\ref{eq:solfit})). The likelihood is modified to account for oscillation-induced spectral distortions and rate time variations and depends then on oscillation parameters, neutrino fluxes, and nuisance parameters describing energy-correlated systematic uncertainties due to the detector energy scale, energy resolution as well as the neutrino energy spectrum:
\begin{multline}
    \mathcal{L} =  \mathrm{e}^{- \left( \sum_{ij} B_{ij} + S\right)} \prod_{i=1}^{N_{\rm{energy}}} \prod_{j=1}^{N_{\rm{MSG}_i}} \prod_{k=1}^{n_{ij}} \\
 \left( B_{ij}\cdot b_{ij} + S \cdot Y_{ij}  \cdot  s_{ij} \cdot \frac{r_i(\cos \theta_{\mathrm{z,solar}})}{r^{\rm{\tiny ave}}_i}   \right),
\label{eq:time1}
\end{multline}
where $r_i(\cos \theta_{\mathrm{z,solar}})$ is the expected solar neutrino event rate as a function of $\cos \theta_{\mathrm{z,solar}}$ 
in the $i$-th energy bin, 
and $r_{i}^{\rm{\tiny ave}}$ shows the average of the event rate over all the solar zenith angle bins. The spectral distortions enter this likelihood via the factors $Y_{ij}$ describing the expected fraction of signal events in the bin labeled with $i$ and $j$.

 A time-independent likelihood $\mathcal{L}_{\mbox{\tiny ave}}$ is defined by replacing the neutrino interaction rates $r_i(\cos \theta_{\mathrm{z,solar}})$ with the time averages $r_{i}^{\mbox{\tiny ave}}$, so
the zenith angle dependence factor $r_i(\cos \theta_{\mathrm{z,solar}})/r_i^\mathrm{\tiny ave}$ is set to 1. The large number of events renders the maximization of this likelihood (which depends on several nuisance parameters) computationally difficult. Therefore, the log-likelihood is formally split into a time-dependent log-likelihood ratio and a time-independent log-likelihood:
\begin{equation}
   \log \mathcal{L}= \log \mathcal{L}- \log \mathcal{L}_{\rm{\tiny ave}}+ \log \mathcal{L}_{\rm{\tiny ave}}.
\end{equation}
The first two terms are identified with the time-variation log-likelihood ratio
$\log \mathcal{L}_{\mbox{\tiny time}}=\log \mathcal{L}- \log \mathcal{L}_{\rm{\tiny ave}}$.
(log of the ratio of time variation over no time variation).

Of course, by definition, $\mathcal{L}_{\mbox{\tiny ave}}$ depends only on the interaction rate and recoil electron spectrum, so to save computation time, it is maximized by minimizing an equivalent 
$\chi^2$. This \emph{spectral} $\chi^2$ is defined as $\chi^2_{\mbox{\tiny spec}}=-2\log \mathcal{L}_{\mbox{\tiny ave}}$ and is then approximated by a binned sum of squared differences of expected and observed event rates divided by the uncertainties (see below in Eq.~\ref{eq:oscchi0}) augmented with several nuisance parameters describing neutrino fluxes and energy-correlation uncertainties. Those nuisance parameters don't affect $\mathcal{L}_{\mbox{\tiny time}}$ as much as $\chi^2_{\mbox{\tiny spec}}$, so the minimization can be factorized: Minimizing $\chi^2_{\mbox{\tiny spec}}$ determines the nuisance parameters, and once best-fit values are obtained 
$\mathcal{L}_{\mbox{\tiny time}}$ can evaluated using just the best-fit values from $\chi^2_{\mbox{\tiny spec}}$. 
This calculation method for the time variation derived from the matter effects in the earth is the same as described in~\cite{Smy:2003jf, Hosaka:2005um}.
The total
equivalent $\chi^2$ of SK is then
\begin{equation}
     \chi^2_{\mbox{\tiny SK}} = \chi^2_{\mbox{\tiny spec}} + \chi^2_{\mbox{\tiny time}},
\label{eq:chi2_all}
\end{equation}
where $\chi^2_{\mbox{\tiny time}}=-2\log \mathcal{L}_{\mbox{\tiny time}}$.
The spectrum part is
\begin{multline}
     \chi^{2}_{\mbox{\tiny spec}}(\beta, \eta)= \sum_i \frac{\big[d_i - (\beta b_i + \eta h_i)\times 
     f_i(\tau, \epsilon, \rho)
     \big]^2}{\sigma^2_i}
 \\ + \tau^2 + \epsilon^2 +  \rho^2,
     \label{eq:oscchi0}
\end{multline}
where $\beta$ and $\eta$ are, respectively, the dimensionless scaling parameter of the $^8$B  and {\it hep} neutrino fluxes, and are the subject of the minimization. The expectations $b_i$ ($h_i$) are the ratios of the oscillated $^8$B ({\it hep}) neutrino interaction rates over the unoscillated combined neutrino interaction rate in MC in energy bin $i$. $d_i$ are the observed ratios with uncertainties $\sigma_i$. The energy-correlated spectral distortion factor $f_i(\tau, \epsilon, \rho$) is controlled by three constrained, dimensionless nuisance parameters: $\tau$ is the systematic deviation of the $^8$B neutrino spectrum (common to all phases); $\epsilon$ is the systematic energy scale deviation and $\rho$ is the systematic energy resolution deviation.
An extra term
\begin{equation}
\Phi = \left( \frac{\beta - 1}{0.20} \times 5.25 \right)^2 + \left(  \frac{\eta - 1 }{15.76} \times 7.88 \right)^2,
    \end{equation}
can be added to represent external constraints on the neutrino fluxes.
The constraint on $\beta$ comes from the measurement of the NC reaction of the SNO experiment,  $(5.25 \pm 0.20) \times 10^{6}~\rm{cm^{-2}\,s^{-1}}$~\cite{Aharmim:2011vm}, 
and the {\it hep} constraint $\eta$ of $(7.88 \pm 15.76)\times10^{3}~\mathrm{cm^{-2}\,s^{-1}}$ 
is based on the standard solar model~\cite{Bahcall:2004fg} while the uncertainty roughly corresponds to the limit reported by SNO~\cite{Aharmim:2020agi}. 

\begin{figure}[tbh]
\begin{center}
\includegraphics[width=\linewidth,clip]{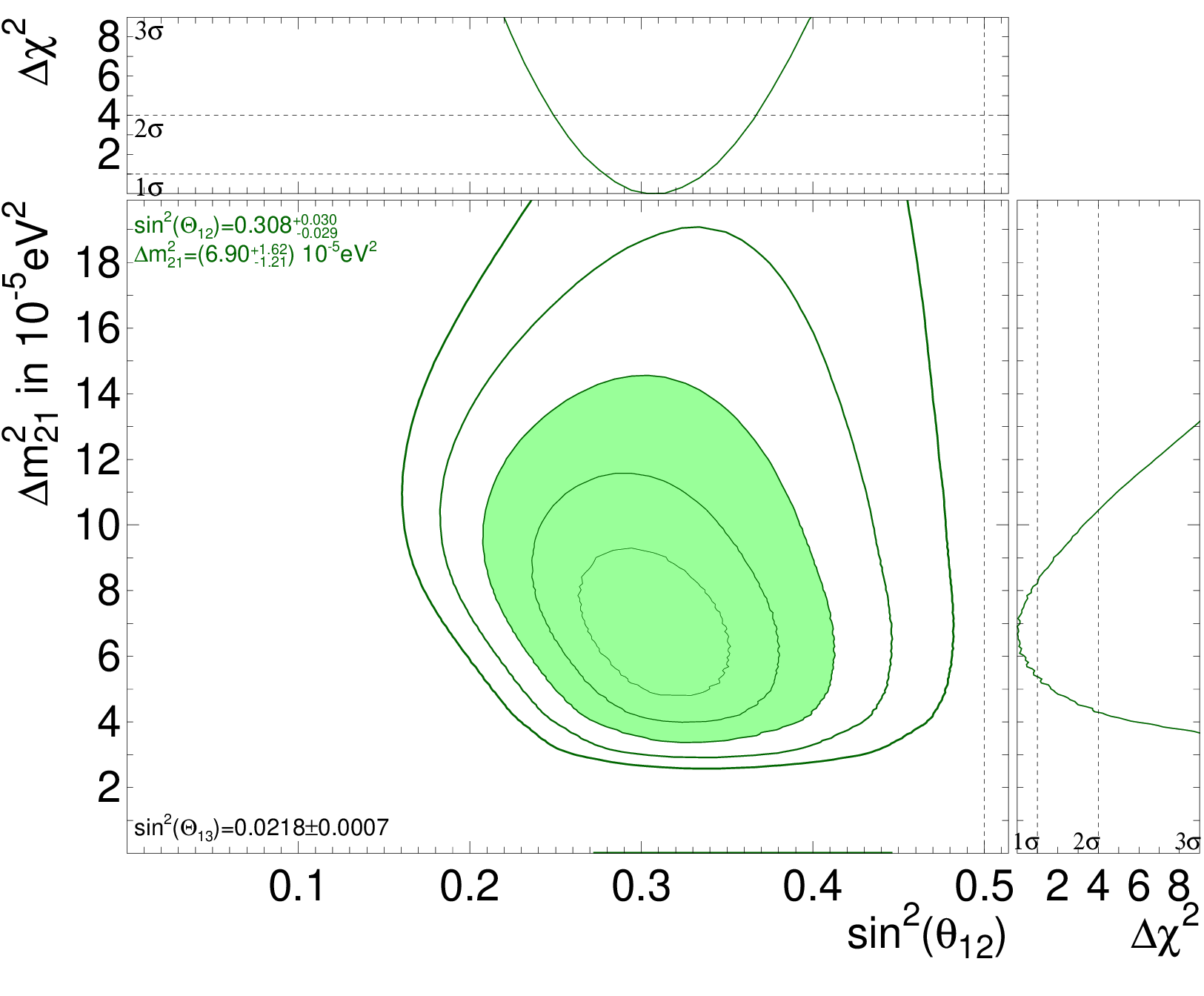}
\end{center}
\caption{(Color online) SK-IV constraints on solar oscillation parameters. The shaded region displays the 3$\sigma$ confidence level. Also shown are $1$--$5\sigma$ allowed areas.}
\label{fig:osc_sk4}
\end{figure}

Using the SK-IV elastic scattering rate, recoil electron spectrum, and day/night variation measurements,
$\sin^{2}\theta_{12}$ and $\Delta m^{2}_{21}$ are determined as $\sin^{2}\theta_{12,\rm{SK-IV}}=0.308^{+0.030}_{-0.029}$ and
$\Delta m^{2}_{21,\rm{SK-IV}}=(6.9^{+1.6}_{-1.2}\times10^{-5})~\mathrm{eV}^{2}$.
Figure~\ref{fig:osc_sk4} shows allowed regions in these parameters for $1$--$5\sigma$. The SK-IV oscillation parameter determination is in excellent agreement with a measurement from KamLAND using anti-neutrinos~\cite{KamLAND:2010fvi}. It also agrees well with other solar neutrino data (including the other SK phases).

\subsection{SK I/II/III/IV Combined Analysis}

\begin{figure}[bth]
    \centering
    \includegraphics[width=\linewidth,clip]{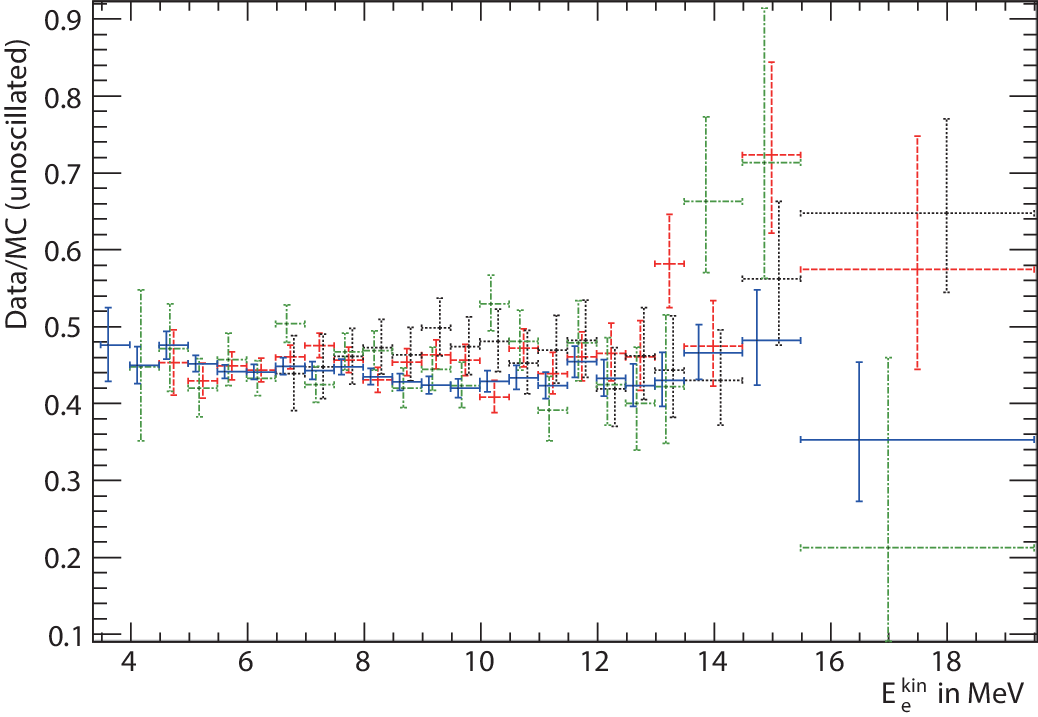}
    \caption{(Color online) Comparison of the Recoil Electron Spectra of SK-IV (blue solid) with SK-I (red dashed), SK-II (black dotted) and SK-III (green dot-dashed).}
    \label{fig:osc_compspec}
\end{figure}

Figure~\ref{fig:osc_compspec} compares the SK-IV spectrum to the spectra from previous phases. The energy scale and resolution of each phase are individually calibrated and evaluated. Within the uncertainties of this calibration, all four phases agree with each other. Not only these energy-correlated but also the energy-uncorrelated systematic uncertainties are evaluated for each phase separately. To combine all phases of SK, the spectral $\chi^2$ is assembled from the spectral $\chi^2$ from each of the four phases of the experiment. 
The $\chi^2$ is defined as in~\cite{PhysRevD.94.052010}. The time variation likelihood ratios $\log \mathcal{L}_{\mbox{\tiny time},p}$ of all four phases $p$ are just added.

When combining SK data from all phases, strong oscillation parameter constraints are obtained from the recoil electron spectrum shape and day/night variation alone without external constraint on the elastic scattering rate (but requiring consistency within uncertainties of the different phases of SK). This is done by removing the external constraint on $\beta$ (based on the NC interaction rate with deuterium measured by SNO) in $\Phi$. 

\begin{figure}[tbh]
\begin{center}
\includegraphics[width=\linewidth,clip]{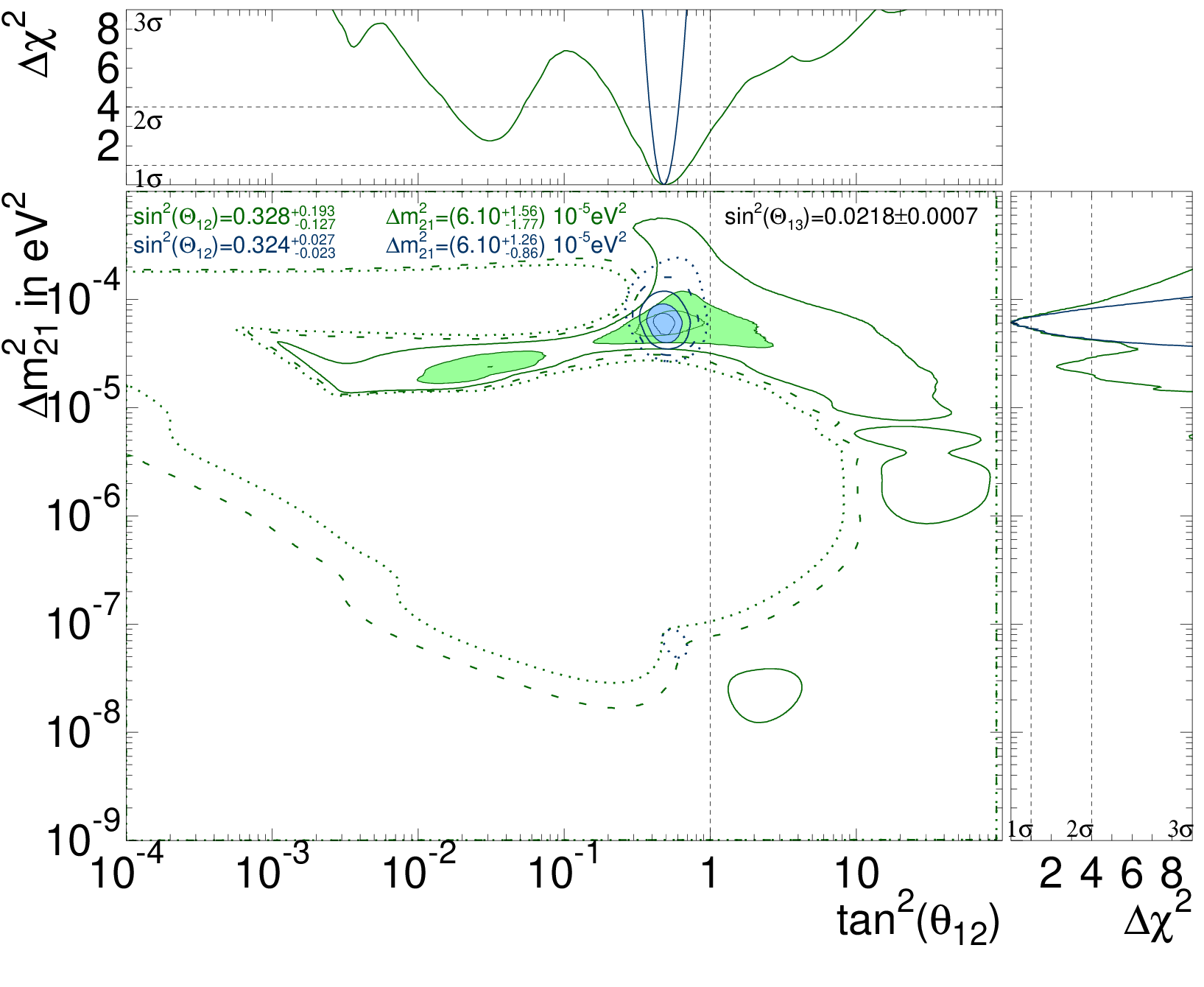}
\end{center}
\caption{(Color online) SK constraints on solar oscillation parameters with~(blue (medium gray)) and without~(green (light gray)) using the absolute elastic scattering rate. The shaded regions are for $2\sigma$ confidence level. Also shown are $1\sigma$ and $3\sigma$ allowed areas. The dotted~(dashed) contours are for $4\sigma~(5\sigma)$ C.L.}
\label{fig:osc_skfreevsconstrained}
\end{figure}

 Figure~\ref{fig:osc_skfreevsconstrained} compares the allowed regions at $1$, $2$, and $3\sigma$ confidence level contours obtained from spectrum and measured day/night rate variation with those using the absolute elastic scattering rate in addition. Even without flux constraint, the no oscillations scenario is excluded by the SK combined analysis at greater than $3\sigma$. Both sets of contours are very consistent, but the addition of the absolute rate of course greatly improves the oscillation constraints: allowed areas are smaller and closed intervals are obtained even at $4\sigma$ and $5\sigma$. Those allowed areas are exclusively in the ``Large Mixing Angle'' region ($\sin^2(\theta_{12})\approx0.3$ and $\Delta m^2_{21}\approx6\cdot10^{-5}$eV$^2$)

\begin{figure}[tbh]
\begin{center}
\includegraphics[width=\linewidth,clip]{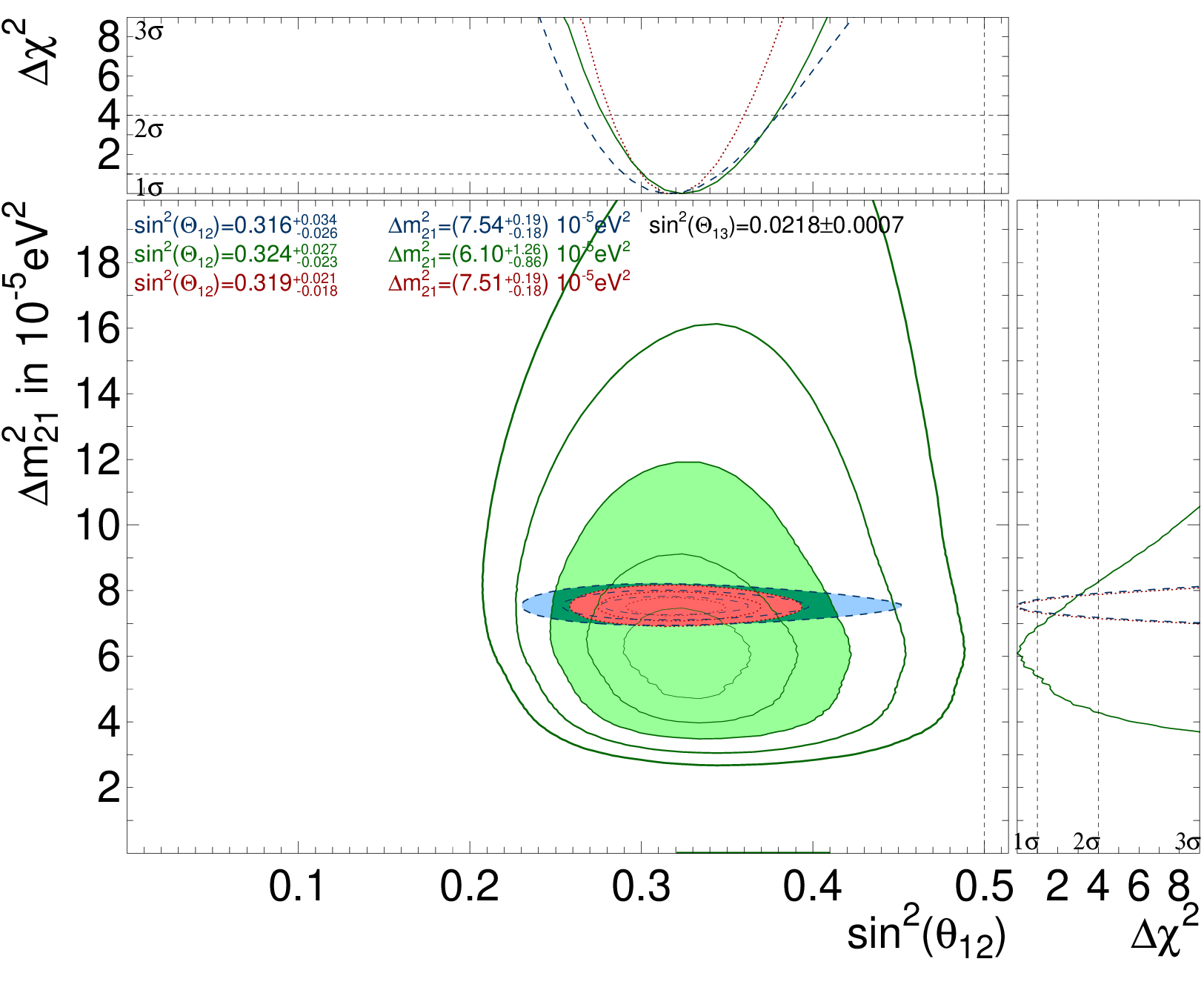}
\end{center}
\caption{
(Color online) Oscillation parameters allowed by SK.
Green (light gray) area is SK contour~($3\sigma$), 
blue (medium gray) area is KamLAND contour~($3\sigma$) 
and red (dark gray) area is SK+KamLAND combined~($3\sigma$).
Green (light gray) solid lines are SK global contours~($1$--$5\sigma$~C.L.),
blue  (medium gray) dashed line: KamLAND contours~($1$--$3\sigma$~C.L.),
and red (dark gray) dotted line: SK + KamLAND contours~($1$--$3\sigma$~C.L.).
A $^8$B ({\it hep}) flux constraint of $(1\pm 0.0381)\times 5.25\times 10^{6}~\mathrm{cm^{-2} \, s^{-1}}$~($(1\pm2)\times 7.88\times10^{3}~\mathrm{cm^{-2}\, s^{-1}}$) is applied.
\label{fig:osc_sk}}
\end{figure}

\begin{figure*}[tbh]
    \centering
    \includegraphics[width=\linewidth,clip]{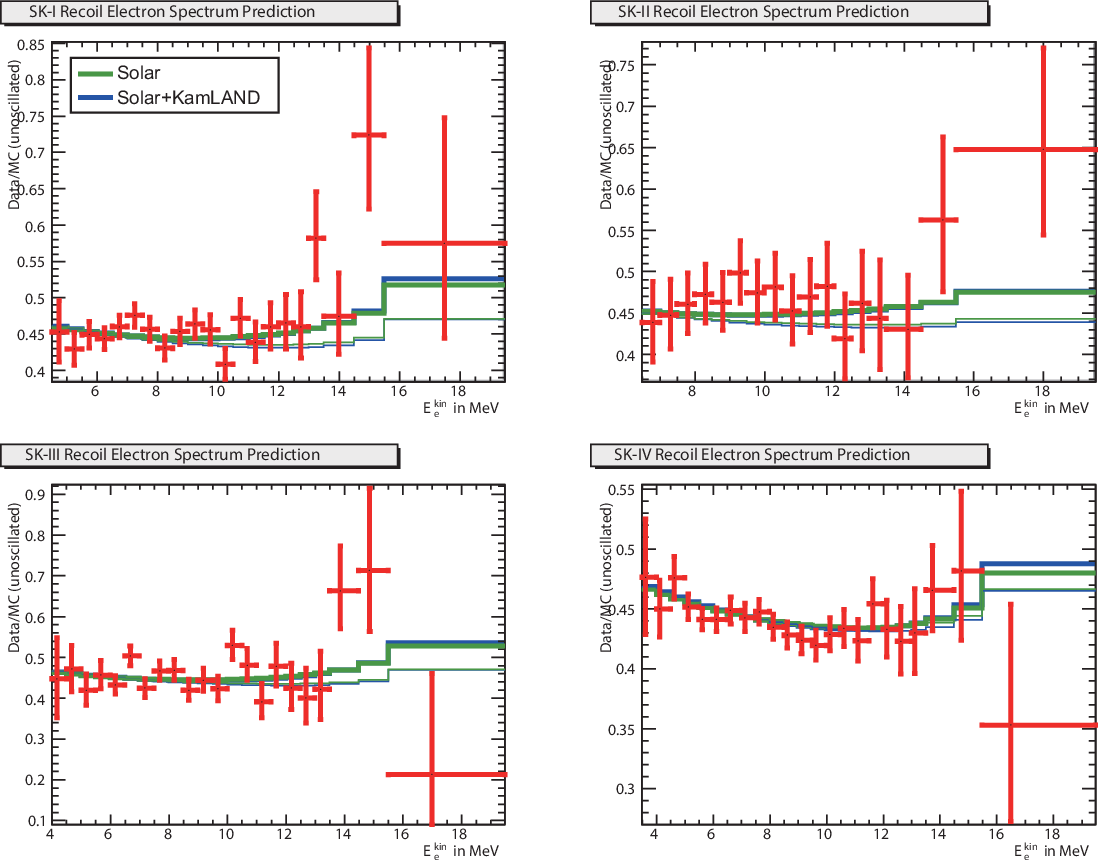}
    \caption{(Color online) Comparison of the Recoil Electron Spectra of SK-I (top left), SK-II (top right), SK-III (bottom left), and SK-IV (bottom right) with MSW predictions. The green(light gray) (blue(medium gray)) histograms are for the solar (solar+KamLAND) best-fit $\Delta m^2_{21}$ value. The thick (thin) lines are distorted (not distorted) by adjusting the energy scale, resolution, and neutrino spectrum within uncertainties.}
    \label{fig:osc_fourspec}
\end{figure*}

\begin{figure}[tbh]
    \centering
    \includegraphics[width=\linewidth,clip]{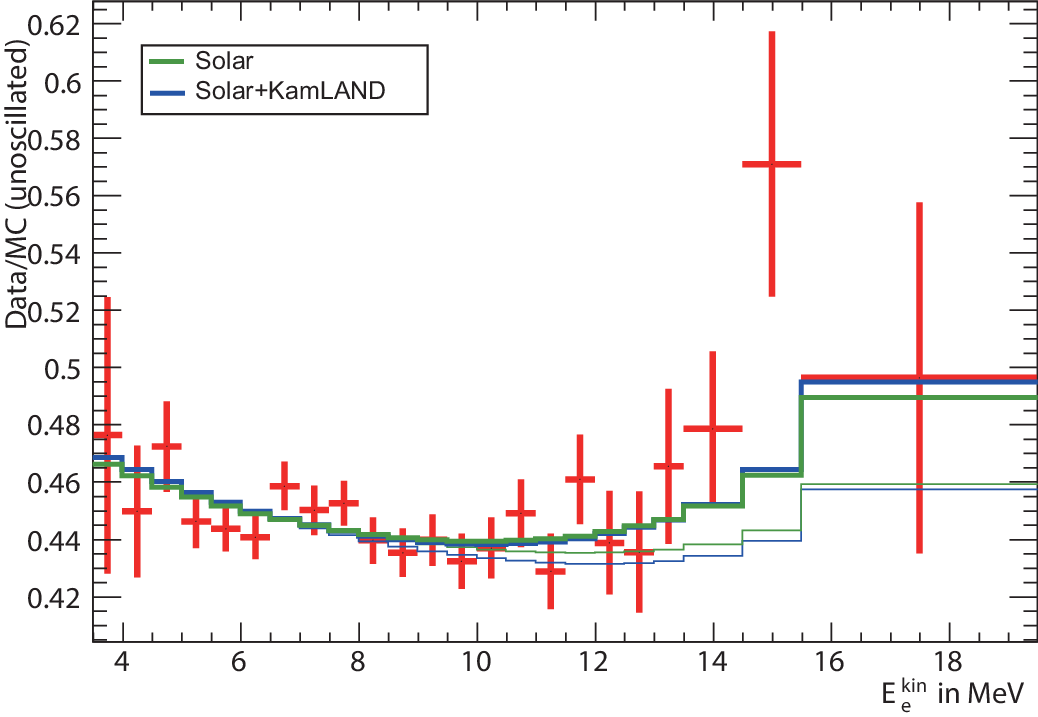}
    \caption{(Color online) Statistical combination of the Recoil Electron Spectra of SK-I, II, III, and IV overlaid with the corresponding combination of the MSW predictions. The green(light gray) (blue(medium gray)) histograms are for the solar (solar+KamLAND) best-fit $\Delta m^2_{21}$ value. The thick (thin) lines are distorted (not distorted) by adjusting the energy scale, resolution, and neutrino spectrum within uncertainties.}
    \label{fig:osc_cspec}
\end{figure}

Using 5805 days of SK data containing more than 100,000 solar neutrino interactions, SK measures the oscillation parameters to be $\sin^{2}\theta_{12,\rm{SK}}=0.324^{+0.027}_{-0.023}$ and $\Delta m^{2}_{21,\rm{SK}}=(6.10^{+1.26}_{-0.86}\times10^{-5})~\mathrm{eV}^{2}$~(Without the total rate: $\sin^{2}\theta_{12}=0.33^{+0.19}_{-0.13}$ and $\Delta m^{2}_{21}=(6.1^{+1.6}_{-1.8}\times10^{-5})~\mathrm{eV}^{2}$).
Figure~\ref{fig:osc_sk} shows, on a linear scale, the allowed regions by combining all data taken during the four SK experimental phases.  When compared to the KamLAND measurements, there is a slight tension~(about $90\%$~C.L. for a $\Delta\chi^2\approx 2.3$) in $\Delta m^{2}_{21}$ as shown in Fig.~\ref{fig:osc_sk}. A significant discrepancy in the oscillation parameters obtained from neutrinos~(SK solar neutrino data) versus anti-neutrinos~(KamLAND reactor anti-neutrino data) would imply CPT violation. The SK and KamLAND data are consistent with CPT conservation within their estimated uncertainties.
Figure~\ref{fig:osc_fourspec} compares the measured with the best-fit MSW spectra. Both the solar best-fit value (see section~\ref{sec:global}) for $\Delta m^2_{21}$ as well as the KamLAND measurement lead to good agreement with the SK spectral data. Note that the different SK phases apply different distortions due to uncertainty in energy scale and resolution. Therefore, a statistical average spectrum can only used for illustration purposes since the predictions of each phase differ, in particular near the $\mathrm{^{8}B}$ endpoint. Figure~\ref{fig:osc_cspec} shows this statistical average together with the MSW predictions.

\subsection{Super-Kamiokande and SNO combined analysis}
\label{sec:osc_sknso}
The neutrino oscillation analysis method to combine data with other experiments is similar to previous SK publications~\cite{Smy:2003jf, Hosaka:2005um, Abe:2016nxk}. Since SNO data also contains solar $^8$B neutrino interactions in a very similar energy range, the combined analysis of SK and SNO plays the most important role in the global analysis. The SNO collaboration has performed a fit of all SNO data to six parameters 
($\Phi_\mathrm{B}$, $c_i$, and $a_i$~\cite{Aharmim:2011vm} )
describing neutrino oscillations in the SNO experiment: three of these parameters are quadratic coefficients of the day-time electron-flavor survival probability~(developed around $10$~MeV: 
$c_0$, $c_1$, and $c_2$
), two more parameters describe a linear approximation of the day/night asymmetry of this survival probability as a function of energy~(also developed around $10$~MeV: 
$a_0$ and $a_1$
). The sixth parameter 
$\Phi_\mathrm{B}$
is the total $\mathrm{^{8}B}$ neutrino flux of all~(active) flavors. SNO published the best-fit values of the six parameters, the fit errors, and the correlation matrix. 
In our analysis, these six parameters are mapped to $s_i$, and
their errors and correlation matrix are assembled in the covariance matrix $V_{\mbox{\tiny SNO}}$. For a given set of oscillation parameter $\theta_{12},\theta_{13},\Delta m^2_{21}$, the day-time survival probability and the day/night asymmetry of the survival probability is calculated as a function of energy. The day-time probability is fitted with a quadratic function, and the day/night asymmetry with a linear function. The fit uses the published relative sensitivity of SNO as a function of energy. From these fit parameters, an expected set of $s_i^{\mbox{\tiny exp}}(\theta_{12},\theta_{13},\Delta m^2_{21})$ is assembled. With $\Delta s_i=s_i-s_i^{\mbox{\tiny exp}}$, the SNO $\chi^{2}$
\[
\chi^{2}_{\mbox{\tiny SNO}}=\sum_{i,j=1}^6 \Delta s_i (V_{\mbox{\tiny SNO}})^{-1}_{ij} \Delta s_j
\]
is added to the SK $\chi^2$,
then the SK+SNO $\chi^{2}$ is defined as
\begin{multline}
     \chi^{2}_{\mbox{\tiny SK,SNO}} = \\
     \underset{\beta, \eta, \tau, \epsilon_p, \rho_p}{\rm{Min}} \left( \sum_{\it{p}\rm{=1}}^{4} 
     \left( \chi^{2}_{\mbox{\tiny spec},\it{p}} + \epsilon^{2}_{\it{p}} + \rho^{2}_{\it{p}} \right) + \tau^{2} + \Phi 
     + \chi^{2}_{\mbox{\tiny SNO}}  \right).
     \label{eq:chi2_sksno}
\end{multline}
In this equation, the $hep$ flux is constrained in $\Phi$ to $(7.9\pm1.2)\times 10^{3}~\mathrm{cm^{-2}\,s^{-1}}$ to be consistent with SNO's assumption in obtaining the SNO parameters while the $\beta$ constraint of $\Phi$ is removed and $\beta$ is identified with the SNO parameter 
$\Phi_\mathrm{B}$.
Then $\beta$, $\eta$, $\epsilon_p$, $\rho_p$, and $\tau$ are minimized.
The definition of parameters is the same as in~\cite{PhysRevD.94.052010}.

\begin{figure}[tbh]
\begin{center}
\includegraphics[width=\linewidth,clip]{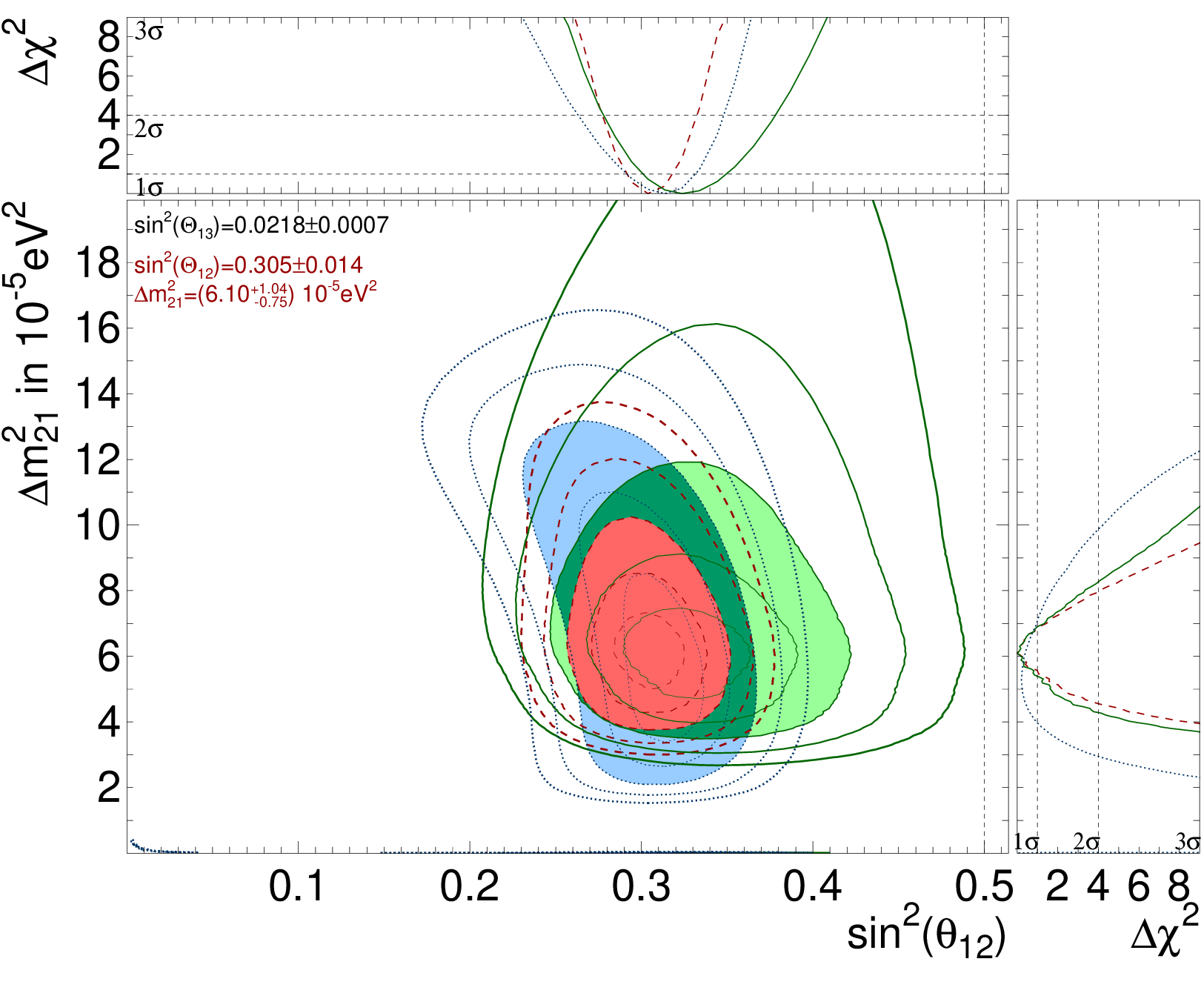}
\end{center}
\caption{(Color online) Oscillation parameters allowed by SK~(green(light gray)) and SNO~(blue(medium gray)) data. The shaded regions are allowed at $3\sigma$ confidence level. Also shown are $1\sigma$, $2\sigma$, $4\sigma$ and $5\sigma$ contours. The SK+SNO combined analysis is shown in red(dark gray).
\label{fig:osc_skvssno}}
\end{figure}

Figure~\ref{fig:osc_skvssno} compares the oscillation constraints of SK data with those of SNO data: SNO data constrains the mixing angle~$\theta_{12}$ better than SK data, while SK data constrains $\Delta m^{2}_{21}$ better than SNO data. Figure~\ref{fig:osc_skvssno} also shows the SK+SNO combined fit, which significantly improves the $\theta_{12}$ constraint compared to the one from just SNO data, while the improvement of $\Delta m^{2}_{21}$ compared to just using SK data is small. The combined fit also eliminates alternate small mixing angle allowed regions~(appearing in the SNO only analysis at $4\sigma$) and small $\Delta m^{2}_{21}$ allowed regions~(appearing in the SNO only analysis at $1\sigma$ and in the SK only analysis at $5\sigma$). This SK+SNO analysis gives the most stringent constraints on $\theta_{12}$ and $\Delta m^2_{21}$ with neutrinos independent of solar model predictions of the solar neutrino fluxes: $\sin^{2}\theta_{12,\rm{SK-SNO}}=0.305\pm0.014$ and $\Delta m^{2}_{21,\rm{SK-SNO}}=(6.10^{+1.04}_{-0.75}\times10^{-5})~\mathrm{eV}^{2}$.

\subsection{Global oscillation analysis}
\label{sec:global}

\begin{figure}[tbh]
\includegraphics[width=9.5cm,clip]{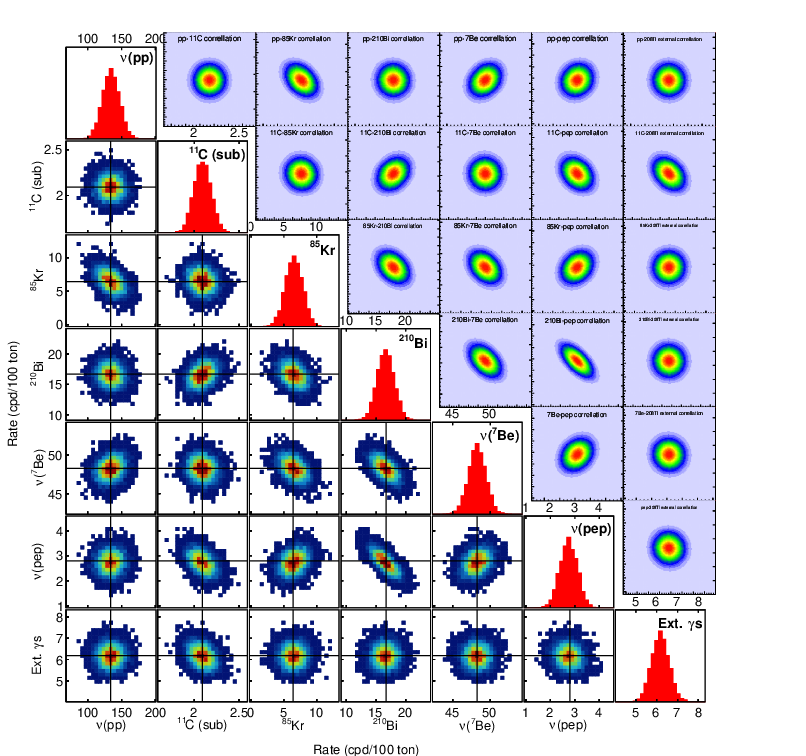}
\caption{(Color online) Borexino spectrum fit parameter pair correlations~(lower left from Ref.~\cite{Agostini:2019}) emulated with a Gaussian covariance~(upper right).
\label{fig:osc_borexino_par_corr}}
\end{figure}

In addition to SNO, other solar neutrino data from Borexino~\cite{Agostini:2019}, the Homestake experiment~\cite{Cleveland:1998nv}, Gallex/GNO~\cite{Altmann:2005ix}, and SAGE~\cite{Abdurashitov:2009tn} are considered.
Unlike in previous publications, Borexino data and radio-chemical solar neutrino data (Homestake, GALLEX/GNO and SAGE) are now treated separately from each other. To analyze Borexino data we extract from~Ref.~\cite{Agostini:2019} $pp$, $pep$, and $^{7}$Be neutrino fluxes, errors and correlations. For the correlations we assume a general Gaussian covariance matrix and match it to the published MC pair correlations of fit parameters in Figure 5 of Ref.~\cite{Agostini:2019}; Figure~\ref{fig:osc_borexino_par_corr} demonstrates this match. These fit parameters describe neutrino interactions and radioactive background rates: Borexino calculates spectra for each neutrino and background species, and (scales by the corresponding parameter) they are fit to the observed Borexino data spectrum. After matching the pair correlations, we then reduce the Gaussian covariance matrix to a smaller matrix that describes just $pp$, $pep$, and $^7$Be neutrino interaction rates by marginalization: the neutrino interaction rate vector is then
\begin{multline}
        r_{\mbox{\tiny Borexino}}=\begin{pmatrix} 
    r_{pp} & r_{pep} & r_{^7\mbox{\tiny Be}}
    \end{pmatrix}
    =\begin{pmatrix} 
    134.0 & 2.43 &48.3
    \end{pmatrix} \\
    \mathrm{counts \, (100~tonne)^{-1}  \, day^{-1}} \notag
\end{multline}

and the inverse covariance matrix is

\begin{multline}
    V^{-1}_{\mbox{\tiny rate}} = \begin{pmatrix}
    0.00647 & -0.0126 & -0.0170 \\
    -0.0126 &  6.36 & -0.501 \\
    -0.017 & -0.501 &  0.784
    \end{pmatrix} \\
    (\mathrm{100~tonne \, day})^{2}. \notag
\end{multline}

The corresponding inferred neutrino fluxes from Borexino data are

\begin{multline}
    \phi_{\mbox{\tiny Borexino}} =  \begin{pmatrix}
    \phi_{pp} & \phi_{pep} & \phi_{^7\mbox{\tiny Be}}
    \end{pmatrix}
     =  \begin{pmatrix} 
    6.10 & 0.0127 & 0.499
    \end{pmatrix} \\
    \times 10^{10}~\mathrm{cm^{-2}\, s^{-1}}. \notag
\end{multline}

These fluxes already take into account the neutrino oscillations using $\sin^{2}\theta_{12}=0.306$, $\sin^{2}\theta_{13}=0.02166$, and $\Delta m^{2}_{21}=7.50\times 10^{-5}~\mathrm{eV}^{2}$~\cite{Agostini:2019}. To calculate neutrino flux predictions for Borexino data for different oscillation parameters, we integrate the neutrino-electron elastic scattering cross section in the region where that particular neutrino species gives the largest contribution to the total Borexino spectrum. We form ratios of the predicted rate with a particular oscillation parameter set over the predicted rate without oscillations. In the case of the ``standard parameters" (meaning the ones assumed by Borexino in~\cite{Agostini:2019}), these ratios are 0.689, 0.614, and 0.638 for $pp$, $pep$, and $^{7}$Be, respectively. We then use solar model predictions with varying oscillation parameters to predict the Borexino event rates and fit those predictions to the observed rates using the Borexino covariance matrix $V_{\mbox{\tiny rate}}$ as well as the solar model neutrino flux 
vector $r_{\mbox{\tiny SSM}}$ and
covariance matrix $V_{\mbox{\tiny SSM}}$~(converted to Borexino interaction rates). 
The components of the flux vector $r$ are the solar neutrino species $pp$, $pep$, $^7$Be, $^8$B, $^{13}$N, $^{15}$O, $^{17}$F, and {\it hep}. 
The {\it relative} solar model covariance matrix~(i.e. constraining the fraction of the neutrino flux deviations) in units of $10^{-3}$ (so that the first fraction is $1.0 \times 10^{-4}$)

{\footnotesize 
\[
\begin{pmatrix}
  0.10 &  0.14 & -0.82 & -1.17 & -1.18 & -1.32 & -0.96 &  0.12 \\
  0.14 &  0.26 & -1.27 & -1.78 & -1.69 & -1.87 & -1.55 &  0.23 \\
 -0.82 & -1.27 & 10.17 & 13.17 &  6.79 &  7.59 &  7.95 & -0.86 \\
 -1.17 & -1.78 & 13.17 & 31.31 & 15.65 & 17.43 & 18.00 & -1.93 \\
 -1.18 & -1.69 &  6.79 & 15.65 & 34.19 & 38.73 & 22.50 & -2.54 \\
 -1.32 & -1.87 &  7.59 & 17.43 & 38.73 & 43.95 & 24.78 & -2.81 \\
 -0.96 & -1.55 &  7.95 & 18.00 & 22.50 & 24.78 & 25.71 & -2.74 \\
  0.12 &  0.23 & -0.86 & -1.93 & -2.54 & -2.81 & -2.74 &  1.17 \\
\end{pmatrix}
\]
}
\noindent 
For $V_{\mbox{\tiny rate}}$ and $r_{\mbox{\tiny Borexino}}$, 0 values are padded for neutrino species that are not $pp$, $pep$ or $^7$Be.

We then fit the Borexino spectrum to the neutrino fluxes constrained by the solar model
with
\begin{multline}
     \chi^{2}_{\mbox{\tiny Borexino}} = 
     \underset{\phi_i}{\rm{Min}} 
     (
        (r - r_{\mbox{\tiny SSM}})^T      \cdot V^{-1}_{\mbox{\tiny SSM}}  \cdot (r - r_{\mbox{\tiny SSM}}) + \\ 
        (r - r_{\mbox{\tiny Borexino}})^T \cdot V^{-1}_{\mbox{\tiny rate}} \cdot (r - r_{\mbox{\tiny Borexino}})
     )
    \notag
\end{multline}

\noindent The minimum $\chi^2$ is then
\begin{align*}
    \chi^2_{\mbox{\tiny Borexino}} &=r_{\mbox{\tiny SSM}}^T \cdot V^{-1}_{\mbox{\tiny SSM}} \cdot r_{\mbox{\tiny SSM}} + r_{\mbox{\tiny Borexino}}^T \cdot V^{-1}_{\mbox{\tiny rate}} \cdot r_{\mbox{\tiny Borexino}} \\ 
    & \quad - r_{\mbox{\tiny best}}^T \cdot V^{-1} \cdot r_{\mbox{\tiny best}} \\
    V^{-1} &=V^{-1}_{\mbox{\tiny SSM}}+V^{-1}_{\mbox{\tiny rate}}  \\
    r_{\mbox{\tiny best}} &= V\cdot \left( 
        V^{-1}_{\mbox{\tiny SSM}}  \cdot r_{\mbox{\tiny SSM}} +
        V^{-1}_{\mbox{\tiny rate}} \cdot r_{\mbox{\tiny Borexino}}
        \right) 
\end{align*}

\noindent which is Borexino's contribution to the global $\chi^2$.
We convert the rate vector and covariance matrix back to the neutrino flux vector and covariance matrix.

The SSM${+}$Borexino neutrino flux vector and covariance are modified by the SK+SNO determination of the $^8$B and $hep$ neutrino fluxes~(and covariance), so that there is no impact of either the SSM $^8$B or {\it hep} neutrino flux uncertainty on the analysis. The radio-chemical covariance matrix $V_{\mbox{\tiny RC}}$ is obtained from the flux covariance matrix $V$ via the cross sections (and errors) of the target isotopes of the radio-chemical experiments~(for some details about the covariance method see Ref.~\cite{Fogli:2002pt}). The radio-chemical rate measurements are then fit via
\begin{equation}
    \chi^2_{\mbox{\tiny RC}} = \sum_{n,m} (R^{\rm{obs}}_n - R^{\rm{exp}}_n) V_{\mbox{\tiny RC},nm}^{-1} (R^{\rm{obs}}_m - R^{\rm{exp}}_m),
\end{equation}
where $R^{\rm{obs}}$ and $R^{\rm{exp}}$ are the observed and expected signal rate, respectively. The indices $n$ and $m$ run over Chlorine and Gallium (Gallex/GNO and SAGE are combined into one $R^{\rm{obs}}$).
The $\chi^2$ of the solar global fit is defined as
\begin{equation}
    \chi^{2}_{\mbox{\tiny solar}} = \chi^{2}_{\mbox{\tiny Borexino}} + \chi^{2}_{\mbox{\tiny SK,SNO}} + \chi^{2}_{\mbox{\tiny time}} + \chi^{2}_{\mbox{\tiny RC}}.
\label{eq:solar_global}
\end{equation}

\begin{figure}[tbh]
\begin{center}
\includegraphics[width=\linewidth,clip]{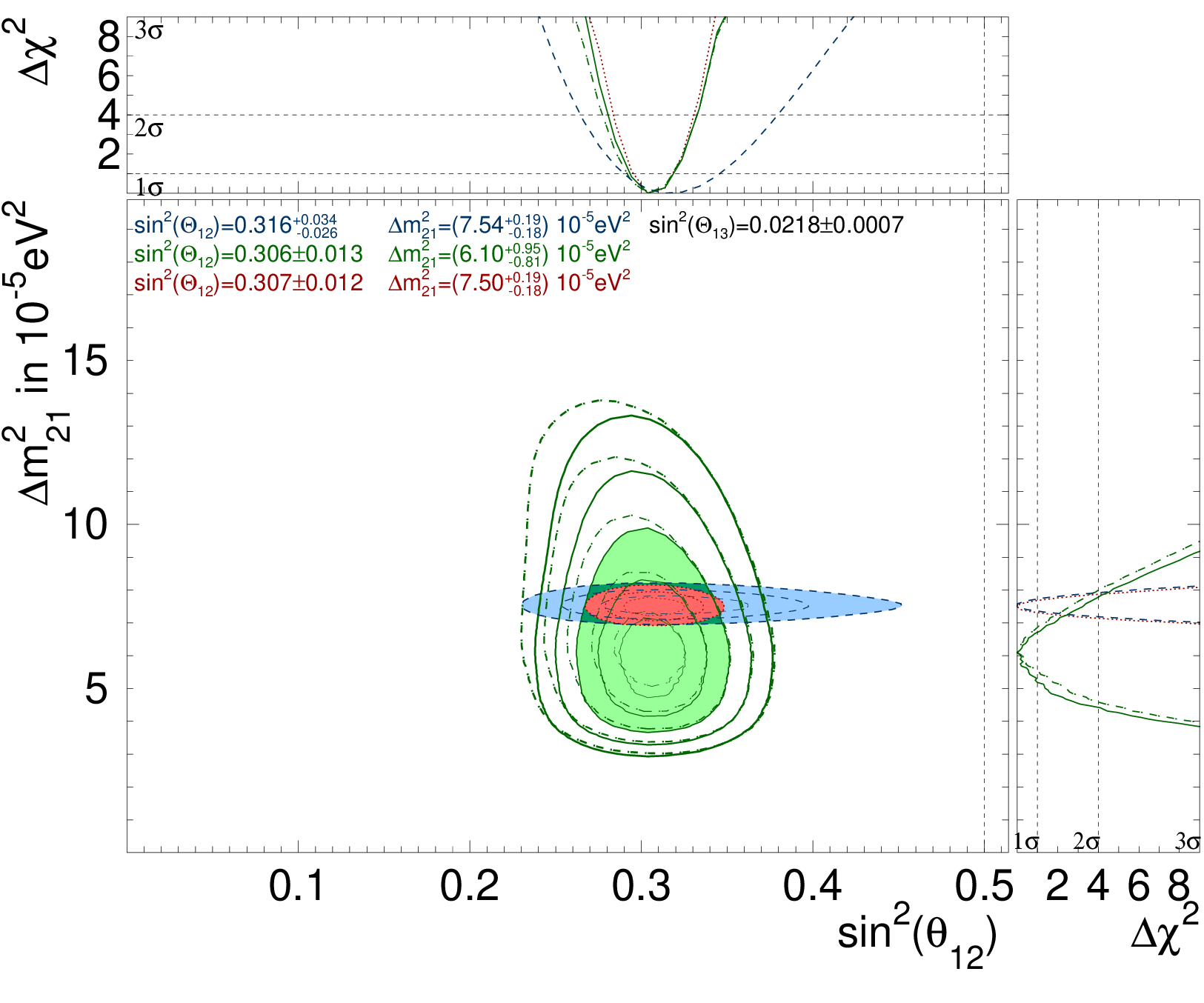}
\end{center}
\caption{
(Color online) $\theta_{12}$ and $\Delta m^2_{21}$ allowed by the global analysis.
The Green(light gray) area is the solar global contour~($3\sigma$), the
blue(medium gray) area is the KamLAND contour~($3\sigma$) 
and the red(dark gray) area is the Solar+KamLAND combined~($3\sigma$).
Green(light gray) solid lines are solar global contours~($1$--$5\sigma$~C.L.),
blue(medium gray) dashed line: KamLAND contours~($1$--$3\sigma$~C.L.),
and red(dark gray) dotted line: Solar + KamLAND contours~($1$--$3\sigma$~C.L.). The dashed-dotted contours are SK+SNO contours for comparison.
\label{fig:global}}
\end{figure}

To extract the best values of $\theta_{12}$ and $\Delta m^2_{21}$ we combine solar experimental neutrino data and KamLAND reactor anti-neutrino data~\cite{Gando:2013nba} (solar+KamLAND) by simple addition of the $\chi^2$ functions, assuming no correlation of the solar and KamLAND results:
\begin{equation}
    \chi^2_{\mbox{\tiny global}} = \chi^2_{\mbox{\tiny Borexino}} + \chi^2_{\mbox{\tiny SK,SNO}} + \chi^2_{\mbox{\tiny time}} 
    + \chi^2_{\mbox{\tiny RC}} + \chi^2_{\mbox{\tiny KamLAND}}.
\label{eq:global}
\end{equation}
Both Eq.~(\ref{eq:solar_global}) and
Eq.~(\ref{eq:global}) are evaluated with an external $\theta_{13}$ constraint of $\sin^2\theta_{13}=0.0218\pm0.0007$.
Figure~\ref{fig:global} shows the oscillation parameters allowed by SK and SNO data, all solar data, KamLAND data, and all solar+KamLAND data.
\begin{figure}[tbh]
\begin{center}
\includegraphics[width=\linewidth,clip]{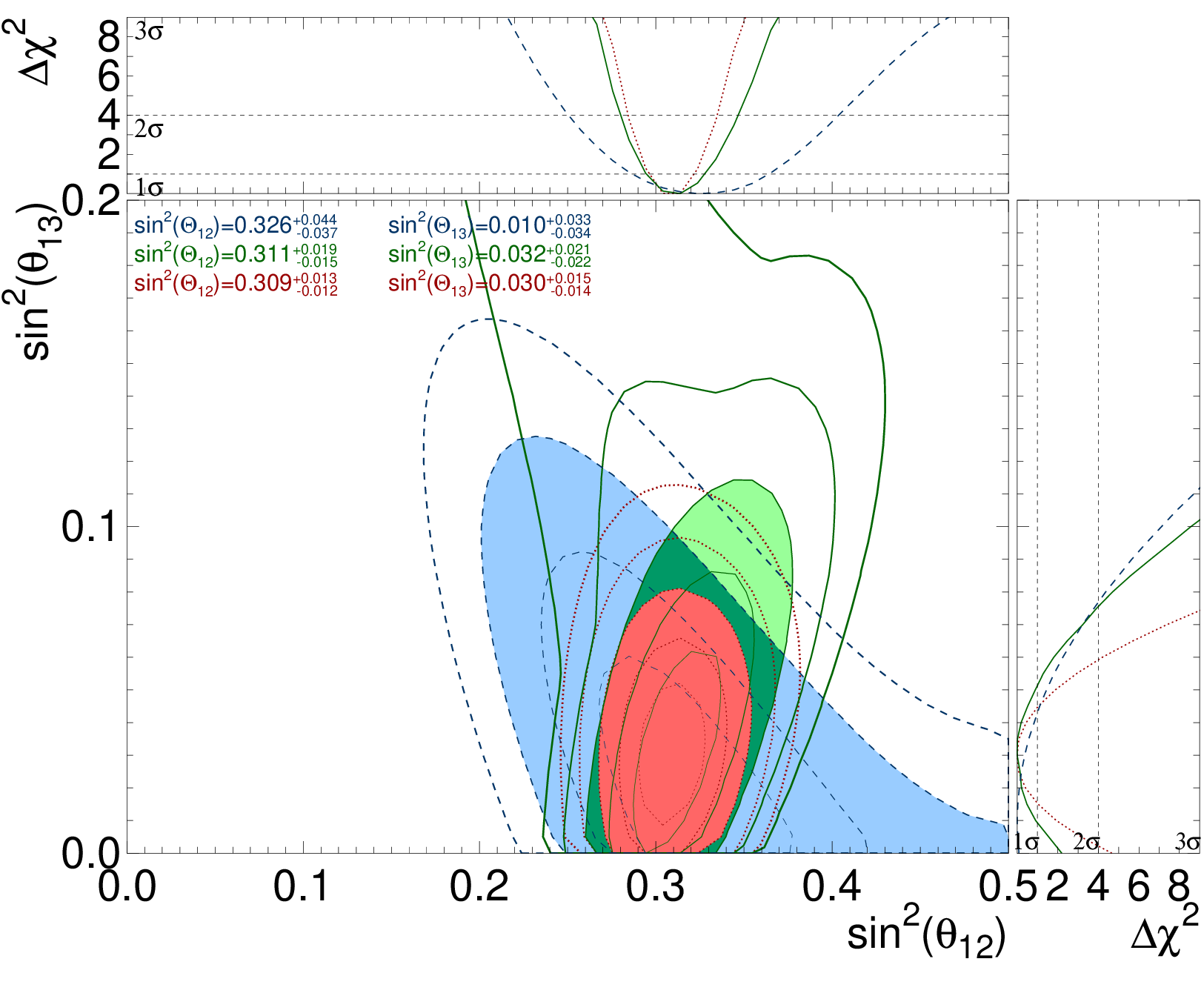}
\end{center}
\caption{
(Color online) $\theta_{12}$ and $\theta_{13}$ allowed by the global analysis.
The Green(light gray) area is the solar global contour~($3\sigma$), the
blue(medium gray) area is the KamLAND contour~($3\sigma$) 
and the red(dark gray) area is the Solar+KamLAND combined~($3\sigma$).
Green(light gray) solid lines are solar global contours~($1$--$5\sigma$~C.L.),
blue(medium gray) dashed line: KamLAND contours~($1$--$4\sigma$~C.L.),
and red(dark gray) dotted line: Solar + KamLAND contours~($1$--$5\sigma$~C.L.). \label{fig:global3f}}
\end{figure}

The best-fit result from the solar global analysis is
\begin{align*}
    \rm{sin^{2} \theta_{12,\mathrm{solar}}} & = 0.306 \pm 0.013, \\
    \Delta m^{2}_{21,\mathrm{solar}} &  = (6.10^{+0.95}_{-0.81})  \times 10^{-5}~\rm{eV}^{2}.
\end{align*}
The best-fit oscillation parameters from all solar experiments and KamLAND are
\begin{align*}
    \rm{sin^2 \theta_{12, \mathrm{global}}} & = 0.307 \pm 0.012, \\
    \Delta m^2_{21,\mathrm{global}} & = (7.50^{+ 0.19}_{-0.18})  \times 10^{-5}~\rm{eV}^{2}.
\end{align*}
The oscillation results of this analysis show a tension of $\Delta m_{21}^2$ between the neutrino 
and anti-neutrino of about $1.5\sigma$ as shown in the right panel in Fig.~\ref{fig:global}, 
and it is slightly stronger for the global solar analysis compared to the SK+SNO analysis.

The global solar neutrino analysis has some sensitivity to $\theta_{13}$ independently from reactor anti-neutrino measurements since the high-energy solar neutrino branches ($^{8}$B and {\it hep}) undergo MSW flavor conversion while the low-energy solar neutrinos ($pp$, $^7$Be, and $pep$) change flavor by averaged vacuum oscillations. Figure~\ref{fig:global3f} shows the resulting contours of the mixing angles $\theta_{12}$ and $\theta_{13}$. The best-fit value from all solar data of
\begin{equation}
    \sin^2\theta_{13}=0.032^{+ 0.021}_{-0.022}, \notag
\end{equation}

\noindent is statistically consistent with zero as well as the reactor anti-neutrino measurements. When combining all solar data with KamLAND data (whose anti-neutrinos are subject to non-averaged vacuum oscillations), the best-fit value is almost unchanged, but the preference for a non-zero value gets somewhat stronger:
\begin{equation}
    \sin^2\theta_{13}=0.030^{+ 0.015}_{-0.014}. \notag
\end{equation}

\section{Day/Night Asymmetry Amplitude Fit}
\label{sec:osc_dn}

\begin{figure}[htp]
    \begin{center}
        \includegraphics[width=\linewidth]{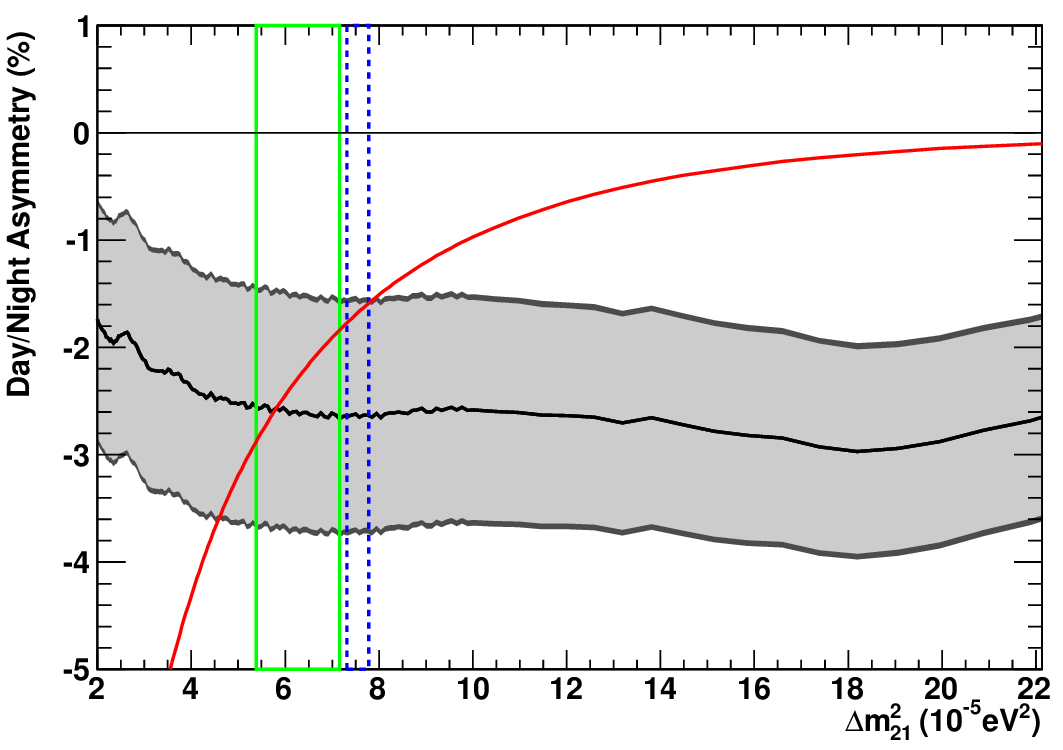}
    \end{center}
    \caption{(Color online) SK-IV day/night amplitude fit dependence on $\Delta m^{2}_{21}$. The black line~(grey band) shows the best-fit value of day/night asymmetry~(its uncertainty). 
    The red solid curve shows the expected day/night
asymmetry.    
    The green solid~(blue dashed) box shows the $1\sigma$ range allowed by solar experiments~(solar experiments and KamLAND).\label{fig:sk4adnvsdm2}}
\end{figure}

Studying the earth matter effect by making separate solar neutrino rate measurements during the day and the night results in comparatively large systematic uncertainties. Those arise from detector asymmetries (the directionality of solar neutrino recoil electrons means that different parts of the detector are illuminated during the day compared to the night) and directional energy scale effects as well as systematics due to the angular distribution of background events. Such systematic errors can be reduced by fitting the amplitude of the rate variation and extracting the corresponding day/night asymmetry from the fit to the amplitude. This fit is done using the likelihood of Eq.~(\ref{eq:time1}). The rate variation (calculated from standard oscillation parameters) $r(\cos\theta_{\rm{z,solar}})$ in each bin is scaled by an amplitude factor $\alpha$ such that the corresponding day/night asymmetry scales with $\alpha$, but the average rate is unchanged~\cite{Smy:2003jf}. Each energy bin rate variation is scaled by the same constant $\alpha$. The corresponding day/night asymmetry for the SK-IV data is
\begin{align}
    A_{\mathrm{D/N}}^{\mbox{\tiny SK-IV, fit}} 
& = -0.0262\pm0.0107\,(\mathrm{stat.})\pm0.0030\,(\mathrm{syst.}).
\end{align}
This fit value is based on the expected zenith variation shape from $\Delta m^2_{21}=6.1\times10^{-5}~\mathrm{eV}^{2}$ with an expected asymmetry of $-0.0238$. Figure~\ref{fig:sk4adnvsdm2} shows the dependence of the fit on $\Delta m_{21}^{2}$.

In the region of interest, the fitted amplitude agrees well with the expected amplitude. The significance of a non-zero day/night asymmetry is $2.4\sigma$. The systematic uncertainty includes energy scale, energy resolution, event selection, the density of electrons from the Earth model~\cite{Dziewonski:1981xy}, and background angular distributions. Table~\ref{tb:sys_dn} summarizes the systematic uncertainties for the day/night amplitude fit.

\begin{table}[htp]
    \begin{center}
    \caption{Summary of systematic uncertainties for the day/night amplitude fit.} 
    \label{tb:sys_dn}
        \begin{tabular}{lc}
        \hline
        \hline
        Item & Systematic uncertainty~[$\%$] \\ \hline
        Energy scale & $0.05\phantom{0}$ \\
        Energy resolution & $0.05\phantom{0}$ \\
        Event selection & $0.10\phantom{0}$ \\
        Earth model & $0.01\phantom{0}$ \\
        Background angular distribution & $0.27\phantom{0}$ \\   \hline
        \hline
        \end{tabular}
    \end{center}
\end{table}

\begin{figure}[htp]
    \begin{center}
        \includegraphics[width=\linewidth]{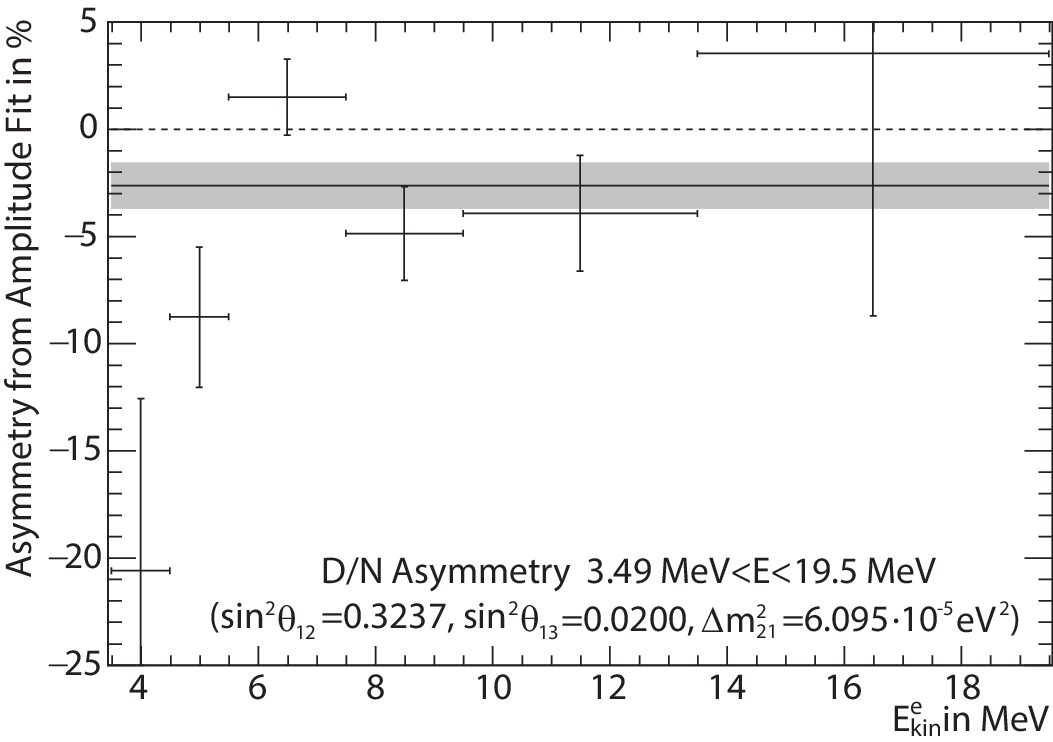}
    \end{center}
    \caption{SK-IV day/night amplitude fit dependence on energy. The grey band shows the combined value.\label{fig:adnvse}}
\end{figure}

The first four contributions are based on Ref.~\cite{Renshaw:2013dzu}, where the dominant last contribution was reevaluated. The amplitude method was improved in three important ways: Firstly, below $7.49$~MeV, the data was split into three MSG regions which improves the stability to fluctuations of intrinsic radioactive background. Secondly, data between $3.49$ and $4.49$~MeV was added, resulting in a combined fit to $39$~data samples~($8\times3$ samples below $7.49$~MeV, and $12$~samples between $7.49$ and $13.49$~MeV, $3$~samples between $13.49$ and $19.49$~MeV). Lastly, the shapes of the angular distribution for the background PDFs vary as a function of solar zenith angle in five regions for day-time events, five regions for event candidate times where solar neutrinos pass only through the Earth's mantle, and another region for times where solar neutrinos pass also through the Earth's core. The amplitude fit allows free variation of the number of background events in each of the eleven regions as well as each of the 39 energy bins~($429$~different background numbers are fit). To estimate the systematic error due to the uncertainty in the angular background PDFs, we use data to measure the background distribution in detector coordinates~(zenith and azimuth of the reconstructed direction). The uncertainty of that measurement is propagated to the calculation of the background PDF~(as a function of $\cos\theta_{\mathrm{Sun}}$), and, subsequently, the amplitude fit result. Due to the larger event statistics as well as these improvements, this important systematic uncertainty was reduced from $0.006$ to $0.0027$. Figure~\ref{fig:adnvse} shows the energy dependence of the asymmetry from amplitude fit using the SK-IV 2970-day data set. The energy region between $3.49$ and $4.49$~MeV differs by $2\sigma$ from the average. However, exclusion of this energy region does not significantly change the best-fit result~($ A_{\mathrm{D/N}}^{\mbox{\tiny SK-IV, fit}}=-0.0241\pm0.0109$ for the higher energy threshold). 

\begin{figure}[htp]
    \begin{center}
    \includegraphics[width=\linewidth]{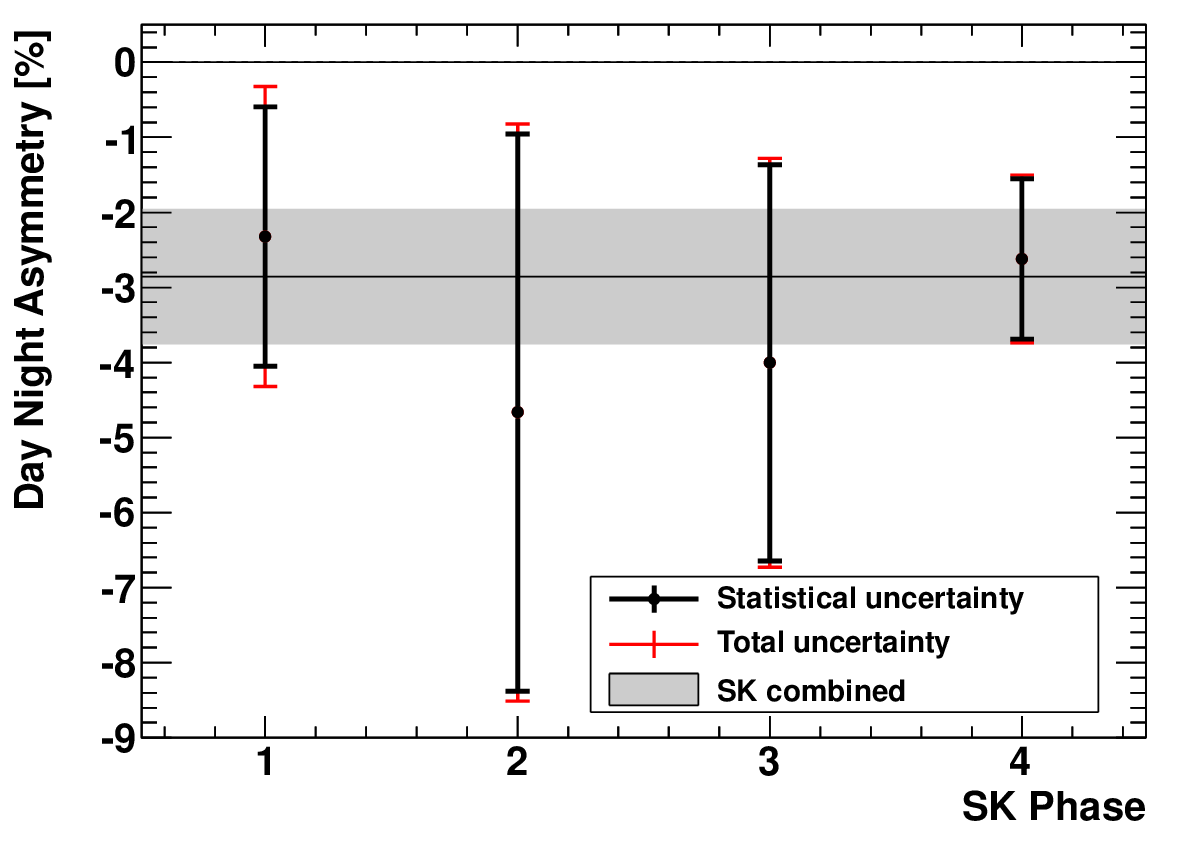}
    \end{center}
    \caption{(Color online) Day/night asymmetries from amplitude fits by the four different SK phases. The black-thick~(red-thin) bar shows the statistical~(total) uncertainty and the grey band shows the combined value. \label{fig:flux_dn}}
\end{figure}

Figure~\ref{fig:flux_dn} compares the results of the day/night flux asymmetry measured at each SK phase. Combining all SK data of day/night flux amplitude fit results in
\begin{align}
    A_{\mathrm{D/N}}^{\mbox{\tiny SK, fit}} 
& = -0.0286\pm0.0085\,(\mathrm{stat.})\pm0.0032\,(\mathrm{syst.}).
\end{align}

\begin{figure}[htp]
    \begin{center}
        \includegraphics[width=\linewidth]{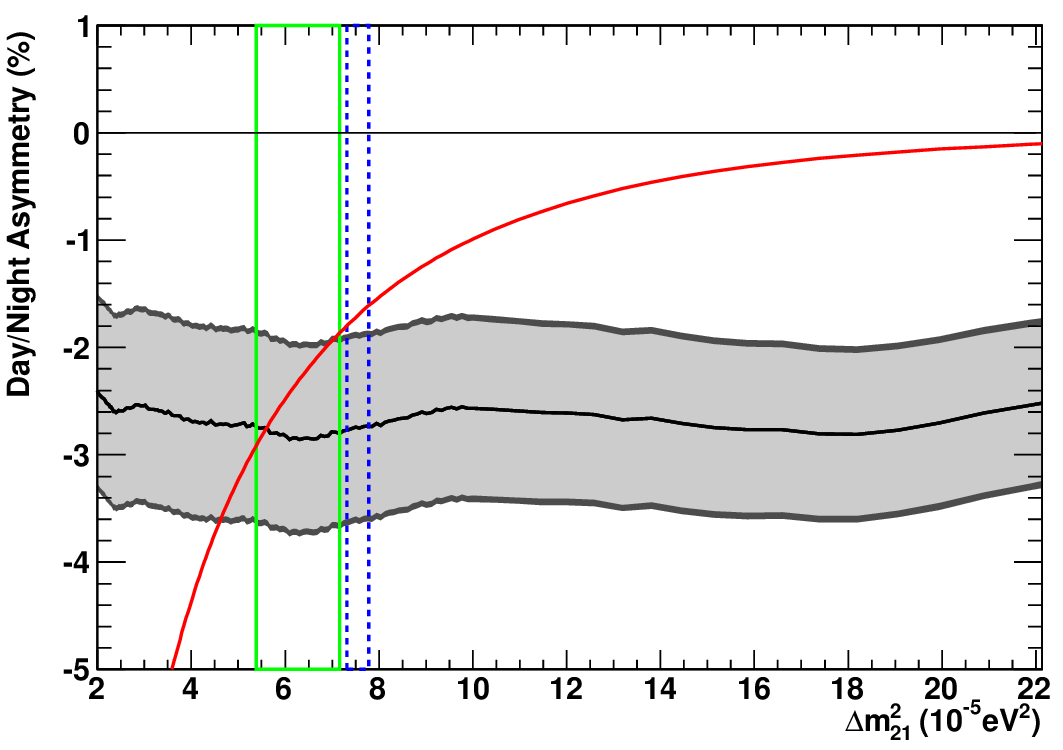}
    \end{center}
    \caption{(Color online) SK-I-IV combined day/night amplitude fit dependence on $\Delta m^{2}_{21}$. 
    The definitions are the same as Fig.~\ref{fig:sk4adnvsdm2}.
    \label{fig:adnvsdm2}}
\end{figure}

\noindent Here, the asymmetry parameter is expressed based on the SK-I energy range, so the expected asymmetry is a bit stronger: $-0.0242$. Since this result differs from zero by $3.2\sigma$, we find evidence for the existence of earth matter effects on solar neutrino oscillation.
The asymmetry parameter depends on the expected zenith angle variation shapes and therefore on the oscillation parameters. The dependence on the mixing angle is negligible. The dependence on $\Delta m^{2}_{21}$ is somewhat stronger. Figure~\ref{fig:adnvsdm2} shows that dependence by analyzing the data taken throughout four SK phases. In particular, at the solar+KamLAND best-fit value of $\Delta m^{2}_{21}=7.5\times10^{-5}~\mathrm{eV}^{2}$, the combined SK day/night amplitude fit corresponds to a slightly smaller asymmetry of $A_{\mathrm{D/N}}^{\mbox{\tiny SK, fit}}=-0.0274\pm0.0083\,(\mathrm{stat})\pm0.0032\,(\mathrm{syst.})$ where an asymmetry of $-0.0172$ is expected. Zero asymmetry differs from this measurement by $3.1\sigma$.
For reference, at the previously favoured~\cite{Renshaw:2013dzu} $\Delta m^{2}=4.8 \times 10^{-5}~\mathrm{eV}^{2}$ we obtain $A_{\mathrm{D/N}}^{\mbox{\tiny SK, fit}}=-0.0273\pm0.0086~\mathrm{(stat.)}\pm0.0032~\mathrm{(syst.)}$ where an asymmetry of $-0.0384$ is expected. Zero asymmetry differs from this measurement by $3.0\sigma$.

\section{Energy spectrum analysis}
\label{sec:osc_spec}

The MSW effect~\cite{Mikheev:1986gs,Wolfenstein:1977ue} of the higher energy $^8$B neutrinos is a unique feature of solar neutrino flavor physics as well as an important test of standard weak interaction theory. It predicts almost complete adiabatic conversion of the electron flavor state at neutrino production in the core of the Sun to the second mass eigenstate when neutrinos leave the Sun. Therefore, the electron flavor survival probability becomes just $\sin^2\theta_{12}$ (in the two neutrino approximation), more or less independent of neutrino energy provided it is far above the ``resonance energy'' at the solar core.
Lower energy $^8$B neutrinos (as well as $pp$, $pep$, and $^7$Be solar neutrinos) on the other hand undergo regular vacuum neutrino oscillations which average out to an energy-independent electron flavor survival probability $P_{ee}$ of $1-\frac{1}{2}\sin^2(2\theta_{12})$. This results in an energy-dependence of that survival probability from a lower value of $\approx 0.3$ at high energy ($>10$ MeV) to a higher value of $\approx 0.6$ at low energy ($<1$ MeV) called the ``upturn''. Each phase of Super-Kamiokande measures the ratio of observed neutrino-electron elastic scattering over the no-flavor change expectation as a function of recoil electron energy (see Figures \ref{fig:osc_compspec} and \ref{fig:osc_fourspec}). The upturn of $P_{ee}$ leads to a distortion of the measured ratio. 
In order to test the presence of such an upturn, $P_{ee}(E_{\nu})$ is parameterized~\cite{Aharmim:2011vm, Abe:2016nxk} in several different ways:
\begin{multline}
\label{eq:quad_fit}
  P_{ee,\mbox{\tiny quad}}(E_{\nu}) 
  = c_0 + c_1 \left( \frac{E_{\nu}}{\rm{MeV}} -10\right) \\
  + c_2 \left( \frac{E_{\nu}}{\rm{MeV}} -10\right)^{2},
\end{multline}
\begin{multline}
\label{eq:cubic_fit}
  P_{ee,\mbox{\tiny cubic}}(E_{\nu})
  = c_0 + c_1 \left( \frac{E_{\nu}}{\rm{MeV}} -10\right) \\
  + c_2 \left( \frac{E_{\nu}}{\rm{MeV}} -10\right)^{2} 
  +c_3 \left( \frac{E_{\nu}}{\rm{MeV}} -10\right)^{3},
\end{multline}
\begin{multline}
\label{eq:exp_fit}
  P_{ee,\mbox{\tiny exp}}(E_\nu) 
  = e_0+\frac{e_1}{e_2} \left(\exp{\left[{e_{2}\left(\frac{E_\nu}{\mathrm{MeV}}-10\right)}\right]}-1\right).
\end{multline}

Using the $P_{ee,\mbox{\tiny par}}(E_{\nu})$~(par is either quad, cubic, or exp), 
the modified expected energy spectrum of the recoil 
electron~$b_{i,\mbox{\tiny par}}$ and $h_{i,\mbox{\tiny par}}$~(corresponding
$b_{i}$ and $h_{i}$ in Sec.~\ref{sec:osc_method}) are made for each experimental phase, 
considering the energy resolution function~(Eq.~(\ref{eq:res_fun})) and the average day/night asymmetry.
Then, the spectral $\chi^{2}$
is calculated by using the 
$b_{i,\mbox{\tiny par}}$ and $h_{i,\mbox{\tiny par}}$.
Finally, the fitted function giving the smallest chi-square~(defined as $\chi^{2}_{\mathrm{min}}$) becomes the best fit.
Among $\Delta \chi^{2} = \chi^{2} -\chi^{2}_{\rm{min}}$ less than $1$, the regions between the maximum and 
the minimum $P_{ee,\mbox{\tiny par}}(E_{\nu})$ at each $E_{\nu}$ are defined as $1\sigma$ region.

\begin{figure}[tbh]
\begin{center}
\includegraphics[width=\linewidth,clip]{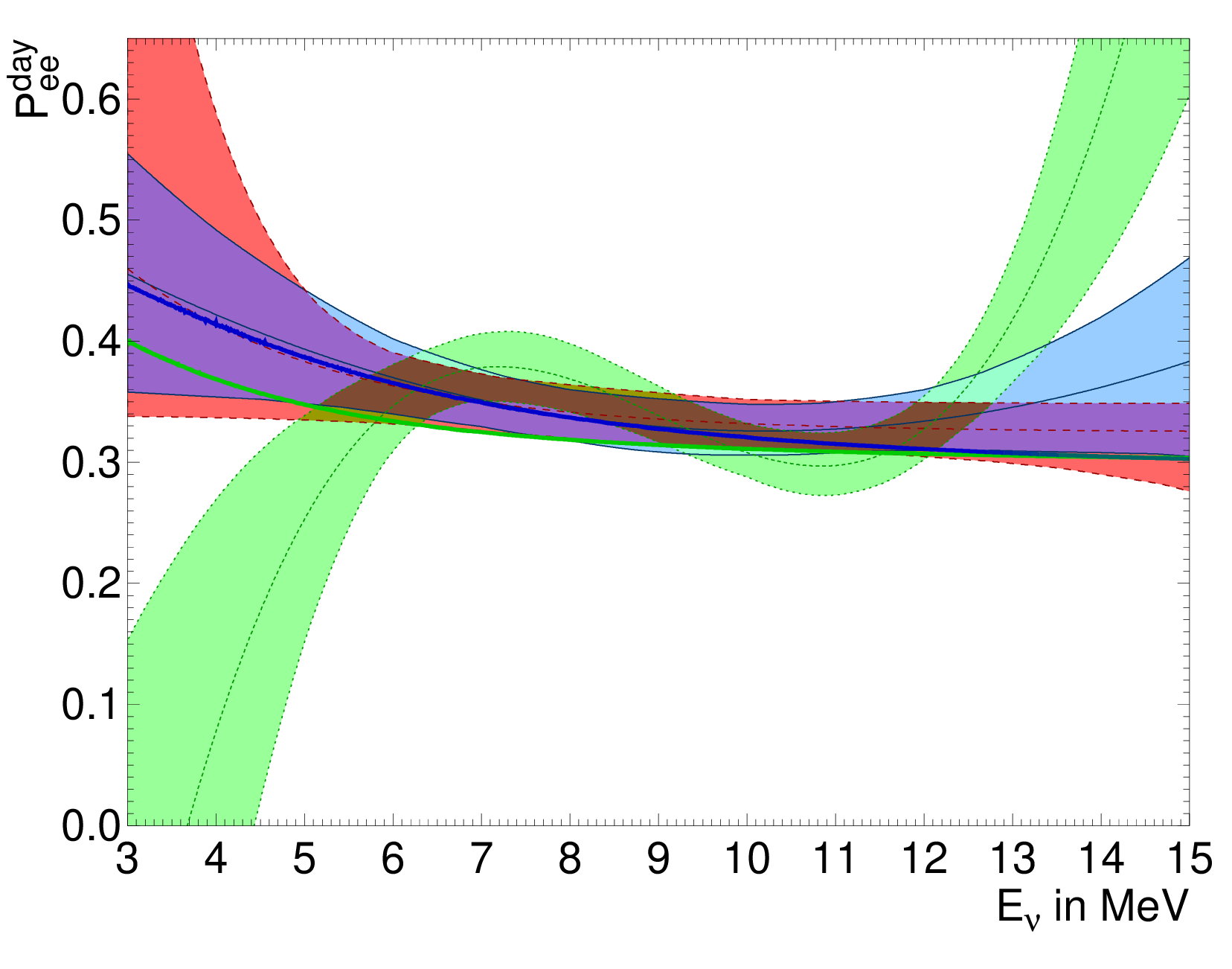}
\end{center}
\caption{(Color online) The electron neutrino survival probability as a function of neutrino energy obtained from all SK data.
The blue(medium gray) (red(dark gray)) region obtained form the $P_{ee,{\mbox{\tiny quad}}}(E_{\nu})$ ($P_{ee,{\mbox{\tiny exp}}}(E_{\nu})$) function. The green(light gray) region is from the $P_{ee,{\mbox{\tiny cubic}}}(E_{\nu})$ function.
The thick blue(medium gray) line is the $P_{ee}(E_{\nu})$ distribution expected from neutrino oscillation at all solar + KamLAND best-fit parameter set,
and the thick green(light gray) line is the $P_{ee}(E_{\nu})$ distribution expected from neutrino oscillation at all solar best-fit parameter set.
\label{fig:pro1}}
\end{figure}

Figure~\ref{fig:pro1} shows the electron neutrino survival probabilities as a function of neutrino energy, obtained using data from all the four SK run periods.

The spectrum $\chi^2$ value for the oscillation parameters $\sin^{2}\theta_{12}=0.304$, $\sin^2\theta_{13}=0.02$, and $\Delta m^{2}_{21}=7.5\times10^{-5}~\mathrm{eV}^{2}$~(near solar+KamLAND data combined best fit: ``sol.+KL'') is $66.50$ with $82$ degrees of freedom. The $\chi^{2}$ decreases to $65.82$ at $\Delta m^{2}_{21}=6.1\times10^{-5}~\mathrm{eV}^{2}$ (near solar data combined best fit: ``solar''), so the SK spectrum only slightly favors the solar data combined best fit by $0.8\sigma$.

As shown in Table~\ref{tab:specpeechi2}, both $\chi^{2}$ values are well-approximated by cubic function $\chi^{2}$ values of $66.34$~(``sol.+KL'') and $65.76$~(``solar'').
The best-fit energy-independent~(``indep'') $P_{ee}=0.336$ has $\chi^{2}=67.20$, so the solar~(solar+KamLAND) best-fit equivalent cubic function is favored by $1.2\sigma$~($0.9\sigma$) over an energy-independent $P_{ee}$. The quadratic approximation is also reasonable: $\chi^{2}=66.34$~(``sol.+KL'') and $\chi^{2}=65.63$~(``solar''). Of course, the best-fit energy-independent $P_{ee}=0.336$ is the same, so the solar~(solar+KamLAND) best-fit equivalent quadratic function is favored by $1.3\sigma$~($0.9\sigma$). The exponential approximation is similar: $\chi^{2}=66.30$ (``sol.+KL'') and $\chi^{2}=65.71$ (``solar'') with similar conclusions. In summary, the SK spectrum measurement favors the existance of an ``upturn'' by $1.2\sigma$. Table~\ref{tab:specpeecoef} shows a summary of the fitting results of the coefficients.
\begin{table}[tbh]
\caption{
$\chi^2$ comparisons for the parameter fits to Eq.~(\ref{eq:quad_fit}), Eq.~(\ref{eq:cubic_fit}), and Eq.~(\ref{eq:exp_fit}).}
\begin{tabular}{l c c c c c c}

\hline\hline
\hline
       & $\chi^2$ & $\Delta\chi^2$ & $c_0$ & $c_1$ & $c_2$ & $c_3$ \cr
\multicolumn{7}{c}{SK cubic fit} \cr
``best''    & $61.15$ & --     & $0.308$ & $-0.025$ & $0.0105$ & $0.00360$ \cr
``indep''   & $67.20$ & $6.04$  & $0.336$ \cr
``sol.+KL'' & $66.34$ & $5.19$ & $0.32037$  & $-0.00594$ & $0.00093$ & $-0.00011$ \cr
``solar''   & $65.76$ & $4.61$ & $0.30904$ & $-0.00375$ & $0.00076$ & $-0.00012$ \cr
\hline
\multicolumn{7}{c}{SK quadratic fit} \cr
``best''    & $64.97$ & --     & $0.326$   & $-0.001$   & $0.0025$ \cr
``indep''   & $67.20$ & $2.23$  & $0.336$ \cr
``sol.+KL'' & $66.34$ & $1.36$ & $0.31939$ & $-0.00711$ & $0.00119$ \cr
``solar''   & $65.63$ & $0.66$  & $0.30791$ & $-0.00509$ & $0.00105$ \cr
\hline
\multicolumn{7}{c}{SK+SNO quadratic fit} \cr
``best''    & $71.63$ & --     & $0.308$ & $-0.004$ & $0.0015$ \cr
``indep''          & $75.83$ & $4.20$ & $0.306$ \cr
\hline
\multicolumn{7}{c}{SK exponential fit} \cr
     & $\chi^{2}$ & $\Delta\chi^{2}$ & $e_{0}$ & $e_{1}$ & $e_{2}$ \cr
``best''     & $65.56$ & --     & $0.332$   & $-0.003$   & $-0.42$ \cr
``indep''    & $67.20$ & $1.64$  & $0.336$ \cr
``sol.+KL''  & $66.30$ & $0.75$  & $0.32053$ & $-0.00622$ & $-0.27074$ \cr
``solar''    & $65.71$ & $0.16$  & $0.30916$ & $-0.00421$ & $-0.31897$ \cr
\hline\hline
\end{tabular}

\label{tab:specpeechi2}
\end{table}
\begin{table}[tbh]
\caption{
The fit coefficients and their correlations for Eq.~(\ref{eq:quad_fit}), Eq.~(\ref{eq:cubic_fit}), and Eq.~(\ref{eq:exp_fit}).}
\begin{tabular}{l c c c }
\hline\hline
Data Set    & $e_0$            & $e_1$              & $e_2$  \cr
\hline
SK          & $0.334\pm0.023$  & $-0.045\pm0.0046$ & $-0.9\pm2.0$ \cr
$e_0$       & $1$              & $0.759$           & $0.130$ \cr
$e_1$       & $0.759$          & $1$               & $0.135$ \cr
$e_2$       & $0.130$          & $0.135$           & $1$ \cr
\hline
Data Set    & $c_0$            & $c_1$              & $c_2$  \cr
\hline
SK          & $0.329\pm0.022$  & $-0.0009\pm0.0058$ & $0.0025\pm0.0026$ \cr
$c_0$       & $1$              & $-0.143$           & $-0.285$ \cr
$c_1$       & $-0.143$         & $1$                & $0.687$ \cr
$c_2$       & $-0.285$         & $0.687$            & $1$ \cr
\hline
SK+SNO      & $0.308\pm0.015$  & $-0.0044\pm0.0034$ & $0.0016\pm0.0017$ \cr
$c_0$       & $1$              & $-0.474$           & $-0.394$ \cr
$c_1$       & $-0.474$         & $1$                & $0.391$ \cr
$c_2$       & $-0.394$         & $0.391$            & $1$ \cr
\hline\hline
\end{tabular}

\vspace*{3mm}

SK cubic:\\

\begin{tabular}{c c c c }
\hline\hline
$c_0$           & $c_1$            & $c_2$             & $c_3$\cr
\hline
$0.310\pm0.024$ & $-0.025\pm0.015$ & $0.0103\pm0.0048$ & $0.0036\pm0.0020$\cr
 $1$            &  $0.265$         & $-0.435$          & $-0.347$\cr
 $0.265$        &  $1$             & $-0.601$          & $-0.919$\cr
$-0.435$        & $-0.601$         &  $1$              & $0.822$\cr
$-0.347$        & $-0.919$         &  $0.822$          & $1$\cr
\hline\hline
\end{tabular}
\label{tab:specpeecoef}
\end{table}

\begin{figure}[tbh]
\begin{center}
\includegraphics[width=\linewidth,clip]{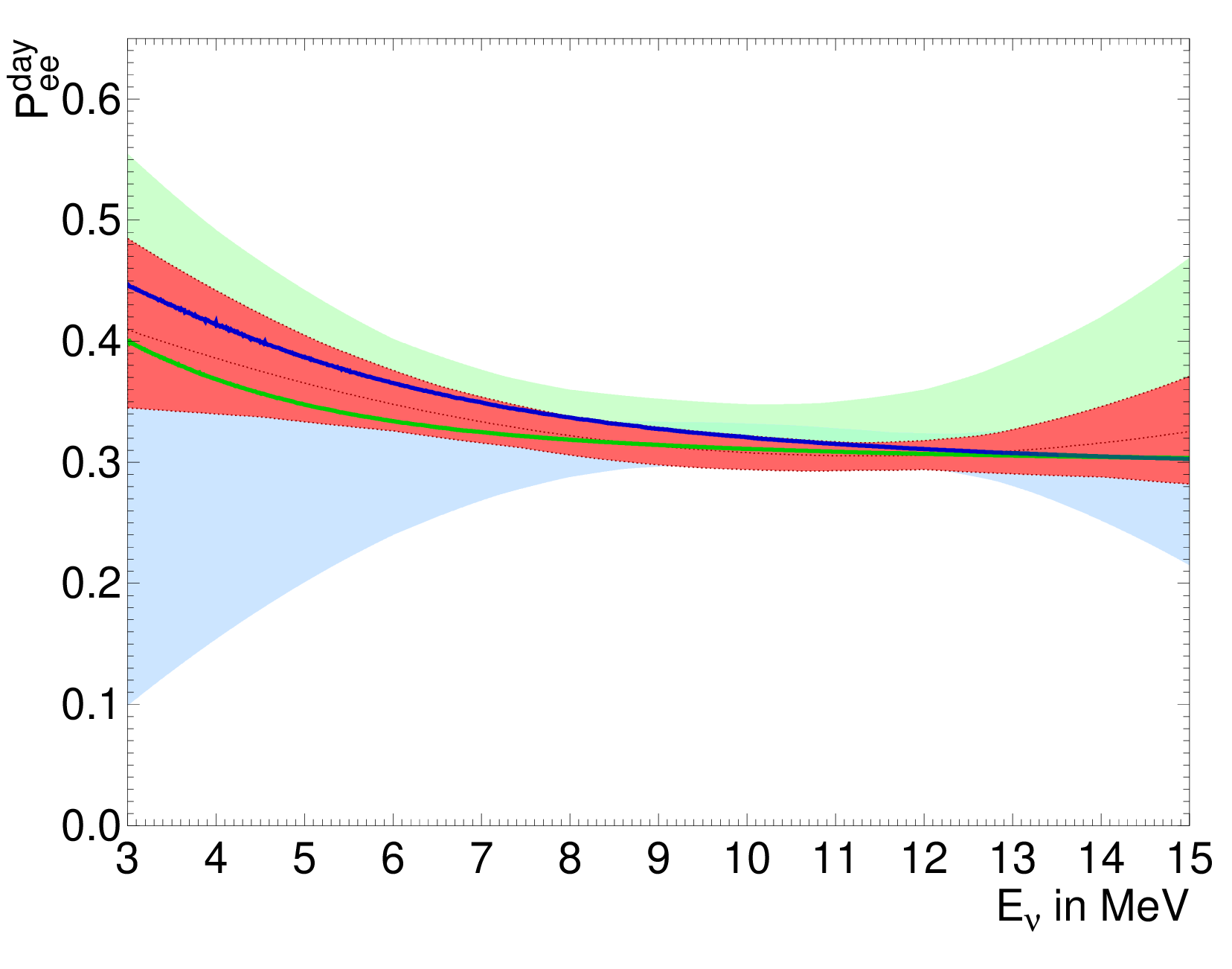}
\end{center}
\caption{(Color online) The electron neutrino survival probability as a function of neutrino energy.
The green(light gray)~(blue(medium gray)) region is obtained from the $P_{ee,{\mbox{\tiny quad}}}(E_{\nu})$ 
function with SK~(SNO) spectrum data.
The red(dark gray) region is obtained from the same function, but with SK+SNO data.
The thick blue(medium gray) and green(light gray) lines are the same as Fig.~\ref{fig:pro1}.
\label{fig:pro2}}
\end{figure}

\begin{figure}[tbh]
\begin{center}
\includegraphics[width=\linewidth,clip]{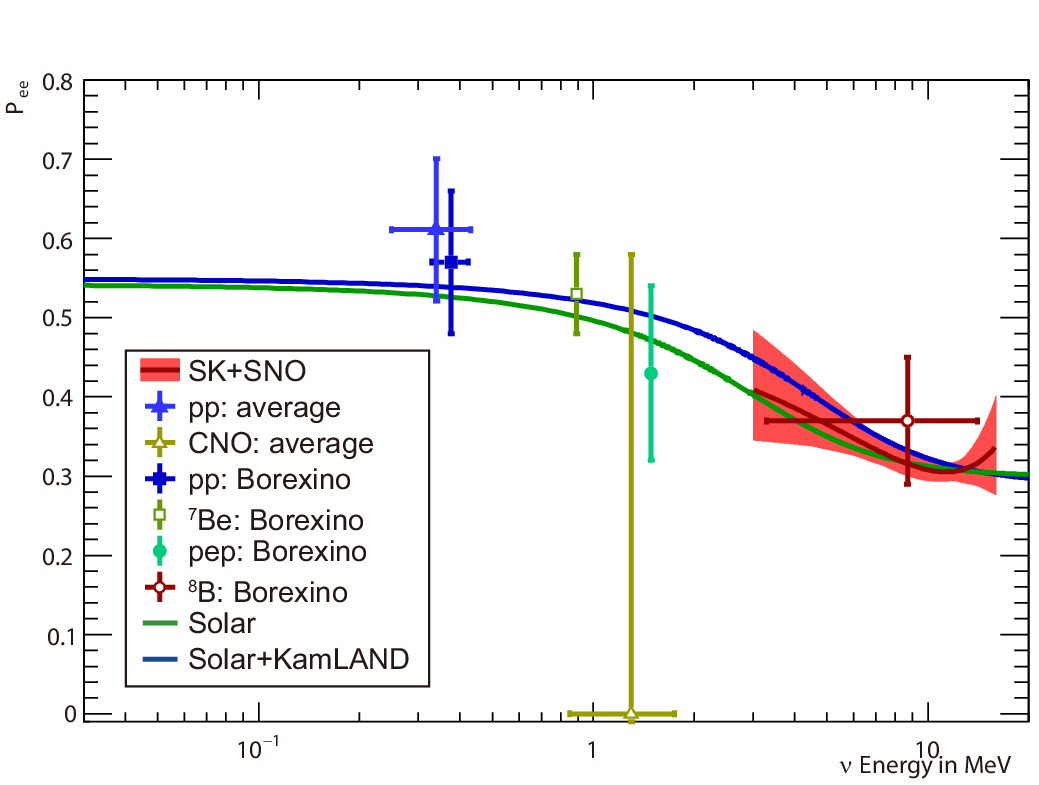}
\end{center}
\caption{(Color online) The electron neutrino survival probability as a function of neutrino energy and solar neutrino measurements.
The red region is obtained from the $P_{ee,{\mbox{\tiny quad}}}(E_{\nu})$ 
function with SK+SNO spectrum data.
The thick blue and green lines are the same as Fig.~\ref{fig:pro1}. 
The light blue (filled triangle) and gold (open triangle) data points are the average $pp$ and CNO neutrino survival probabilities inferred from the radio-chemical solar neutrino data as well as SK+SNO and Borexino $^7$Be measurements, respectively. 
The dark blue (filled square) and dark red (open circle) data points represent the average $pp$ and $^8$B neutrino survival probabilities, respectively.
The green (open square) and turquoise (filled circle) data points represent the $^7$Be and {\it pep} neutrino survival probabilities from Borexino data, respectively.
\label{fig:pee_global}}
\end{figure}

The SNO constraint on the ``upturn'' is obtained and combined with the SK results as follows:
from the SNO parameters $s_i$, the first three are calculated from the quadratic fit parameters $c_0$, $c_1$, and $c_2$ of $P_{ee,\mbox{\tiny quad}}$ while
with day/night asymmetry fitting parameters $a_0$ and $a_1$ are set to the oscillation best fit. The SK+SNO $\chi^{2}$ is then the same as in Eq.~(\ref{eq:chi2_sksno}).
Figure~\ref{fig:pro2} shows the electron neutrino survival probability distributions as a function of neutrino energy
obtained from Eq.~(\ref{eq:quad_fit}) with SK and SNO data.
The SK+SNO combined result on quadratic $P_{ee}$ coefficients favor a distorted spectrum by $2.1\sigma$. Figure~\ref{fig:pee_global} shows the SK+SNO combined result in the context of other solar neutrino survival probability measurements (assuming the standard solar model predictions of the unoscillated neutrino fluxes). The SK+SNO result fits in well with the other data as well as the MSW+neutrino oscillation prediction.

\section{Conclusion}

The fourth phase of the Super-Kamiokande~(SK) has measured 
the solar neutrinos from September $2008$ to May $2018$. 
The total operation period of SK covers almost two solar activity cycles of $23$ and $24$.

In order to obtain precise solar neutrino measurements, several improvements to the analysis are applied in SK-IV.
The data acquisition threshold was lowered in May 2015, as a result the trigger efficiency 
in the $3.49$--$3.99$~MeV region is 
significantly improved.
The improved energy reconstruction method reduces the systematic uncertainties caused by the position dependence of the energy scale.
The spallation backgrounds are further reduced by considering neutron clustering events using Wide-band Intelligent Trigger~(WIT) data.
In the solar neutrino signal extraction, the multiple scattering goodness~(MSG) parameter is newly considered.
For the neutrino oscillation analysis, we have newly incorporated Borexino's pp, pep, and $^7$Be observation data into our global analysis.
Based on these improvements, precise solar neutrino measurements are performed with additional observation data after our previous analysis.

The observed number of the solar neutrino events in $3.49$--$19.49$~MeV region in SK-IV (total live time $2970$~days) becomes:
\begin{eqnarray}
  65,443^{+390}_{-388}\,(\mathrm{stat.})\pm925\,(\mathrm{syst.}). \notag
\end{eqnarray}
Then, the measured solar $\mathrm{^{8}B}$ neutrino flux in SK-IV is
\begin{eqnarray}
  (2.314 \pm 0.014\, \rm{(stat.)} \pm 0.040 \, \rm{(syst.)}) \times 10^{6}~\mathrm{cm^{-2}\,s^{-1}}, \notag
\end{eqnarray}
assuming a pure electron neutrino flavor component without neutrino oscillation.
The flux combined with all the SK phases of the $5805$~days data set is
\begin{eqnarray}
  (2.336 \pm 0.011\, \rm{(stat.)} \pm 0.043 \, \rm{(syst.)}) \times 10^{6}~\mathrm{cm^{-2}\,s^{-1}}. \notag
\end{eqnarray}
The solar neutrino rate measurements in SK are fully consistent with a constant solar neutrino flux emitted by the Sun.

The SK-IV time variation data fit results in a day/night asymmetry of
\[
A^{\rm{SK-IV, fit}}_{\rm{D/N}}=-0.0262\pm0.0107\rm{(stat.)}\pm0.0030\rm{(syst)},
\]
while a fit to all SK data gives
\[
A^{\rm{SK, fit}}_{\rm{D/N}}=-0.0286\pm0.0085\rm{(stat.)}\pm0.0032\rm{(syst)}.
\]
This is a $3.2\sigma$ direct evidence for the existence of earth matter effects on solar neutrino oscillation. The fit assumes $\Delta m^2_{21}=6.1\times10^{-5}$eV$^2$, but similar conclusions hold over the region of interest.

SK-IV data measures the solar oscillation parameters to be
\begin{align*}
    \sin^2\theta_{12,\rm{SK-IV}} & =0.308^{+0.030}_{-0.029}\\
    \Delta m^2_{21,\rm{SK-IV}}   & =6.9^{+1.6}_{-1.2}\times10^{-5}\rm{eV}^2,
\end{align*}
the SK combined result from $5805$~days of data is
\begin{align*}
  \sin^2\theta_{12,\rm{SK}} & = 0.324^{\, +0.027}_{\, -0.023} \\
  \Delta m^{2}_{21,\rm{SK}} & = (6.10 ^{\, +1.26}_{\, -0.86}) \times 10^{-5}~\rm{eV}^{2}.
\end{align*}

The solar model neutrino flux prediction independent measurement of the oscillation parameters by SK and SNO are
\begin{align*}
    \sin^2 \theta_{12,\rm{SK-SNO}} & = 0.305 \pm 0.014 \\
    \Delta m^{2}_{21,\rm{SK-SNO}} & = (6.10^{+ 1.04}_{-0.75})  \times 10^{-5}~\rm{eV}^{2}.
\end{align*}
The SK measurement of the recoil electron spectrum from $^8$B neutrino-electron elastic scattering mildly favors an ``upturn'' by $1.2\sigma.$ Combined with SNO spectral measurements, a distorted spectrum is favored by $2.1\sigma$.

The oscillation parameters from all solar experiments (including SK) are
\begin{align*}
    \sin^2 \theta_{12,\rm{solar}} & = 0.306 \pm 0.013 \\
    \Delta m^{2}_{21,\rm{solar}} & = (6.10^{+ 0.95}_{-0.81})  \times 10^{-5}~\rm{eV}^{2}.
\end{align*}
The best-fit oscillation parameters from all solar experiments and KamLAND are
\begin{align*}
\sin^2 \theta_{12,\rm{global}} & = 0.307\pm0.012 \\
\Delta m^{2}_{21,\rm{global}} & = (7.50^{+ 0.19}_{-0.18})  \times 10^{-5}~\rm{eV}^{2}.
\end{align*}
A tension of $\Delta m_{21}^2$ between all solar experiments and KamLAND persists at about $1.5\sigma$.
Further precise measurements may shed light on this tension in the future.

\section*{Acknowledgments}

We gratefully acknowledge the cooperation of the Kamioka Mining and Smelting Company.
The Super-Kamiokande experiment has been built and operated from funding by the
Japanese Ministry of Education, Culture, Sports, Science and Technology; the U.S.
Department of Energy; and the U.S. National Science Foundation. Some of us have been
supported by funds from the National Research Foundation of Korea (NRF-2009-0083526
and NRF 2022R1A5A1030700) funded by the Ministry of Science,
Information and Communication Technology (ICT); the Institute for
Basic Science (IBS-R016-Y2); and the Ministry of Education (2018R1D1A1B07049158,
2021R1I1A1A01042256, 2021R1I1A1A01059559); the Japan Society for the Promotion of Science; the National
Natural Science Foundation of China under Grants No.11620101004; the Spanish Ministry of Science,
Universities and Innovation (grant PID2021-124050NB-C31); the Natural Sciences and
Engineering Research Council (NSERC) of Canada; the Scinet and Westgrid consortia of
Compute Canada; the National Science Centre (UMO-2018/30/E/ST2/00441 and
UMO-2022/46/E/ST2/00336) and the Ministry of Education and Science (2023/WK/04),
Poland; the Science and Technology Facilities Council (STFC) and
Grid for Particle Physics (GridPP), UK; the European Union's
Horizon 2020 Research and Innovation Programme under the Marie Sklodowska-Curie grant
agreement no.754496; H2020-MSCA-RISE-2018 JENNIFER2 grant agreement no.822070,
H2020-MSCA-RISE-2019 SK2HK grant agreement no. 872549; and
European Union's Next Generation EU/PRTR grant CA3/RSUE2021-00559.

\appendix

\section{Energy reconstruction improvements} \label{sec:app_erec}

In this analysis, the energy reconstruction method was improved to reduce
time and position dependence of the energy scale. Here we describe
in detail
major improvements that we introduced to the energy reconstruction after our
previous publication.

\subsection{Correction for gain shift}

\begin{figure*}[htp]
    \begin{center}
        \includegraphics[width=\linewidth]{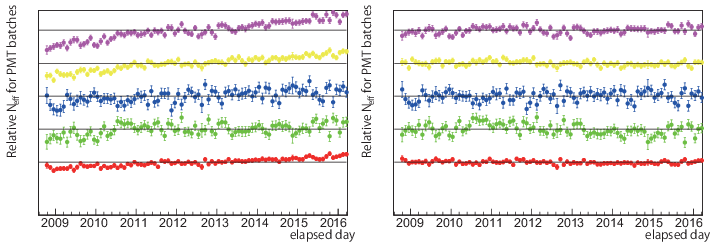}
    \end{center}
    \caption{(Color online) Relative size of effective number of hits without~(left) and with~(right) the correction of the gain drift. Different colored points represent PMTs for different production years (From bottom to top: 1992-1995 (red), 1996-1997 (green), 2003 (blue), 2004 (yellow), 2005 (purple)). The gap between adjacent horizontal lines corresponds to a 5 \% deviation. \label{fig:neff_tvariation_pmtgroups}}
\end{figure*}

As described in Sec.~\ref{sec:gain}, a constant upward drift of the PMT gain was observed. This effectively decreased the hit detection threshold at the front-end electronics, QBEEs~\cite{Nishino:2009zu}, and affected the global energy scale as well as the position dependence.
This effect was evaluated using decay electrons from stopping muons inside the detector. An empirical correction factor, $1/(1+0.226 G_i)$ was applied, where $G_i$ is the relative gain of the $i$-th PMT. This minimizes the time variation of the light yield from decay-electron events after correcting for light attenuation.
Figure~\ref{fig:neff_tvariation_pmtgroups} shows the relative size of the effective number of hits for PMTs, grouped by their production time before and after correction. This demonstrates that this method effectively corrects the observed time dependence of the number of hits for each PMT group. 
This correction of the time variation was cross-checked with an independent set of calibration data taken with the Ni-Cf source. 
We confirmed that the new correction of the gain drift successfully removes the time dependence of the energy scale.

\subsection{Correction for effective PMT coverage}

\begin{figure}[htp]
    \begin{center}
        \includegraphics[width=\linewidth]{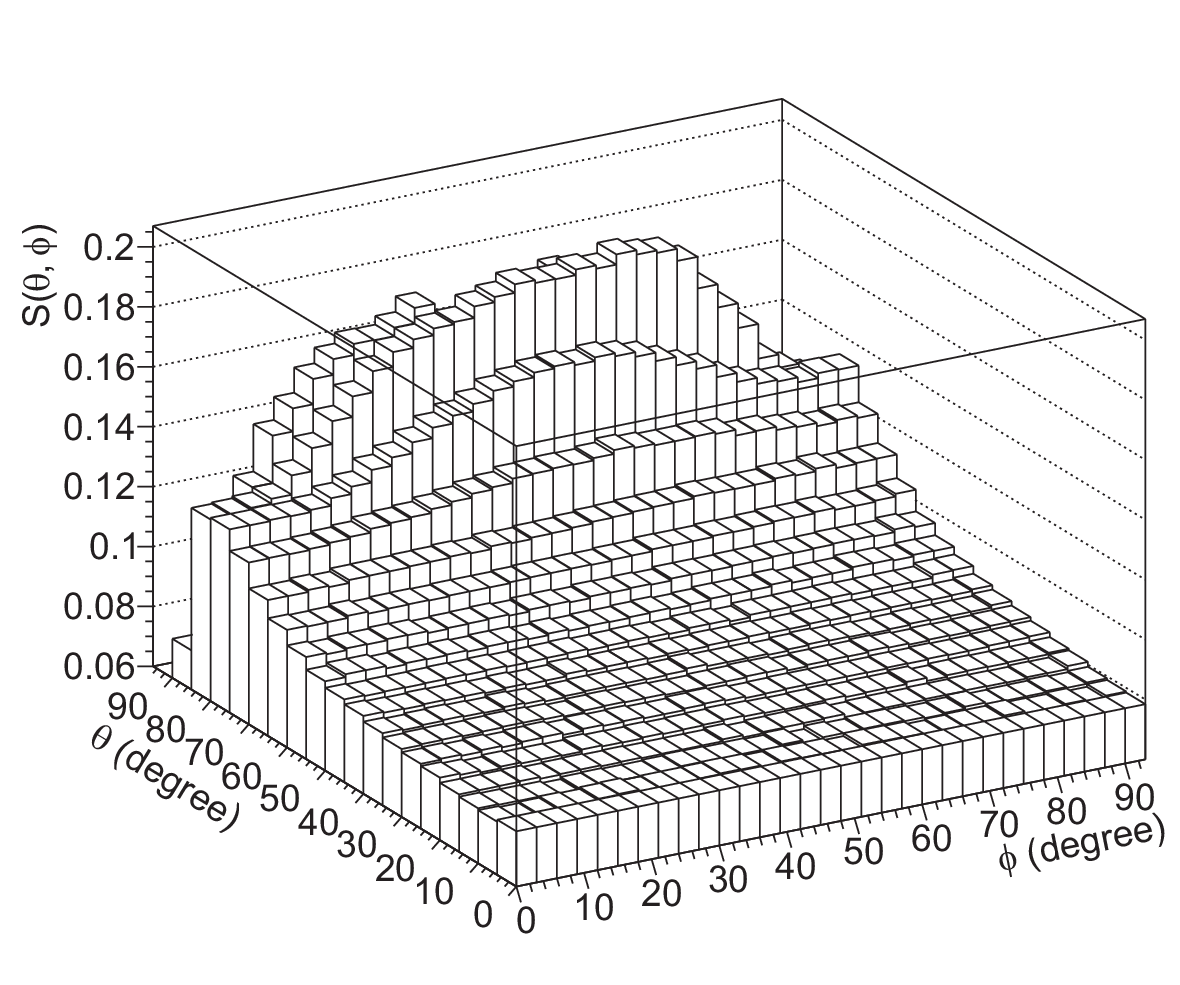}
        \includegraphics[width=\linewidth]{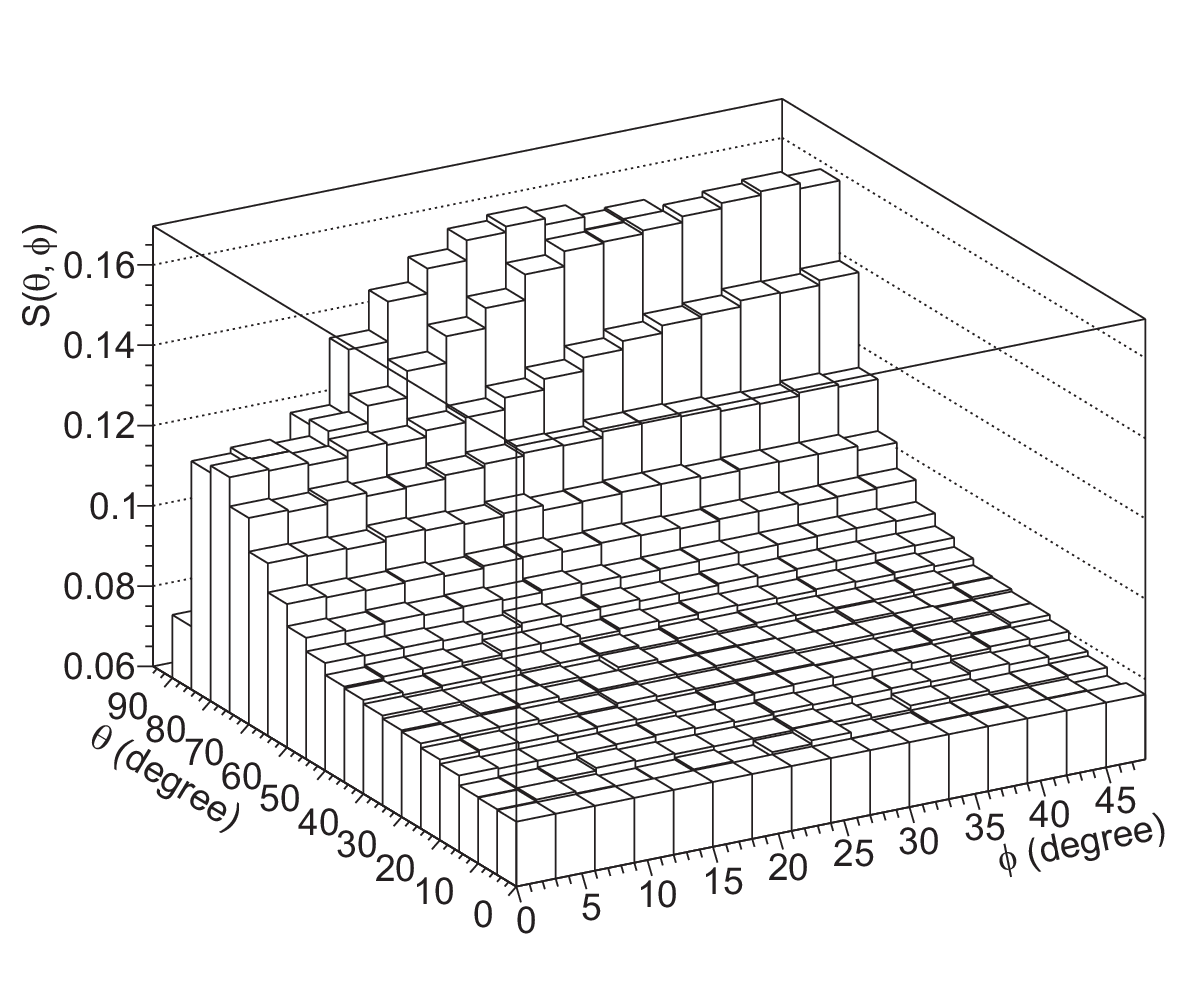}
    \end{center}
    \caption{Values of $S(\theta,\phi)$ for the barrel (upper panel) and top/bottom (lower panel) PMTs. \label{fig:skcoverage} }
\end{figure}

Because of the curved structure of the PMT sphere, the effective surface area of the PMTs varies significantly depending on the photon incident angle. The PMT coverage is approximately $40\%$ for normal incident photons and increases as the incident angle gets shallower. 
The correction of this effect is implemented into the calculation of $N_\mathrm{eff}$ as the $(S(0,0)/S(\theta,\phi))$ term. In the previous analyses, this effect was estimated by a simple simulation of PMT geometrical structure shape arranged on a flat surface. In this analysis, this was improved with a \textsc{Geant4} based simulation that includes the light propagation in the acrylic cover and the PMT glass structures. In addition, the PMT arrangement on the curved structure was also considered for the correction for the barrel PMTs.
Figure~\ref{fig:skcoverage} shows the $S(\theta,\phi)$ values used for this analysis.

\subsection{Correction for light attenuation in water}
\label{appendix:water_quality}
Correction for light attenuation in water was also improved with a more accurate representation of the water properties including its position dependence. 
As described in Ref.~\cite{Abe:2013gga}, the interaction cross section of optical photon in water as a function of the wavelength is characterized with the following three components: absorption ($\sigma_{\mathrm{abs}}$), symmetric scattering  ($\sigma_{\mathrm{sym}}$) and asymmetric forward scattering  ($\sigma_{\mathrm{asym}}$).
The absorption cross section, $\sigma_{\mathrm{abs}}$, is further modeled as a function of time ($t$) and vertical position in the detector ($z$) as,
\begin{multline}
        \sigma_{\mathrm{abs}}(\lambda, z, t) \\ =
    \begin{cases}
        \sigma_{\mathrm{abs}}^{0}(\lambda, t) (1+\beta (t)\cdot z) & (z \geq \mathrm{-11~m}) \\
        \sigma_{\mathrm{abs}}^{0}(\lambda, t) (1 - \beta (t)\cdot 11) &  (z  \leq \mathrm{-11~m})
    \end{cases}.
\end{multline}
Here, the time-dependence of the position-averaged absorption cross section, $ \sigma_{\mathrm{abs}}^{0}(\lambda, t)$ is inferred from the attenuation length measured with the decay-electron sample~\cite{Abe:2016nxk}. 
The coefficient for the $z$ dependence of the absorption, $\beta(t)$, is computed using an empirical function of TBA in percent measured with the auto-Xe calibration system and water attenuation length in centimeters ($L$) as,
\begin{equation}
    \beta(t) = \beta_{0}(L) + \beta_{1}(L) \cdot TBA(t),
\end{equation}
where,
\begin{eqnarray}
\beta_{1}(L) & = & \left(1.04 \times 10^{-10}\right) L^{2} 
    - \left(3.30 \times 10^{-6}\right) L \notag \\
     & + & 0.0203, \\
\beta_{0}(L) & = & 2.42 \times \beta_{1}(L) + 0.00306.
\end{eqnarray}
The coefficients in the above formula were derived using simulated hits from a Ni-Cf source placed at the detector center.

The survival probability of photons which originate from a vertex position $(v_{x},v_{y},v_{z})$ with a given wavelength $\lambda$ 
is estimated by solving the following differential equation.
\begin{equation}
\frac{dN(r)}{dr} = - N(r) (\sigma_{\mathrm{abs}}(\lambda,z,t) 
    + C_{\mathrm{scat}} (\sigma_{\mathrm{sym}}(\lambda) + \sigma_{\mathrm{asym}}(\lambda)),
    \label{eq:wt_diff}
\end{equation}
where  $r = \sqrt{(x-x_{v})^2+(y-y_{v})^2+(z-z_{v})^{2}}$ is the distance from the vertex position along the photons' path, $N(r)$ is number of photons surviving at $r$,
and $C_{\mathrm{scat}}$ is an empirical correction factor to effectively characterize photon loss due to scattering.
Based on a simulation study of mono-energetic electrons uniformly distributed in the detector, $C_{\mathrm{scat}} = 0.6$ is chosen as it minimizes the position dependence of the overall energy scale.

When a photon path is contained either within $z \geq -11$~m or $z \leq -11$~m, 
Eq.~\eqref{eq:wt_diff} can be analytically solved.
The survival probability after integrating over the path length of $r_i$ for the i-th PMT, $p(r_i,\lambda,t)$, is described as,
\begin{equation}
        p(r_i,\lambda,t) = N(r_i,\lambda)/N(0,\lambda) 
        = \exp \left[-r_i\cdot \sigma_{\mathrm{eff}}(r_i,\lambda,t)\right],
\end{equation}
where,
\begin{widetext}
\begin{equation}
    \sigma_{\mathrm{eff}}(r_i,\lambda,t) =
    \begin{cases}
        \sigma_{\mathrm{abs}}^{0}(\lambda,t) \left[ 1 + \beta(t) \cdot \left(z_{v} + \frac{1}{2}(z_i-z_{v})\right) \right]
        +  C_{\mathrm{scat}} (\sigma_{\mathrm{sym}}(\lambda) + \sigma_{\mathrm{asym}}(\lambda))
        & (z_i, z_v \geq \mathrm{-11~m}) \\
        \sigma_{\mathrm{abs}}^{0}(\lambda, t) (1 - \beta (t) \cdot 11) 
        +  C_{\mathrm{scat}} (\sigma_{\mathrm{sym}}(\lambda) + \sigma_{\mathrm{asym}}(\lambda))
        &  (z_i, z_v  \leq \mathrm{-11~m})
\end{cases}.
\end{equation}
\end{widetext}
When the path crosses the $z=-11$~m boundary, this survival probability is calculated separately for  $z \geq -11$~m and $z \leq -11$~m regions. Then they are multiplied together to evaluate the attenuation over the entire path.

Finally, the photon survival probability for the $i$-th PMT, $P_{i}(t)$, is evaluated by integrating 
$p(r_{i},\lambda,t)$ over the wavelength, as,
\begin{equation}
    P_{i} (t)
    = \int_{\lambda_{\mathrm{min}}}^{\lambda_{\mathrm{max}}}
    w_{0}(\lambda) p(r_{i},\lambda,t)
    d\lambda,
\end{equation}
where $w_{0}(\lambda)$ is a weight function that represents the product of the Cherenkov light emission spectrum and the PMT's photon detection efficiency.
In the previous study, this $P_{i} (t)$ was simply $P_{i}^{\mathrm{prev}} (t) = \exp[-r_{i}/L(t)]$ where $L(t)$ is water attenuation length evaluated from the decay-electron sample. The new definition used for this analysis significantly reduces the position dependence of the energy scale. Combined with other improvements described in this section the variation of the energy scale across the detector was improved, characterized by a reduction in the standard deviation from about $1.7\%$ in the previous analyses to about $0.5\%$ level in this analysis.

\section{Signal extraction}\label{sec:app_solfit}

The observed number of solar neutrino signal events are extracted from the final data sample
using an extended maximum likelihood function fit.
The distribution of $\theta_{\rm{Sun}}$ ( which is the angle between the reconstructed direction and the direction from the Sun at that time)
is used from $3.49$~MeV to $19.49$~MeV in kinetic energy.
The likelihood function is defined as Eq.~(\ref{eq:solfit}),
which is formed from the following quantities:
\begin{description}
\item[$N_{\rm{energy}}$]
Total number of energy bins \newline
  The kinetic energy from $3.49$~MeV to $19.49$~MeV is divided into twenty-three bins
  ($N_{\rm{energy}} = 23$). The width of the energy bin is defined as follows:  \newline
  \[
  \left\{ \begin{array}{ll}
    \rm{0.5 \, MeV /\,bin}   & (  \,\, \,3.49  \leq E  <  13.49 \, \rm{MeV}) \\
    \rm{1.0 \, MeV /\,bin}   & ( 13.49  \leq E  <  15.49 \, \rm{MeV}) \\
    \rm{4.0 \, MeV /\,bin}   & ( 15.49  \leq E  <  19.49 \, \rm{MeV})
  \end{array} \right.
  \]

\item[$N_{\mathrm{MSG}_i}$]
Total number of MSG bins \newline
Below an electron kinetic energy of $7.49$~MeV, the MSG parameter (g$_{\rm{MS}}$), which runs from $0.0$ to $1.0$, is 
divided into three bins in each energy bin~($N_{\mathrm{MSG}_i} = 3$). 
In higher energy bins, no separation by the MSG parameter is applied~($N_{\mathrm{MSG}_i} = 1$). 
The width of the MSG bins are defined as follows:  \newline
  \[
  \left\{ \begin{array}{ll}
    \rm{0.00 \, < \, {g}_{\mathrm{MS}}\, <\, 0.35} & ( E  <  7.49 \, \rm{MeV} )\\
    \rm{0.35 \, < \, {g}_{\mathrm{MS}}\, <\, 0.45} & ( E  <  7.49 \, \rm{MeV} )\\
    \rm{0.45 \, < \, {g}_{\mathrm{MS}}\, <\, 1.00} & ( E  <  7.49 \, \rm{MeV} )\\
    \rm{0.00 \, < \, {g}_{\mathrm{MS}}\, <\, 1.00} & ( E  \geq 7.49 \, \rm{MeV} )\\
  \end{array} \right.
  \]

\item[$n_{ij}$] 
Total number of events in the $i$-th energy bin and in the $j$-th MSG bin.
\item[$b_{ij}$]
The background probability density function for $k$-th event in the $i,j$-th bin.
This is obtained from the $\phi$ and $\theta$ distribution of data. For the day/night amplitude fit, this function is evaluated separately in five different solar zenith angle regions for the day and six different regions for the night.

\item[$s_{ij}$]
The signal probability density function evaluated for the $k$-th event in the $i,j$-th bin.
This function is extracted by the
solar neutrino MC simulation.

\item {$Y_{ij}$:}
The fraction of signal events in the $j$-th MSG bin
  in the $i$-th energy bin based on the MC simulation.
  This is obtained from the signal energy spectrum, 
  which is the energy distribution of the solar neutrino MC final sample.
  The signal fraction is calculated from the number of signal events of each energy and MSG 
  bin divided by the total number of signal events.
\item {$B_{ij}$:}
Free parameter corresponding to the number of background events
  in the $i$-th energy bin and the $j$-th MSG bin.
  
\item {$S$:}
Free parameter corresponding to the total number of solar
  neutrino events in all energy and MSG bins.
\end{description}

The fitting parameters, $S$ and $B_{ij}$, represent the number of signal and background events, respectively. These parameters are obtained by maximizing the likelihood function.

The recoil electrons from solar neutrinos undergo multiple Coulomb scattering in water.
This multiple scattering distorts the pattern of Cherenkov rings.
For the higher energy region, the true energy range of the spallation background and solar neutrino events is similar, 
so multiple scattering has a similar impact on signal and background events.
But when comparing events in a lower reconstructed energy region, the impact of multiple scattering is large 
for beta-decay backgrounds, such as bismuth-214~\cite{Nakano:2019bnr}, because their true energy is often lower than the solar neutrino signal.
Therefore we have introduced the MSG binning in Eq.~(\ref{eq:solfit}).

\begin{figure*}[p]
		\centering
\includegraphics[width=1.0\textwidth]{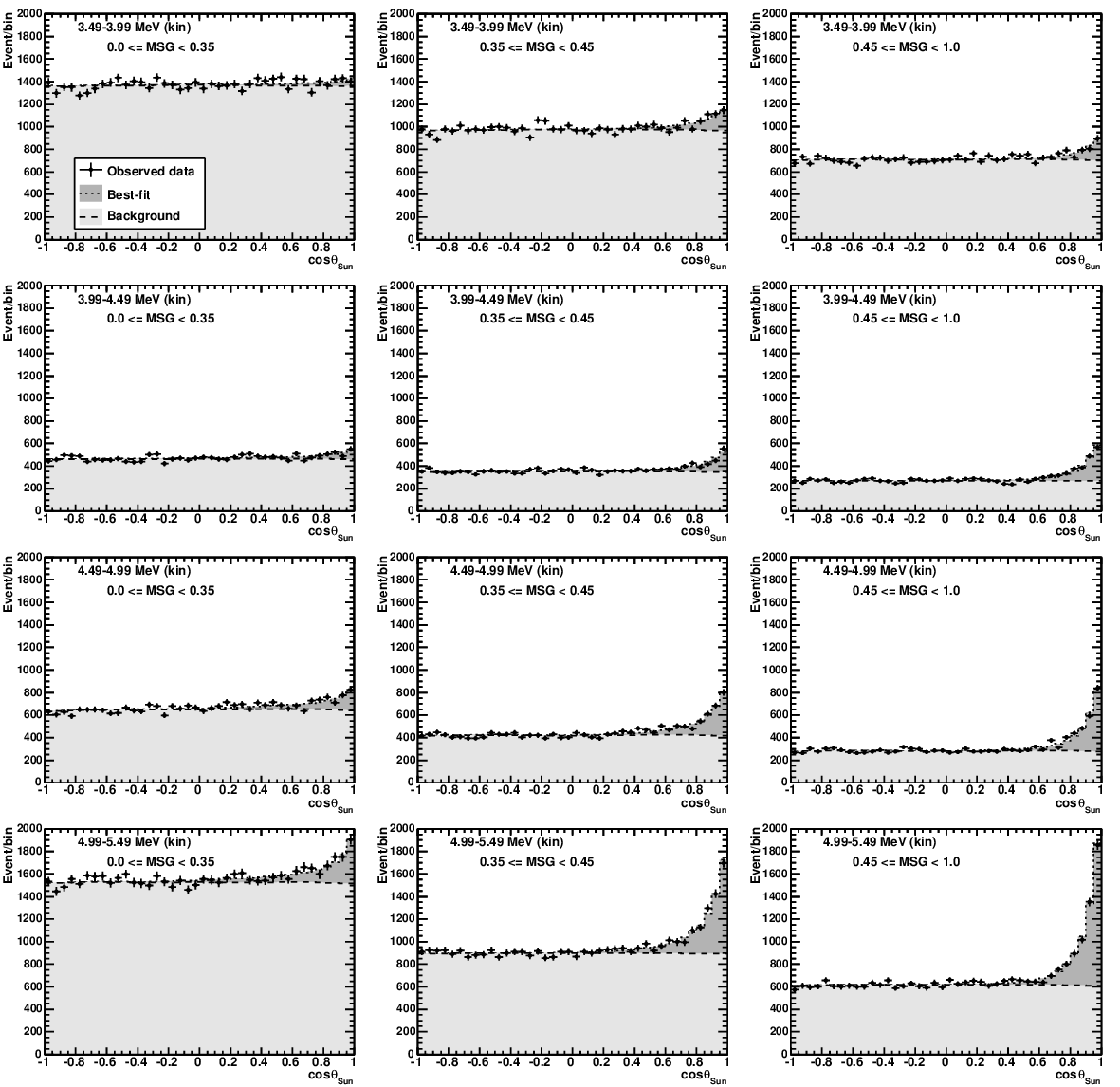}
	  \caption{Solar angle distribution for the MSG sub-groups below $5.49$~MeV. 
	Below $4.99$~MeV, the tight fiducial volume cut is applied. 
	\label{fig:solar_dist_msg1}}
\end{figure*}

\begin{figure*}[p]
		\centering
	\includegraphics[width=1.0\textwidth]{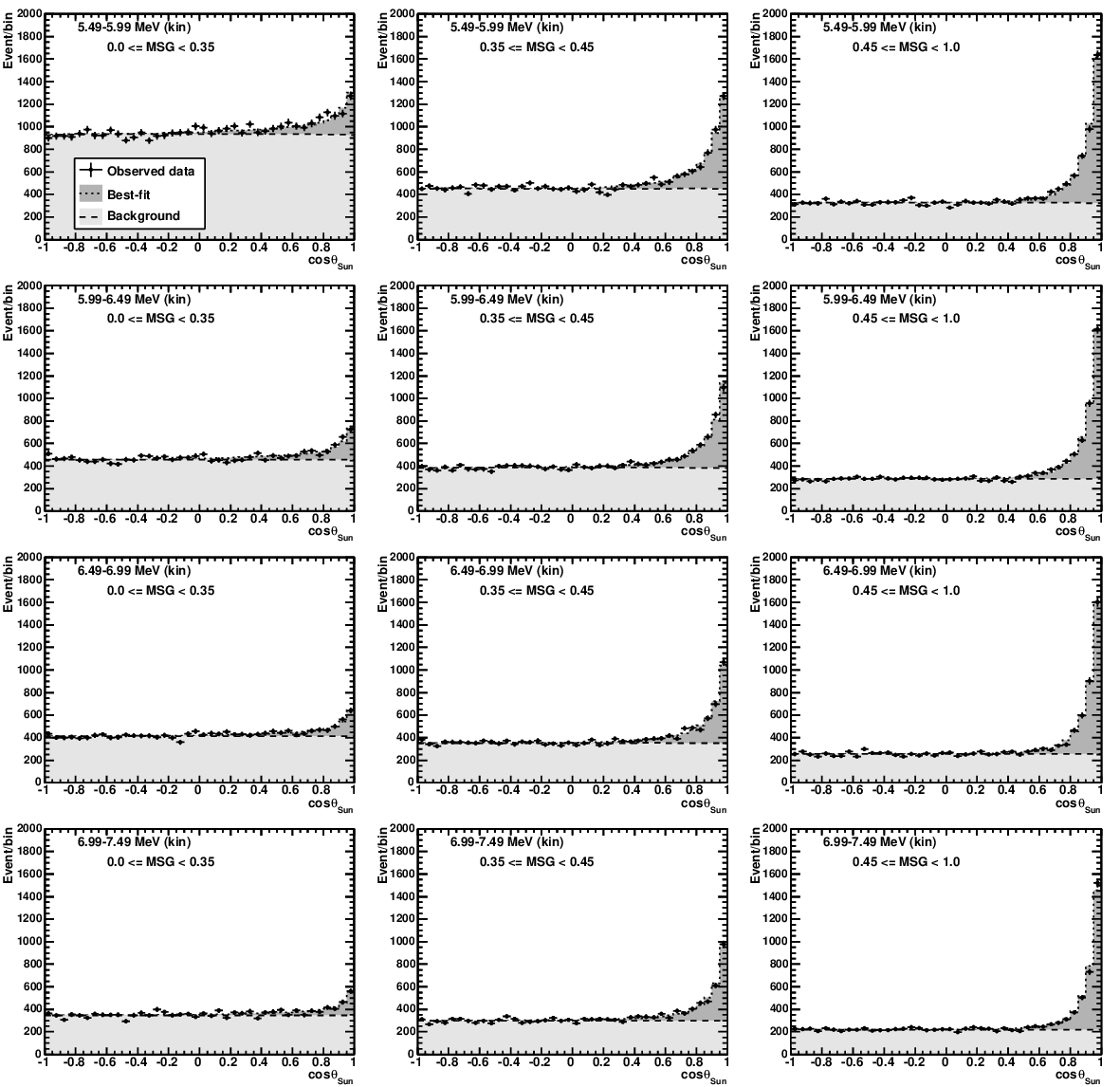}
    \caption{Solar angle distribution for the MSG sub-groups above $5.49$~MeV and below $7.49$~MeV.  
	\label{fig:solar_dist_msg2}}
\end{figure*}

Figure~\ref{fig:solar_dist_msg1} and Figure~\ref{fig:solar_dist_msg2} show the solar angle distributions of MSG sub-groups. 
As expected, the peak corresponding to the solar neutrino signal is more prominent in the high MSG-value bins but much smaller in the low MSG-value bins.
This tendency is very clear in the low-energy samples.

\begin{figure*}[p]
		\centering
	\includegraphics[width=1.0\textwidth]{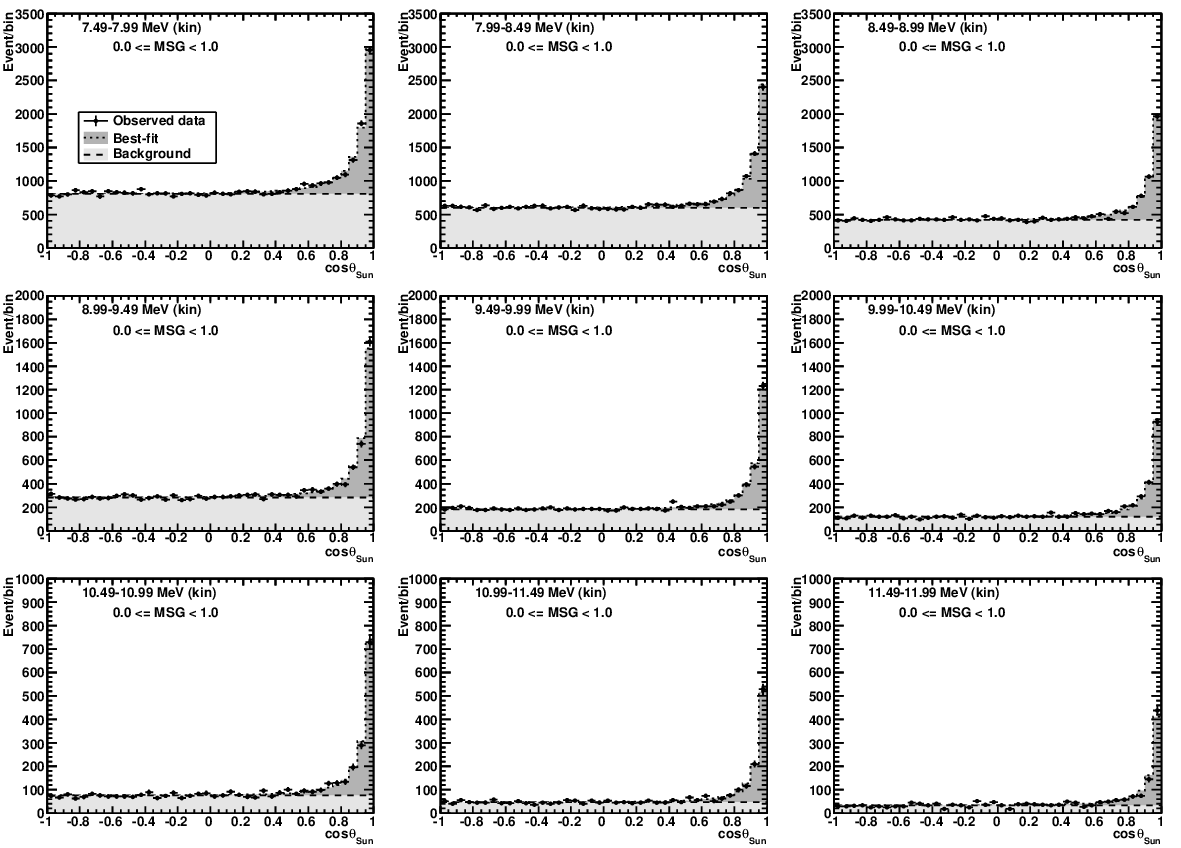}
  	\caption{Solar angle distribution for the energy region above $7.49$~MeV and below $12.49$~MeV. 
	In this energy region, categorization by the MSG parameter is not used.
	\label{fig:solar_dist_nomsg1}}
\end{figure*}

\begin{figure*}[p]
		\centering
		\includegraphics[width=1.0\textwidth]{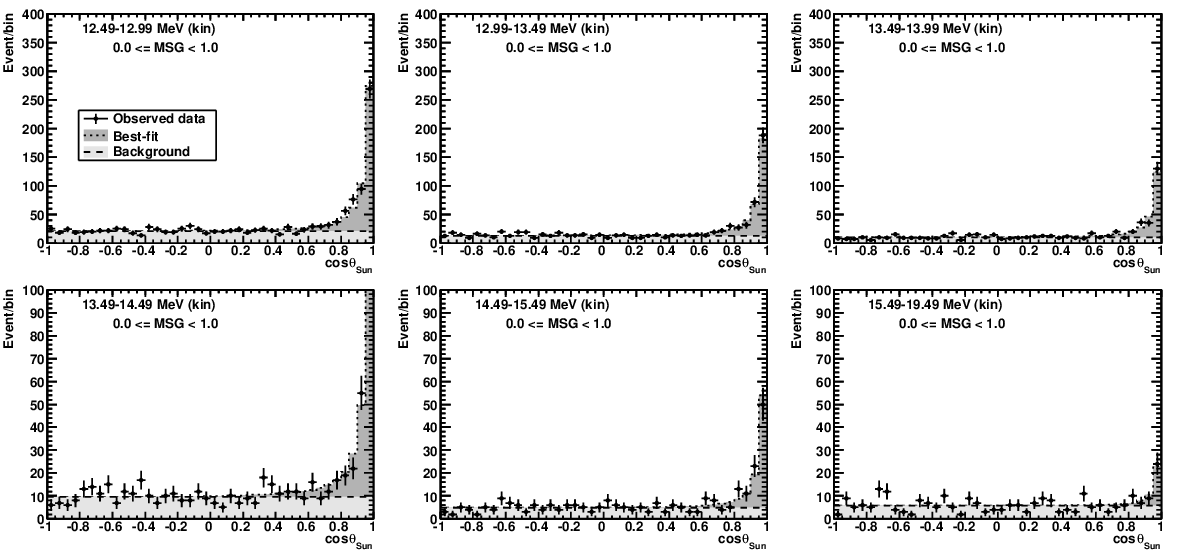}
	\caption{Solar angle distribution for the energy region above $12.49$~MeV. 
	In this energy region, categorization by the MSG parameter is not used.
	\label{fig:solar_dist_nomsg2}}
\end{figure*}

Figure~\ref{fig:solar_dist_nomsg1} and Figure~\ref{fig:solar_dist_nomsg2} also show the solar angle distribution of the observed events above $7.49$~MeV, 
where the MSG sub-group is not used.

\section{Supplementary explanation of the solar neutrino results}
\label{sec:app_results}

\subsection{Spectrum results from SK}

The expected rate of the solar neutrino interactions throughout the detector is estimated by considering the solar neutrino flux and the detector performance. The observed rate and the expected rate are summarized in Table~\ref{tab:obs-rate}. The ratios of the observed rate to the expected rate are summarized in Table~\ref{tab:data-mc}.

\begin{table*}[h]
\caption{
The observed and expected event rates in each
 energy bin at 1~AU. 
 The unit of the rates is event/kton/year.
 The errors are statistical errors only. The
 $\mathrm{^{8}B}$ and $hep$ solar neutrino fluxes are assumed to be
 $5.25\times10^{6}$~$\mathrm{cm^{-2}s^{-1}}$ and $7.88\times10^{3}$~$\mathrm{cm^{-2}s^{-1}}$, respectively. 
 A correction is applied for the signal efficiency shown in Fig.~\ref{fig:mc_reduc_eff}. The solar zenith angle~$(\theta_{\mathrm{z,solar}})$ is defined in Sec.~\ref{sec_dn}.}  
\label{tab:obs-rate}
\begin{center}
\begin{tabular}{l@{\quad}ccc@{\quad}cr} \hline \hline
Energy & \multicolumn{3}{c}{Observed rate} & \multicolumn{2}{c}{Expected rate} \\
{}[MeV (kin)]& ALL & DAY & NIGHT & $\mathrm{^{8}B}$ & $hep$ \\
& $ -1 \leq \cos\theta_{z,solar} \leq 1 $   & $ -1 \leq \cos\theta_{z,solar} \leq 0 $  & $  0 < \cos\theta_{z,solar} \leq 1 $  & &\\
\hline
$3.49$--$3.99$ & $  94.1^{ +8.4}_{ -8.3}$ & $ 100.2^{+13.3}_{-13.0}$ & $  89.0^{+10.8}_{-10.6}$ & $  197.2$ & $  0.347$ \\
$3.99$--$4.49$ & $  82.6^{ +3.9}_{ -3.9}$ & $  77.5^{ +6.0}_{ -5.8}$ & $  86.7^{ +5.3}_{ -5.1}$ & $  183.1$ & $  0.336$ \\
$4.49$--$4.99$ & $  80.5^{ +2.4}_{ -2.4}$ & $  76.9^{ +3.6}_{ -3.5}$ & $  83.4^{ +3.3}_{ -3.2}$ & $  168.6$ & $  0.324$ \\
$4.99$--$5.49$ & $  69.7^{ +1.5}_{ -1.4}$ & $  67.2^{ +2.1}_{ -2.1}$ & $  72.0^{ +2.0}_{ -2.0}$ & $  154.0$ & $  0.313$ \\
$5.49$--$5.99$ & $  61.4^{ +1.1}_{ -1.1}$ & $  62.7^{ +1.6}_{ -1.6}$ & $  60.1^{ +1.6}_{ -1.5}$ & $  138.5$ & $  0.299$ \\
$5.99$--$6.49$ & $  54.4^{ +1.0}_{ -1.0}$ & $  54.8^{ +1.5}_{ -1.4}$ & $  54.1^{ +1.4}_{ -1.4}$ & $  122.7$ & $  0.283$ \\
$6.49$--$6.99$ & $  48.3^{ +0.9}_{ -0.9}$ & $  48.0^{ +1.3}_{ -1.3}$ & $  48.5^{ +1.3}_{ -1.3}$ & $  107.5$ & $  0.267$ \\
$6.99$--$7.49$ & $  41.1^{ +0.8}_{ -0.8}$ & $  41.4^{ +1.2}_{ -1.2}$ & $  40.9^{ +1.1}_{ -1.1}$ & $   92.7$ & $  0.250$ \\
$7.49$--$7.99$ & $  35.3^{ +0.7}_{ -0.7}$ & $  34.0^{ +1.1}_{ -1.0}$ & $  36.5^{ +1.0}_{ -1.0}$ & $   78.7$ & $  0.233$ \\
$7.99$--$8.49$ & $  28.7^{ +0.6}_{ -0.6}$ & $  27.8^{ +0.9}_{ -0.9}$ & $  29.5^{ +0.9}_{ -0.9}$ & $   65.6$ & $  0.215$ \\
$8.49$--$8.99$ & $  23.2^{ +0.6}_{ -0.5}$ & $  23.0^{ +0.8}_{ -0.8}$ & $  23.3^{ +0.8}_{ -0.7}$ & $   53.9$ & $  0.197$ \\
$8.99$--$9.49$ & $  18.4^{ +0.5}_{ -0.5}$ & $  18.3^{ +0.7}_{ -0.7}$ & $  18.5^{ +0.7}_{ -0.6}$ & $   43.2$ & $  0.180$ \\
$9.49$--$9.99$ & $  14.3^{ +0.4}_{ -0.4}$ & $  13.6^{ +0.6}_{ -0.6}$ & $  15.0^{ +0.6}_{ -0.6}$ & $   34.0$ & $  0.162$ \\
$9.99$--$10.49$ & $  11.3^{ +0.4}_{ -0.3}$ & $  11.7^{ +0.5}_{ -0.5}$ & $  10.9^{ +0.5}_{ -0.5}$ & $   26.2$ & $  0.145$ \\
$10.49$--$10.99$ & $8.60^{+0.30}_{-0.29}$ & $7.99^{+0.43}_{-0.41}$ & $9.16^{+0.43}_{-0.41}$ & $  19.75$ & $  0.128$ \\
$10.99$--$11.49$ & $6.17^{+0.25}_{-0.24}$ & $6.20^{+0.36}_{-0.34}$ & $6.14^{+0.35}_{-0.33}$ & $  14.44$ & $  0.112$ \\
$11.49$--$11.99$ & $4.76^{+0.21}_{-0.20}$ & $4.87^{+0.31}_{-0.29}$ & $4.66^{+0.30}_{-0.28}$ & $  10.35$ & $  0.097$ \\
$11.99$--$12.49$ & $3.13^{+0.17}_{-0.16}$ & $2.96^{+0.25}_{-0.23}$ & $3.30^{+0.25}_{-0.23}$ & $   7.16$ & $  0.083$ \\
$12.49$--$12.99$ & $2.07^{+0.14}_{-0.13}$ & $1.81^{+0.20}_{-0.18}$ & $2.33^{+0.21}_{-0.19}$ & $   4.84$ & $  0.071$ \\
$12.99$--$13.49$ & $1.39^{+0.12}_{-0.11}$ & $1.34^{+0.17}_{-0.15}$ & $1.45^{+0.17}_{-0.15}$ & $   3.15$ & $  0.059$ \\
$13.49$--$14.49$ & $1.54^{+0.12}_{-0.11}$ & $1.55^{+0.18}_{-0.16}$ & $1.54^{+0.17}_{-0.15}$ & $   3.22$ & $  0.088$ \\
$14.49$--$15.49$ & $0.57^{+0.08}_{-0.07}$ & $0.59^{+0.12}_{-0.10}$ & $0.57^{+0.11}_{-0.09}$ & $   1.13$ & $  0.056$ \\
$15.49$--$19.49$ & $0.18^{+0.05}_{-0.04}$ & $0.16^{+0.08}_{-0.06}$ & $0.21^{+0.08}_{-0.06}$ & $   0.46$ & $  0.064$ \\
\hline\hline
\end{tabular}
\end{center}
\end{table*}

\begin{table*}[h]
\caption{Elastic scattering rate ratios and energy-uncorrelated
 uncertainties~(statistical plus systematic) for each SK
 phase. \label{tab:data-mc}} 
\centerline{\begin{tabular}{lc@{\qquad}c@{\qquad}c@{\qquad}c}
\hline\hline
Energy~[MeV~(kin)] & SK-I & SK-II & SK-III & SK-IV \\ \hline
$3.49$--$3.99$   & ---                        & ---                        & ---                        &  $0.476^{+0.049}_{-0.048}$  \\
$3.99$--$4.49$   & ---                        & ---                        & $0.448^{+0.100}_{-0.096}$ &  $0.450\pm0.024$  \\
$4.49$--$4.99$   & $0.453^{+0.043}_{-0.042}$ & ---                        & $0.472^{+0.058}_{-0.056}$ &  $0.476\pm0.018$  \\
$4.99$--$5.49$   & $0.430^{+0.023}_{-0.022}$ & ---                        & $0.420^{+0.039}_{-0.037}$ &  $0.452\pm0.011$  \\
$5.49$--$5.99$   & $0.449{\pm}0.018$         & ---                        & $0.457^{+0.035}_{-0.034}$ &  $0.442\pm0.009$  \\
$5.99$--$6.49$   & $0.444{\pm}0.015$         & ---                        & $0.433^{+0.023}_{-0.022}$ &  $0.442\pm0.010$  \\
$6.49$--$6.99$   & $0.461^{+0.016}_{-0.015}$ & $0.439^{+0.050}_{-0.048}$ & $0.504^{+0.025}_{-0.024}$ &  $0.448\pm0.011$  \\
$6.99$--$7.49$   & $0.476{\pm}0.016$         & $0.448^{+0.043}_{-0.041}$ & $0.424^{+0.024}_{-0.023}$ &  $0.443^{+0.012}_{-0.011}$  \\
$7.49$--$7.99$   & $0.457^{+0.017}_{-0.016}$ & $0.461^{+0.037}_{-0.036}$ & $0.467^{+0.024}_{-0.023}$ &  $0.448\pm0.010$  \\
$7.99$--$8.49$   & $0.431^{+0.017}_{-0.016}$ & $0.473^{+0.036}_{-0.035}$ & $0.469^{+0.026}_{-0.025}$ &  $0.435\pm0.010$  \\
$8.49$--$8.99$   & $0.454^{+0.018}_{-0.017}$ & $0.463^{+0.036}_{-0.034}$ & $0.420^{+0.026}_{-0.025}$ &  $0.429\pm0.011$  \\
$8.99$--$9.49$   & $0.464{\pm}0.019$         & $0.499^{+0.038}_{-0.037}$ & $0.444^{+0.029}_{-0.027}$ &  $0.424^{+0.012}_{-0.011}$  \\
$9.49$--$9.99$   & $0.456^{+0.021}_{-0.020}$ & $0.474^{+0.038}_{-0.036}$ & $0.423^{+0.031}_{-0.029}$ &  $0.420^{+0.012}_{-0.012}$  \\
$9.99$--$10.49$  & $0.409{\pm}0.021$         & $0.481^{+0.041}_{-0.039}$ & $0.529^{+0.037}_{-0.035}$ &  $0.429\pm0.014$  \\
$10.49$--$10.99$ & $0.472^{+0.025}_{-0.024}$ & $0.452^{+0.043}_{-0.040}$ & $0.481^{+0.041}_{-0.037}$ &  $0.433^{+0.016}_{-0.015}$  \\
$10.99$--$11.49$ & $0.439^{+0.028}_{-0.026}$ & $0.469^{+0.046}_{-0.043}$ & $0.391^{+0.044}_{-0.040}$ &  $0.424^{+0.018}_{-0.017}$  \\
$11.49$--$11.99$ & $0.460^{+0.033}_{-0.031}$ & $0.482^{+0.052}_{-0.048}$ & $0.479^{+0.055}_{-0.049}$ &  $0.456^{+0.021}_{-0.020}$  \\
$11.49$--$12.49$ & $0.465^{+0.039}_{-0.036}$ & $0.419^{+0.054}_{-0.049}$ & $0.425^{+0.061}_{-0.053}$ &  $0.433^{+0.024}_{-0.023}$  \\
$12.49$--$12.99$ & $0.461^{+0.048}_{-0.043}$ & $0.462^{+0.063}_{-0.057}$ & $0.400^{+0.073}_{-0.061}$ &  $0.421^{+0.029}_{-0.027}$  \\
$12.99$--$13.49$ & $0.582^{+0.064}_{-0.057}$ & $0.444^{+0.070}_{-0.062}$ & $0.422^{+0.093}_{-0.074}$ &  $0.434^{+0.037}_{-0.034}$  \\
$13.49$--$14.49$ & $0.475^{+0.059}_{-0.052}$ & $0.430^{+0.066}_{-0.059}$ & $0.663^{+0.110}_{-0.093}$ &  $0.465^{+0.037}_{-0.034}$  \\
$14.49$--$15.49$ & $0.724^{+0.120}_{-0.102}$ & $0.563^{+0.100}_{-0.087}$ & $0.713^{+0.201}_{-0.150}$ &  $0.483^{+0.066}_{-0.058}$  \\
$15.49$--$19.49$ & $0.575^{+0.173}_{-0.130}$ & $0.648^{+0.123}_{-0.103}$ & $0.212^{+0.248}_{-0.122}$ &  $0.349^{+0.100}_{-0.080}$  \\
\hline\hline
\end{tabular}}
\end{table*}

\clearpage

\bibliography{001_references}

\end{document}